\documentclass[twocolumn,tighten,trackchanges]{aastex62}
\usepackage{CJKutf8}
\usepackage{amsmath}
\usepackage{enumitem}
\usepackage{booktabs,pbox}
\usepackage{multirow}

\graphicspath{{./figures/}{../figures/}}

\DeclareMathOperator{\sech}{sech}

\newcommand{\sbullet}{\mathbin{\vcenter{\hbox{\scalebox{0.7}{$\bullet$}}}}}

\newcommand{\SfourG}{\mbox{$\mathrm{S^4G}$}}

\newcommand{\HI}{\mbox{\textsc{Hi}}}

\newcommand{\CO}[2]{\mbox{$\mathrm{CO}\,(#1\text{--}#2)$}}

\newcommand{\sbsc}[1]{_\mathrm{#1}}
\newcommand{\spsc}[1]{^\mathrm{#1}}
\newcommand{\brkt}[1]{\left<#1\right>}
\newcommand{\brktkpc}[1]{\left<#1\right>_\mathrm{1kpc}}

\newcommand{\ICOpc}{I\sbsc{CO,\,\theta pc}}
\newcommand{\ICOkpc}{I\sbsc{CO,\,1kpc}}

\newcommand{\fCO}{f\sbsc{CO}}
\newcommand{\fCOpc}{f\sbsc{CO,\,\theta pc}}

\newcommand{\sigCOpc}{\sigma\sbsc{CO,\,\theta pc}}

\newcommand{\alphaCO}{\alpha\sbsc{CO}}
\newcommand{\Sigmol}{\Sigma\sbsc{mol}}
\newcommand{\Sigmolpc}{\Sigma\sbsc{mol,\,\theta pc}}
\newcommand{\Sigmolkpc}{\Sigma\sbsc{mol,\,1kpc}}

\newcommand{\Sigmolavgkpc}{\brktkpc{\Sigma\sbsc{mol,\,\theta pc}}}
\newcommand{\sigmol}{\sigma\sbsc{mol}}
\newcommand{\sigmolpc}{\sigma\sbsc{mol,\,\theta pc}}

\newcommand{\sigmolavgkpc}{\brktkpc{\sigma\sbsc{mol,\,\theta pc}}}

\newcommand{\rhomol}{\rho\sbsc{mol}}
\newcommand{\Rcloud}{R\sbsc{cloud}}
\newcommand{\Pturb}{P\sbsc{turb}}
\newcommand{\Pturbpc}{P\sbsc{turb,\,\theta pc}}
\newcommand{\Pturbavg}{\brkt{P\sbsc{turb,\,\theta pc}}}
\newcommand{\Pturbavgkpc}{\brktkpc{P\sbsc{turb,\,\theta pc}}}

\newcommand{\SigSFR}{\Sigma\sbsc{SFR}}
\newcommand{\SigSFRkpc}{\Sigma\sbsc{SFR,\,1kpc}}
\newcommand{\Sigatom}{\Sigma\sbsc{atom}}
\newcommand{\Sigatomkpc}{\Sigma\sbsc{atom,\,1kpc}}

\newcommand{\Siggaskpc}{\Sigma\sbsc{gas,\,1kpc}}
\newcommand{\Sigstar}{\Sigma\sbsc{\star}}
\newcommand{\Sigstarkpc}{\Sigma\sbsc{\star,\,1kpc}}
\newcommand{\rhostar}{\rho\sbsc{\star}}
\newcommand{\rhostarkpc}{\rho\sbsc{\star,\,1kpc}}
\newcommand{\Hstar}{H\sbsc{\star}}
\newcommand{\Rstar}{R\sbsc{\star}}
\newcommand{\Reff}{R\sbsc{e}}
\newcommand{\rgal}{r\sbsc{gal}}

\newcommand{\sigatom}{\sigma\sbsc{atom}}
\newcommand{\siggasz}{\sigma\sbsc{gas,\,z}}
\newcommand{\PDE}{P\sbsc{DE}}
\newcommand{\PDEkpc}{P\sbsc{DE,\,1kpc}}
\newcommand{\PDEavg}{\brkt{P\sbsc{DE,\,\theta pc}}}
\newcommand{\PDEavgkpc}{\brktkpc{P\sbsc{DE,\,\theta pc}}}

\newcommand{\uI}{\mbox{$\rm MJy\,sr^{-1}$}}
\newcommand{\uIco}{\mbox{$\rm K\,km\,s^{-1}$}}
\newcommand{\uV}{\mbox{$\rm km\,s^{-1}$}}
\newcommand{\uM}{\mbox{$\rm M_{\odot}$}}
\newcommand{\uSig}{\mbox{$\rm M_{\odot}\,pc^{-2}$}}

\newcommand{\uSigSFR}{\mbox{$\rm M_{\odot}\,yr^{-1}\,kpc^{-2}$}}
\newcommand{\uP}{\mbox{$k_\mathrm{B}\rm\,K\,cm^{-3}$}}

\shorttitle{Dynamical Equilibrium in the Molecular ISM}
\shortauthors{Sun, Leroy, Ostriker, and the PHANGS Collaboration}

\begin{document}


\title{Dynamical Equilibrium in the Molecular ISM in 28 Nearby Star-Forming Galaxies}


\correspondingauthor{Jiayi Sun}
\email{sun.1608@osu.edu}

\newcommand{\OSU}{\affil{Department of Astronomy, The Ohio State University, 140 West 18th Avenue, Columbus, Ohio 43210, USA}}
\newcommand{\Princeton}{\affil{Department of Astrophysical Sciences, Princeton University, Princeton, NJ 08544 USA}}
\newcommand{\CNRS}{\affil{CNRS, IRAP, 9 Av. du Colonel Roche, BP 44346, F-31028 Toulouse cedex 4, France}}
\newcommand{\Toulouse}{\affil{Universit\'{e} de Toulouse, UPS-OMP, IRAP, F-31028 Toulouse cedex 4, France}}
\newcommand{\UAlberta}{\affil{Department of Physics, University of Alberta, Edmonton, AB T6G 2E1, Canada}}
\newcommand{\MPE}{\affil{Max-Planck-Institut f\"{u}r extraterrestrische Physik, Giessenbachstra{\ss}e 1, D-85748 Garching, Germany}}
\newcommand{\MPIA}{\affil{Max-Planck-Institut f\"{u}r Astronomie, K\"{o}nigstuhl 17, D-69117, Heidelberg, Germany}}
\newcommand{\UGent}{\affil{Sterrenkundig Observatorium, Universiteit Gent, Krijgslaan 281 S9, B-9000 Gent, Belgium}}
\newcommand{\ARI}{\affil{Astronomisches Rechen-Institut, Zentrum f\"{u}r Astronomie der Universit\"{a}t Heidelberg,\\ M\"{o}nchhofstra\ss e 12-14, D-69120 Heidelberg, Germany}}
\newcommand{\ITA}{\affil{Institut f\"{u}r Theoretische Astrophysik, Zentrum f\"{u}r Astronomie der Universit\"{a}t Heidelberg,\\ Albert-Ueberle-Str 2, D-69120 Heidelberg, Germany}}
\newcommand{\ESO}{\affil{European Southern Observatory, Karl-Schwarzschild Stra{\ss}e 2, D-85748 Garching bei M\:{u}nchen, Germany}}
\newcommand{\OAN}{\affil{Observatorio Astron\'{o}mico Nacional (IGN), C/Alfonso XII, 3, E-28014 Madrid, Spain}}
\newcommand{\UBonn}{\affil{Argelander-Institut f\"ur Astronomie, Universit\"at Bonn, Auf dem H\"ugel 71, 53121 Bonn, Germany}}
\newcommand{\ULyon}{\affil{Univ Lyon, Univ Lyon 1, ENS de Lyon, CNRS, Centre de Recherche Astrophysique de Lyon UMR5574,\\ F-69230 Saint-Genis-Laval, France}}
\newcommand{\CfA}{\affil{Harvard-Smithsonian Center for Astrophysics, 60 Garden
Street, Cambridge, MA 02138, USA}}
\newcommand{\UCSD}{\affil{Center for Astrophysics and Space Sciences, Department of Physics,  University of California,\\ San Diego, 9500 Gilman Drive, La Jolla, CA 92093, USA}}
\newcommand{\Caltech}{\affil{Infrared Processing and Analysis Center (IPAC), California Institute of Technology, Pasadena, CA 91125, USA}}
\newcommand{\Carnegie}{\affil{Observatories of the Carnegie Institution for Science, 813 Santa Barbara Street, Pasadena, CA 91101, USA}}
\newcommand{\UWyoming}{\affil{Department of Physics \& Astronomy, University of Wyoming, Laramie, WY 82071, USA}}
\newcommand{\UChile}{\affil{Departamento de Astronom\'{i}a, Universidad de Chile, Camino del Observatorio 1515, Las Condes, Santiago, Chile}}
\newcommand{\ANU}{\affil{Research School of Astronomy and Astrophysics, Australian National University, Canberra, ACT 2611, Australia}}
\newcommand{\IRAM}{\affil{Institut de Radioastronomie Millim\'{e}trique (IRAM), 300 Rue de la Piscine, F-38406 Saint Martin d'H\`{e}res, France}}
\newcommand{\ObsParis}{\affil{Sorbonne Universit\'{e}, Observatoire de Paris, Universit\'{e} PSL, CNRS, LERMA, F-75014, Paris, France}}
\newcommand{\UMaryland}{\affil{Department of Astronomy, University of Maryland, College Park, MD 20742, USA}}

\author[0000-0003-0378-4667]{Jiayi~Sun \begin{CJK*}{UTF8}{gbsn}(孙嘉懿)\end{CJK*}}
\OSU

\author[0000-0002-2545-1700]{Adam~K.~Leroy}
\OSU

\author[0000-0002-0509-9113]{Eve~C.~Ostriker}
\Princeton

\author[0000-0002-9181-1161]{Annie~Hughes}
\CNRS
\Toulouse

\author[0000-0002-5204-2259]{Erik~Rosolowsky}
\UAlberta

\author{Andreas~Schruba}
\MPE

\author[0000-0002-3933-7677]{Eva~Schinnerer}
\MPIA

\author[0000-0003-4218-3944]{Guillermo~A.~Blanc}
\Carnegie
\UChile

\author[0000-0001-5310-467X]{Christopher~Faesi}
\MPIA

\author[0000-0002-8804-0212]{J.~M.~Diederik~Kruijssen}
\ARI

\author[0000-0002-6118-4048]{Sharon~Meidt}
\UGent

\author[0000-0003-4161-2639]{Dyas~Utomo}
\OSU

\author[0000-0003-0166-9745]{Frank~Bigiel}
\UBonn

\author[0000-0002-5480-5686]{Alberto~D.~Bolatto}
\UMaryland

\author[0000-0002-5635-5180]{M\'{e}lanie~Chevance}
\ARI

\author[0000-0003-2551-7148]{I-Da~Chiang \begin{CJK*}{UTF8}{bsmi}(江宜達)\end{CJK*}}
\UCSD

\author[0000-0002-5782-9093]{Daniel~Dale}
\UWyoming

\author[0000-0002-6155-7166]{Eric~Emsellem}
\ESO
\ULyon

\author[0000-0001-6708-1317]{Simon~C.~O.~Glover}
\ITA

\author[0000-0002-3247-5321]{Kathryn~Grasha}
\ANU

\author[0000-0001-9656-7682]{Jonathan~Henshaw}
\MPIA

\author{Cinthya~N.~Herrera}
\IRAM

\author[0000-0002-9165-8080]{Maria~Jesus~Jimenez-Donaire}
\CfA

\author[0000-0002-2278-9407]{Janice~C.~Lee}
\Caltech

\author[0000-0003-3061-6546]{J\'er\^ome~Pety}
\IRAM
\ObsParis

\author[0000-0002-0472-1011]{Miguel~Querejeta}
\ESO
\OAN

\author{Toshiki~Saito}
\MPIA

\author[0000-0002-4378-8534]{Karin~Sandstrom}
\UCSD

\author[0000-0003-1242-505X]{Antonio~Usero}
\OAN


\begin{abstract}

    We compare the observed turbulent pressure in molecular gas, $P_\mathrm{turb}$, to the required pressure for the interstellar gas to stay in equilibrium in the gravitational potential of a galaxy, $P_\mathrm{DE}$.
    To do this, we combine arcsecond resolution CO data from PHANGS-ALMA with multi-wavelength data that traces the atomic gas, stellar structure, and star formation rate (SFR) for 28 nearby star-forming galaxies.
    We find that $P_\mathrm{turb}$ correlates with, but almost always exceeds the estimated $P_\mathrm{DE}$ on kiloparsec scales. This indicates that the molecular gas is over-pressurized relative to the large-scale environment.
    We show that this over-pressurization can be explained by the clumpy nature of molecular gas; a revised estimate of $P_\mathrm{DE}$ on cloud scales, which accounts for molecular gas self-gravity, external gravity, and ambient pressure, agrees well with the observed $P_\mathrm{turb}$ in galaxy disks.
    We also find that molecular gas with cloud-scale ${P_\mathrm{turb}}\approx{P_\mathrm{DE}}\gtrsim{10^5\,k_\mathrm{B}\,\mathrm{K\,cm^{-3}}}$ in our sample is more likely to be self-gravitating, whereas gas at lower pressure appears more influenced by ambient pressure and/or external gravity.
    Furthermore, we show that the ratio between $P_\mathrm{turb}$ and the observed SFR surface density, $\Sigma_\mathrm{SFR}$, is compatible with stellar feedback-driven momentum injection in most cases, while a subset of the regions may show evidence of turbulence driven by additional sources.
    The correlation between $\Sigma_\mathrm{SFR}$ and kpc-scale $P_\mathrm{DE}$ in galaxy disks is consistent with the expectation from self-regulated star formation models.
    Finally, we confirm the empirical correlation between molecular-to-atomic gas ratio and kpc-scale $P_\mathrm{DE}$ reported in previous works.

\end{abstract}




\section{Introduction} \label{sec:intro}

Molecular clouds host a significant fraction of the molecular gas mass in the interstellar medium (ISM), and all star formation activity in galaxies. Understanding how the properties of molecular clouds change in response to the galactic environment is crucial for building a successful theory for star formation.

Early studies of giant molecular clouds (GMCs) in the Milky Way conjectured that they had ``universal'' properties, in the sense that all GMCs exist at the same surface density and follow the same size--linewidth scaling relation \citep[e.g.,][]{Larson_1981,Solomon_etal_1987}. \citet{Bolatto_etal_2008} also reached a similar conclusion after studying a sample of GMCs in some of the nearest galaxies.
However, subsequent observational studies have instead suggested that cloud properties may change systematically as a function of environment. In a careful re-analysis of the \citet{Solomon_etal_1987} Milky Way clouds, \citet{Heyer_etal_2009} demonstrated that Milky Way GMCs vary systematically in their surface density and size--linewidth parameter. This finding has been confirmed by more recent works analyzing GMCs in the Milky Way \citep{Rice_etal_2016,Miville-Deschenes_etal_2017,Colombo_etal_2019}.
Based on observations from the PdBI Arcsecond Whirlpool Survey \citep[PAWS;][]{Pety_etal_2013,Schinnerer_etal_2013}, \citet{Hughes_etal_2013a}, \citet{Hughes_etal_2013b}, and \cite{Colombo_etal_2014a} showed strong variations in GMC properties within the inner $\sim4$\,kpc of M51 and among M51, M33, and the Large Magellanic Cloud (LMC). Similar results were found by subsequent works studying GMCs in various types of galaxies in the local universe \citep[e.g.,][]{DonovanMeyer_etal_2013,Leroy_etal_2015,Rebolledo_etal_2015,Utomo_etal_2015,Freeman_etal_2017,Egusa_etal_2018,Hirota_etal_2018,Schruba_etal_2019}.

In a comprehensive work studying molecular gas properties with uniform treatment across 15 nearby galaxies, \citet{Sun_etal_2018} showed that the cloud-scale surface density and velocity dispersion of molecular gas vary systematically both within and among galaxies. The correlation of these two quantities implies a narrow range of virial parameter and a wide spread in the internal turbulent pressure in molecular gas. Quantitatively, \citet{Sun_etal_2018} found the cloud-scale turbulent pressure of molecular gas to span $\sim 4{-}5$ orders of magnitude.
An important next step for the field is to understand the physics that drive the changes in the turbulent gas pressure in different galactic environments.

One promising scenario is that the mean pressure in the ISM sets the internal pressure in molecular clouds. This idea has been considered in various forms to explain observed molecular cloud properties since early CO studies \citep[][]{Keto_Myers_1986,Elmegreen_1989,Bertoldi_McKee_1992,Heyer_etal_2001,Oka_etal_2001,Rosolowsky_Blitz_2005}. More recently, following the observations of \citet{Heyer_etal_2009}, \citet{Field_etal_2011} presented a simple model for how the observed variations in the line width and surface density could be linked to pressure in the ambient medium.
The work of \citet{Heyer_etal_2009} and \citet{Field_etal_2011} led to widespread appreciation of the potential role of external pressure and adoption of the size--linewidth--surface density parameter space as a crucial diagnostic of the link between molecular clouds and environment \citep[e.g.,][and for recent synthesis works see \citealt{Colombo_etal_2019}; \citealt{Schruba_etal_2019}]{Hughes_etal_2013a,Johnson_etal_2015,Leroy_etal_2015,Utomo_etal_2015}. The \citet{Field_etal_2011} formalism to consider clouds interacting with external pressure has also been revisited by subsequent works \citep[e.g.,][]{Meidt_2016,Schruba_etal_2019}, and further generalized to include the effect of the external, galactic gravitational potential and in-plane gas motions \citep[e.g.,][]{Meidt_etal_2018,Meidt_etal_2019}.

Expanding on the work of \citet{Field_etal_2011}, \citet{Hughes_etal_2013a} estimated the ambient pressure in the ISM based on hydrostatic equilibrium, and made a synthetic, direct comparison of cloud internal pressure to the ambient pressure for GMCs in M51, M33, and the LMC. They observed a significant correlation between internal pressure and external pressure. Based on this finding, \citet{Hughes_etal_2013a} proposed that one can use the ambient pressure to predict the internal pressure in GMCs. This hypothesis is appealing because it allows one to predict GMC properties based on the large-scale mass distribution in a galaxy, which is readily available from low resolution observations or numerical simulations.

The \citet{Hughes_etal_2013a} hypothesis implies a deep connection between the properties of GMCs and the galaxy disks that they inhabit. In any long-lived, stable galaxy disk, we expect the mean pressure in the ISM to be set by the weight of the ISM in the galaxy's gravitational potential \citep[e.g.,][]{Elmegreen_1989}. If the internal pressure of GMCs is in turn set by this mean ISM pressure, then GMC properties result directly from the large-scale properties of the galaxy disk.

A natural next step to understand what sets molecular cloud properties is to compare GMC turbulent pressure to the mean ISM pressure across a range of galactic environments. If the results of \citet{Hughes_etal_2013a} apply generally, then the mean ISM pressure, which can be estimated from the large-scale disk structure, can be used to predict the properties of the local GMC population.

Furthermore, if the picture of \citet{Field_etal_2011} and related works holds, then individual GMCs exist in a state of balance between internal pressure, ambient pressure, self-gravity, and external potential. In this case, a version of dynamical equilibrium is reached -- at least in a population-averaged sense -- on cloud scales. In a recent observational study, \citet{Schruba_etal_2019} have shown that GMCs in the Milky Way and seven nearby galaxies indeed appear to achieve this equilibrium state when averaged over the population.

We seek to investigate whether the observed turbulent pressure in the molecular ISM can be explained by dynamical equilibrium considerations across multiple spatial scales, and to understand the interplay between internal pressure, ambient pressure, self-gravity, and the external potential in the molecular ISM in a wide range of galactic environments.
Carrying out these cloud-scale tests requires a) sampling diverse environments across a well-selected and homogeneously observed galaxy sample, b) achieving sufficient angular resolution to reach cloud scales ($\lesssim$100~pc), and c) having the rich supporting multi-wavelength data needed to characterize the galactic environments in which molecular gas resides.
Until now, the lack of a uniform set of sensitive, high resolution, wide-field maps of the molecular gas distribution across a large, diverse sample of galaxies has prevented such an investigation.

In this paper, we use data from the new PHANGS-ALMA survey\footnote{``\underline{P}hysics at \underline{H}igh \underline{A}ngular resolution in \underline{N}earby \underline{G}alaxie\underline{S} with \underline{ALMA}''. For more information, see \url{www.phangs.org}.} \citep[A.~K.~Leroy et al., in preparation; also see presentation of the pilot sample in ][]{Sun_etal_2018}. The core data product from PHANGS-ALMA is $\sim 100$~pc resolution CO~(2-1) mapping, which captures the ``cloud-scale'' properties of the molecular ISM. The PHANGS-ALMA galaxy sample also have rich supporting multi-wavelength data, including the atomic (\HI) gas mass, stellar mass, and star formation rate estimates. We use these data to measure the cloud-scale turbulent pressure in the molecular gas, estimate the local mean ISM pressure, and compare them at multiple spatial scales.

As emphasized above, this paper builds on a number of previous works. Using a larger, more homogeneous sample, we aim to explain the observed variations in cloud-scale gas properties in \citet{Sun_etal_2018}, to explicitly test the hypotheses of \citet{Field_etal_2011} and \citet{Hughes_etal_2013a}, and to extend the consideration of some of the same topics covered by \citet{Schruba_etal_2019}.
Our calculation of the mean ISM pressure adapts from \citet{Elmegreen_1989} and many subsequent works \citep[e.g.,][]{Wong_Blitz_2002,Blitz_Rosolowsky_2004,Blitz_Rosolowsky_2006,Leroy_etal_2008}. We particularly build on the dynamical equilibrium model developed by \citet{Ostriker_etal_2010} and \citet{Ostriker_Shetty_2011}.

The structure of this paper is as follows. We describe our galaxy sample and data sources in Section~\ref{sec:data}, and data reduction methodology in Section~\ref{sec:method}. Then we provide detailed explanations of the key pressure estimates in Section~\ref{sec:expectation}. We present our main results in Section~\ref{sec:result}, and discuss their associated systematic uncertainties in Section~\ref{sec:systematics}. We further put them in the broader context of ISM evolution and star formation in Section~\ref{sec:discussion}. We summarize our findings in Section~\ref{sec:summary}.


\section{Data} \label{sec:data}

\begin{deluxetable}{lcrcrr}
\tabletypesize{\small}
\tablecaption{Galaxy Sample\label{tab:sample}}
\tablewidth{0pt}
\colnumbers
\tablehead{
\colhead{Galaxy} &
\colhead{Dist.} &
\colhead{Incl.} &
\colhead{$\log M_\star$} &
\colhead{$\Reff$} &
\colhead{$R_\star$} \\
\colhead{} &
\colhead{[$\mathrm{Mpc}$]} &
\colhead{[$\deg$]} &
\colhead{[$\uM$]} &
\colhead{[$\arcsec$]} &
\colhead{[$\arcsec$]}
}
\startdata
NGC~628 & 9.77 & 8.7 & 10.24 & 89.1 & 69.3 \\
NGC~1087 & 14.4 & 40.5 & \phn9.82 & 31.9 & 26.4 \\
NGC~1300 & 26.1 & 31.8 & 10.85 & 69.2 & 62.3 \\
NGC~1792 & 12.8 & 64.7 & 10.34 & 49.8 & 33.3 \\
NGC~2903 & 8.47 & 67.0 & 10.42 & 77.9 & 54.9 \\
NGC~3351 & 10.0 & 45.1 & 10.34 & 62.5 & 58.0 \\
NGC~3511 & 9.95 & 75.0 & \phn9.68 & 66.3 & 37.9 \\
NGC~3521 & 11.2 & 69.0 & 10.79 & 65.7 & 39.1 \\
NGC~3627 & 10.6 & 56.5 & 10.70 & 64.5 & 54.0 \\
NGC~4293 & 16.0 & 65.0 & 10.46 & 51.4 & 50.1 \\
NGC~4298 & 16.8 & 59.6 & 10.09 & 36.3 & 26.2 \\
NGC~4321 & 15.2 & 39.1 & 10.69 & 73.9 & 61.1 \\
NGC~4424 & 16.4 & 58.2 & \phn9.91 & 31.9 & 27.6 \\
NGC~4457 & 15.6 & 17.4 & 10.43 & 14.9 & 45.4 \\
NGC~4496A & 14.9 & 55.3 & \phn9.54 & 42.4 & 31.1 \\
NGC~4535 & 15.8 & 42.1 & 10.51 & 79.8 & 57.0 \\
NGC~4536 & 15.2 & 64.8 & 10.27 & 40.2 & 50.7 \\
NGC~4540 & 16.8 & 38.3 & \phn9.82 & 21.5 & 15.3 \\
NGC~4571 & 14.9 & 31.9 & 10.05 & 48.6 & 35.2 \\
NGC~4689 & 16.8 & 39.0 & 10.25 & 48.5 & 35.9 \\
NGC~4694 & 16.8 & 60.7 & \phn9.91 & 18.7 & 31.9 \\
NGC~4731 & 12.4 & 64.0 & \phn9.47 & 79.0 & 73.5 \\
NGC~4781 & 15.3 & 56.4 & \phn9.88 & 32.8 & 25.5 \\
NGC~4826 & 4.36 & 58.6 & 10.21 & 68.2 & 56.2 \\
NGC~4951 & 12.0 & 70.5 & \phn9.59 & 25.9 & 16.1 \\
NGC~5042 & 12.6 & 51.4 & \phn9.62 & 43.2 & 35.9 \\
NGC~5068 & 5.16 & 27.0 & \phn9.32 & 81.1 & 59.7 \\
NGC~5134 & 18.5 & 22.7 & 10.32 & 23.2 & 23.2 \\
\enddata
\tablecomments{
(2) Distance (from the Extragalactic Distance Database; \citealt{Tully_etal_2009});
(3) inclination angle \citep[from PHANGS;][]{Lang_etal_2019};
(4) logarithmic global stellar mass (from $z$0MGS; \citealt{Leroy_etal_2019});
(5) effective radius (from PHANGS; A.~K.~Leroy et al., in preparation);
(6) stellar disk scale length (from \SfourG; \citealt{Salo_etal_2015});
}
\vspace{-3em}
\end{deluxetable}

We study 28 galaxies selected from the PHANGS-ALMA parent sample (A.~K.~Leroy et al., in preparation). All of these galaxies are nearby, massive, and actively forming stars (see Table~\ref{tab:sample}). We select these targets because: (a) their PHANGS-ALMA \CO21\ data reach a linear resolution of 120~pc or better\footnote{This matches the lowest resolution level in \citet{Sun_etal_2018}.} (in terms of beam full width half maximum, FWHM), and (b) they have a complete set of multi-wavelength supporting data, including new and archival \HI\ 21~cm line maps, \textit{Spitzer} IRAC 3.6~\micron\ images, GALEX near-UV data, and WISE mid-IR data.
Combined together, these data allow us to characterize molecular gas properties on cloud scales, and to put them in the context of their local ISM, star formation, and galactic stellar disk environment.

\subsection{Cloud-scale Resolution CO Data} \label{sec:data:CO}

We use PHANGS-ALMA \CO21 data (internal release version 3.4) to trace molecular gas distribution and kinematics across the star-forming disks in all our sample galaxies \citep[A.~K.~Leroy et al., in preparation; also see][]{Sun_etal_2018}.
These CO observations target the actively star-forming area in each of these galaxies, typically covering out to 5--15 kpc in galactic radius.
They include data from both the 12-m array and the Morita Atacama Compact Array (ACA; consisting of the 7-m array and four 12-m total power antennas), and therefore should recover emission on all spatial scales.
The 1$\sigma$ noise level in the data is $0.2\text{--}0.3$~K in a $2.5\,\uV$ channel.
The angular resolution (i.e., beam FWHM) ranges between $1\farcs0$--$1\farcs8$, corresponding to 25--120~pc linear resolution at the targets' distances.
This allows us to resolve the molecular gas distribution and kinematics at physical scales comparable to the typical size of GMCs (i.e., ``cloud scales'').

To homogenize the CO data, we convolve all the data cubes to two different linear resolutions, 60~pc and 120~pc, whenever possible. This allows us to control for resolution-related systematics by comparing results derived at two different linear scales. All CO data in our sample can be matched to 120~pc resolution, whereas only a subset of them (6 galaxies) reach 60~pc.

We create CO line intensity ($I\sbsc{CO}$; or moment-0) maps and line effective width\footnote{This line width metric is also referred to as ``equivalent width'' in \citet{Heyer_etal_2001} and subsequent works. It is defined as $I\sbsc{CO}/(\sqrt{2\pi}\,T\sbsc{peak,CO})$, where $T\sbsc{peak,CO}$ is the CO line peak temperature. See \citet{Sun_etal_2018} for more details.} ($\sigma\sbsc{CO}$) maps from the matched resolution CO data cubes. We build these maps by analyzing only significant CO detections in the cube (selected by the ``strict'' signal masks as described in A.~K.~Leroy et al., in preparation). This strategy ensures high signal-to-noise in the derived CO line intensity and line width maps, but discards a fraction of CO flux existing at low signal-to-noise \citep[see e.g., table~2 in][]{Sun_etal_2018}. Below in Section~\ref{sec:method:fCO}, we quantify this effect by measuring a CO flux recovery fraction.

For more details regarding the CO data reduction, readers are referred to A.~K.~Leroy et al. (in preparation) and \citet[which adopts a similar data reduction method]{Sun_etal_2018}.

\subsection{Kpc-scale Resolution Supporting Data} \label{sec:data:kpc}

To measure the distribution of gas, stars, and star formation activity in each of our target galaxies, we assemble a multi-wavelength supporting dataset from a variety of sources. These supporting data typically have much coarser angular resolution than the PHANGS-ALMA CO data, corresponding to linear scales of hundreds to a thousand parsecs.

We use \HI\ 21~cm emission data to trace the atomic gas distribution. We include new VLA data from the PHANGS-VLA project (covering 11 targets in our sample; PI: D.~Utomo) and the EveryTHINGS project (NGC~4496A; PI: K.~Sandstrom), as well as existing data from VIVA \citep[7 targets;][]{Chung_etal_2009}, THINGS \citep[6 targets;][]{Walter_etal_2008}, VLA observations associated with HERACLES \citep[NGC~4321 and NGC~4536;][]{Leroy_etal_2013a}, and an individual ATCA observation \citep[NGC~1792;][]{Murugeshan_etal_2019}. Most of these \HI\ data have native angular resolution of $15\text{--}25\arcsec$, which corresponds to linear scales of $0.5\text{--}2$~kpc at the distances of the targets. Typical column density sensitivities are $0.5\text{--}2\times10^{20}\;\mathrm{cm}^{-2}$ at 3$\sigma$ after integrating over the $10\text{--}20\;\uV$ line width.

We use \textit{Spitzer} IRAC 3.6~\micron\ data to trace the stellar mass distribution. These data come from the \textit{Spitzer} Survey of Stellar Structure in Galaxies \citep[\SfourG;][]{Sheth_etal_2010}. For most targets in our sample, we use the ICA 3.6~\micron\ maps \citep{Querejeta_etal_2015}, for which the emission from dust has been subtracted. For the remaining three targets (NGC~4571, NGC~4689, and NGC~5134), we use the raw 3.6~\micron\ maps as these galaxies have global $[3.6]-[4.5]$ colors compatible with old stellar population \citep[see section~4.2 in][]{Querejeta_etal_2015}. All these maps are masked to remove foreground stars. These data have a native angular resolution of $\sim2\arcsec$, which corresponds to linear scales of hundreds of parsecs for our targets.

We combine GALEX near-UV (NUV) data and WISE mid-IR (MIR) data to derive kpc-scale star formation rate (SFR) estimates. We use the post-processed GALEX NUV images and WISE band-3 images from the $z=0$ Multi-wavelength Galaxy Synthesis \citep[$z$0MGS;][]{Leroy_etal_2019}. These images are corrected for background, aligned for astrometry, and masked to remove foreground stars.
Almost all targets in our sample have both GALEX and WISE coverage (except for NGC~4689, for which only WISE is available). We use the data at fixed angular resolution of $7.5\farcs$, which corresponds to linear scales $\lesssim1$~kpc.

We homogenize the resolution of these supporting data to a common 1~kpc linear scale. As described above, most data have native resolutions better than 1~kpc, and thus can be directly convolved to this coarser resolution. For some of the \HI\ data, however, the native resolution is coarser. As the resolution of these \HI\ data is not crucial for our analysis (see discussion in Section~\ref{sec:expectation:PDE-cloud}), we keep these data at their native resolution in the following analysis.

To complement these supporting data, we also convolve the PHANGS-ALMA CO data to 1~kpc resolution. This provides a tracer of the large-scale molecular gas distribution. The convolution is performed on the CO data cubes, and from these convolved cubes we derive a set of low resolution CO line intensity maps. This strategy takes advantage of the better surface brightness sensitivity at coarser resolution. In the resulting kpc-scale CO intensity maps, we detect emission at high signal-to-noise over almost every sightline, and expect to recover essentially all emission within the observation footprint.

Altogether, these data provide us with spatially resolved information about the (molecular and atomic) gas distribution, the stellar disk structure, and the local star formation rate, all of which are measured on a matched 1~kpc spatial scale.


\section{Methods} \label{sec:method}

The PHANGS-ALMA CO data probe the molecular gas distribution and kinematics on 60--120~pc scales (cloud scales), whereas the multi-wavelength supporting data characterize the galactic environment on 1~kpc scale. We conduct a cross-spatial-scale analysis to bridge the gap between these two spatial scales. The methodology that we adopt here has been developed and used in a series of previous works \citep[see e.g.,][]{Ossenkopf_MacLow_2002,Sandstrom_etal_2013,Leroy_etal_2013a,Leroy_etal_2016,Gallagher_etal_2018b,Utomo_etal_2018}.

We first divide the PHANGS-ALMA CO observation footprint into hexagonal apertures. They form a hexagonal tiling in the plane of the sky, with a 1~kpc spacing (corresponding to $8''$--$40''$) between the centers of adjacent apertures. These kpc-sized, hexagonal apertures are the fundamental units of our analysis.

The key step of this cross-spatial-scale analysis is to characterize the statistics of the many independent cloud-scale measurements within each kpc-sized aperture.
In this work, we use a CO intensity-weighted averaging scheme to quantify the ensemble average of a given cloud-scale measurement in each aperture.
As an example, we calculate the ensemble average of CO line width $\sigma\sbsc{CO}$ within a kpc-sized aperture $A$ via
\begin{equation}
    \brktkpc{\sigma\sbsc{CO,\,\theta pc}} = \frac{\int_A \sigma\sbsc{CO,\,\theta pc}\, I\sbsc{CO,\,\theta pc}\, \mathrm{d}S}{\int_A I\sbsc{CO,\,\theta pc}\, \mathrm{d}S}~.
    \label{eq:average}
\end{equation}
\noindent Here $\sigma\sbsc{CO,\,\theta pc}$ is the CO line width measured on $\theta=60$~pc and $120$~pc scales, $I\sbsc{CO,\,\theta pc}$ is the line intensity measured on the same scale (i.e., the statistical ``weight''). The ``$\brkt{}\sbsc{1kpc}$'' symbol denotes a CO intensity-weighted average over the kpc-sized aperture.

This CO intensity-weighted averaging scheme preserves information on the gas distribution on cloud scales \citep[see][]{Leroy_etal_2016}.
By using CO line intensity (or equivalently CO flux, given a fixed beam size) as the statistical weight, it prevents the averaged value from being ``diluted'' by areas with no CO detections.
As long as the CO flux detection fraction in an aperture remains reasonably high (which is the case within most apertures in our sample; see Section~\ref{sec:method:fCO} below, and plots in Section~\ref{sec:result}), the CO intensity-weighted average value does not suffer from a strong dilution effect.
Conversely, due to the generally low area coverage fraction of CO detection in our data, a direct, area-weighted average would include many more non-detections than detections, and yield significantly diluted average values.

To supplement these ensemble averages of cloud-scale CO measurements, we include all the kpc-scale resolution supporting data into this analysis, by directly sampling these maps at the center of each hexagonal aperture. The matched 1~kpc spacing between the centers of adjacent apertures ensures that there is little correlation between measurements in adjacent apertures.

Our cross-spatial-scale analysis produces a rich multi-wavelength database for every target in our sample. This database has been used in a previous publication \citep{Herrera_etal_2019}, and the current work uses it as the foundation for carrying out the calculations presented below in Section~\ref{sec:expectation}. The full database, as well as the correlations between the basic observables will be published in a companion paper (J.~Sun et al., in preparation).

\subsection{CO Flux Recovery Fraction} \label{sec:method:fCO}

As discussed above, for our CO intensity-weighted averaging scheme to work optimally, the CO observations should be sensitive enough to detect a significant fraction of CO flux at cloud-scale resolution in each aperture. This is indeed the case for the majority of the apertures in our sample (see below and plots in Section~\ref{sec:result}). When this is not the case, however, the averaged values will suffer from larger uncertainties and reflect only properties of the brightest clouds.

To control for these effects due to the finite sensitivity of CO observations, we quantify a CO flux recovery fraction, $\fCOpc$, within each 1~kpc aperture. We do this by comparing, within the footprint of each aperture, the total flux included in the 60--120~pc resolution CO line intensity map (i.e., within the ``strict'' signal masks, see Section~\ref{sec:data:CO}) to the total flux in the corresponding CO data cube. We estimate the latter quantity by summing up the cube within a wide, high completeness signal mask (referred to as the ``broad'' mask in A.~K.~Leroy et al., in preparation).

Our calculation shows that for the majority ($\sim$70\%) of the kpc-scale apertures in our sample, the CO flux recovery fraction on 120~pc scales, $f\sbsc{CO,\,120pc}$, is higher than 50\%. Since the CO intensity-weighted averages are calculated from only the \textit{detected} emission, a recovery fraction of $\fCO > 50\%$ means that the majority of the CO emission in that kpc-scale aperture is included in the averaging. First, this assures that the derived averages in these apertures have reasonably small statistical error ($\le\sqrt{2}$~times larger than the case of infinite sensitivity given homoscedastic individual measurements). Second, and most importantly, in the cases where the undetected molecular gas has systematically different properties than the detected, the intensity-weighted averages in these high $\fCO$ apertures are also much less susceptible to systematic effects due to sampling biases.

Hereafter, when presenting results derived from this CO intensity-weighted averaging approach, we represent the data in darker/lighter colors to denote higher/lower $f\sbsc{CO}$.
Measurements shown in darker colors are thus more representative of the overall cloud population, whereas those shown in lighter colors only characterize the brightest clouds in the kpc-sized aperture.

\subsection{CO Flux from Various Morphological Regions} \label{sec:method:region}

Our weighted averaging scheme is a way to quantify the mean molecular gas properties in each kpc-sized aperture. This kpc aperture size is large enough that the CO flux within each aperture might come from various morphological regions of a galaxy (e.g., bulges, bars). In this work, for each of the averaging apertures, we also keep track of the fractional CO flux contribution from different morphological regions.

In detail, we first identify areas covered by morphological structures like bulges and bars (when applicable) in every galaxy. To identify the bulge regions, we use the \SfourG\ structural decomposition results presented by \citet[there referred to as \SfourG\ pipeline 4]{Salo_etal_2015}. These results are based on two-dimensional structural decomposition of \textit{Spitzer} IRAC 3.6~\micron\ images with GALFIT3.0 \citep{Peng_etal_2010}. To identify bar regions, we instead use results from the visual identification in \citet{Herrera-Endoqui_etal_2015}, which have higher quality than the \SfourG\ pipeline 4 results.

Within each kpc-sized aperture, we calculate the fractional contribution in the total CO flux from the bulge and bar regions. Whenever an aperture includes non-zero CO flux coming from bulge or bar regions, we classify it as a ``bulge/bar'' aperture\footnote{Using a non-zero threshold (e.g., 10\%) would only change the number of ``bulge/bar'' apertures by a very small amount.}. Otherwise, we classify it as a ``disk'' aperture. When presenting our results, we show measurements in bulge/bar apertures in orange colors, and those in disk apertures in blue colors.

\subsection{Converting Observables to Physical Quantities} \label{sec:method:phys}

We use the PHANGS-ALMA CO data and multi-wavelength supporting data to estimate physical properties of the molecular gas and its ambient galactic environment. Here we detail our methods to convert direct observables (e.g., CO intensity) into physical quantities (e.g., molecular gas surface density) in each kpc-sized aperture in our sample.
Table~\ref{tab:def} lists the key physical quantities we derive, including both average molecular cloud properties and environment characteristics.
These physical quantities are the basis of all following calculations detailed in Section~\ref{sec:expectation}.

\subsubsection{Cloud-scale Molecular Gas Properties} \label{sec:method:phys:mol}

Following \citet{Sun_etal_2018}, we estimate cloud-scale molecular gas surface density ($\Sigmolpc$) and velocity dispersion ($\sigmolpc$) from the observed \CO21\ line intensity ($\ICOpc$) and effective width ($\sigCOpc$), via
\begin{align}
    \Sigmolpc &= \frac{\alphaCO}{R_{21}}\,\ICOpc~, \label{eq:Sigmol}\\
    \sigmolpc &= \sigCOpc~. \label{eq:sigmol}
\end{align}
\noindent In Equation~\ref{eq:Sigmol}, $R_{21} = 0.7$ is the adopted \CO21\ to \CO10\ line ratio \citep[D.~Chatzigiannakis et al., in preparation]{Leroy_etal_2013a,Saintonge_etal_2017}, whereas $\alphaCO$ is a CO-to-H$_2$ conversion factor\footnote{Throughout this paper, we use the term ``conversion factor'' and the symbol $\alphaCO$ to refer to the ratio of molecular gas mass to \CO10\ line luminosity (or equivalently, the ratio of mass surface density to line intensity). By definition, this includes the mass contribution from heavy elements.}, whose value varies aperture-by-aperture.

We adopt the following prescription to predict the values of $\alphaCO$ for each aperture in our sample.
Similar to the calibration suggested by \citet{Accurso_etal_2017} and adopted in the xCOLD GASS survey \citep{Saintonge_etal_2017}, we predict $\alphaCO$ via
\begin{equation}
    \alphaCO = 4.35\,Z'^{-1.6}\;\uSig\,(\uIco)^{-1}~.
\label{eq:alphaCO}
\end{equation}
\noindent Here $Z'$ is the local gas phase abundance normalized to the solar value appropriate for the \citet{Pettini_Pagel_2004} metallicity calibration [$12+\mathrm{log\,(O/H)}=8.69$].

Using Equation~\ref{eq:alphaCO} to predict $\alphaCO$ requires knowing $Z'$ for every kpc-sized aperture across our sample. However, metallicity measurements existing in the literature only cover a subset of our targets, and they are derived using heterogeneous calibration methods.
To ensure a homogeneous coverage of the entire sample, we instead predict $Z'$ in a uniform, empirical way.
Using a mass-metallicity relation reported by \citet[see table~1 therein]{Sanchez_etal_2019}, we first predict $Z'$ at one effective radius ($\Reff$) in each galaxy based on the galaxy global stellar mass (see Table~\ref{tab:sample}). We then extend our prediction to cover the entire galaxy assuming a universal radial metallicity gradient of $-0.1\;\mathrm{dex}/\Reff$ \citep{Sanchez_etal_2014}.
Combining this locally predicted $Z'$ with Equation~\ref{eq:alphaCO}, we have a predicted $\alphaCO$ value for every aperture in our sample.

Our choices on the prescriptions for predicting $Z'$ and $\alphaCO$ could affect many of the measured molecular gas properties in this work.
To quantify the systematic effects associated with these choices, in Section~\ref{sec:systematics:alphaCO} we consider three alternative $\alphaCO$ prescriptions, and compare the quantitative results with those derived based on our ``fidual'' prescriptions.

\subsubsection{Kpc-scale Environment Characteristics} \label{sec:method:phys:kpc}

In each kpc-sized aperture, we estimate the physical properties of the large-scale galactic environment from the multi-wavelength supporting data described above in Section~\ref{sec:data:kpc}.

\begin{itemize}[itemsep=1em,leftmargin=1em,parsep=0pt,partopsep=0pt]

\item[$\sbullet$] \textit{Molecular Gas Surface Density:} We estimate the kpc-scale molecular gas surface density $\Sigmolkpc$ from the \CO21 intensity $\ICOkpc$, via
\begin{equation}
    \Sigmolkpc = \frac{\alphaCO}{R_{21}}\,\ICOkpc\cos{i}~.
    \label{eq:Sigmolkpc}
\end{equation}
\noindent This conversion is similar to Equation~\ref{eq:Sigmol}, except that here we also add a ``$\cos{i}$'' term to correct for the projection effect due to galaxy inclination\footnote{We do not apply this inclination correction in Equation~\ref{eq:Sigmol}, because at 60--120~pc scales the geometry of molecular gas structure is no longer well approximated by a thin disk \citep[see][]{Sun_etal_2018}.} (see Table~\ref{tab:sample}).

\item[$\sbullet$] \textit{Atomic Gas Surface Density:} We estimate the kpc-scale atomic gas surface density $\Sigatomkpc$ from the observed \HI\ 21~cm line intensity $I\sbsc{21cm,\,1kpc}$, via
\begin{equation}
    \frac{\Sigatomkpc}{\uSig} = 1.97\times10^{-2}\;\frac{I\sbsc{21cm,\,1kpc}}{\uIco}\,\cos{i}~.
    \label{eq:Sigatomkpc}
\end{equation}
\noindent This conversion includes the mass contribution from heavy elements.

\item[$\sbullet$] \textit{Stellar Mass Surface Density:} We estimate the kpc-scale stellar mass surface density $\Sigstarkpc$ from the (dust-corrected) 3.6~\micron\ specific surface brightness $I_\mathrm{3.6,\,1kpc}$, via\footnote{Note that this differs from the adopted conversion in \citet{Querejeta_etal_2015}, which is based on a higher mass-to-light ratio of $Y_{3.6} = 0.6\,M_\odot/L_\odot$ \citep{Meidt_etal_2014}. See discussions in \citet{Leroy_etal_2019}.}
\begin{equation}
    \frac{\Sigstarkpc}{\uSig} = 3.3\times10^2\;\frac{I\sbsc{3.6,\,1kpc}}{\uI}\,\cos{i}~.
    \label{eq:Sigstarkpc}
\end{equation}
\noindent This conversion assumes a 3.6~\micron\ mass-to-light ratio of $Y_{3.6} = 0.47\,M_\odot/L_\odot$ \citep{McGaugh_Schombert_2014}.

\item[$\sbullet$] \textit{SFR Surface Density:} We estimate the kpc-scale SFR surface density $\SigSFRkpc$ from the combined GALEX NUV and WISE 12~\micron\ (band 3) data, following the calibration suggested by \citet{Leroy_etal_2019}:
\begin{align}
    \frac{\SigSFRkpc}{\uSigSFR}
    =&\, \left(8.9\times10^{-2}\;\frac{I\sbsc{NUV,\,1kpc}}{\uI}\right. \nonumber\\
    &+ \left. 4.1\times10^{-3}\;\frac{I\sbsc{12,\,1kpc}}{\uI}\right)\,\cos{i}~.
    \label{eq:SigSFRkpc}
\end{align}
\noindent These prescriptions are broadly consistent with stellar initial mass functions suggested by \citet{Chabrier_2003} and \citet{Kroupa_Weidner_2003}.

\end{itemize}


\section{Pressure Estimates} \label{sec:expectation}

In this work, we test the hypothesis that dynamical equilibrium holds in the ISM. Given that molecular gas is usually found near the disk mid-plane \citep{Heyer_Dame_2015}, it is interesting to compare a direct measurement of the internal pressure in the molecular gas to the predicted mid-plane pressure from dynamical equilibrium models. We use the term ``dynamical equilibrium pressure'' ($\PDE$) to refer to the latter quantity throughout this work.

Estimating $\PDE$ from observations is challenging. Many previous studies \citep[e.g.,][]{Blitz_Rosolowsky_2004,Blitz_Rosolowsky_2006,Leroy_etal_2008,Hughes_etal_2013a} treat the gas disk as a smooth, single-phase fluid, assuming no substructure below kpc scales (this is the typical resolution of \HI\ surveys targeting nearby galaxies). In reality, turbulence, shocks, and gravitational instabilities create a rich multi-scale structure in the ISM. The cold and dense molecular phase, in particular, is highly structured on small scales, which leads to enhanced gas self-gravity in denser regions. As a result of this small-scale structure, the total weight of the ISM is higher than one would infer assuming a smooth disk geometry, and a greater pressure is needed to balance this weight. Estimates of $\PDE$ that assume a smooth disk neglect this enhancement in gas self-gravity. While they might be able to reflect the mean pressure over a large portion of the ISM disk, these estimates of $\PDE$ in many previous works represent only lower limits on the expected pressure within molecular clouds.

Building on these previous works, here we present estimates for $\PDE$ that also take into account the presence of small-scale substructure in the molecular ISM. This is possible because of the new, high resolution PHANGS-ALMA CO maps. They allow us to estimate the weight of molecular gas in its local gravitational potential on cloud scales, which is the same spatial scale on which we measure molecular gas internal pressure. We then combine our estimate for the weight of molecular gas with the weight of the volume-filling atomic gas in the combined gas and stellar potential. This leads to a modified, cloud-scale equilibrium pressure, which accounts for both the dense, clumpy molecular phase and the diffuse, smooth atomic phase.

We note that similar approaches have been adopted to explain molecular cloud properties in the Milky Way \citep[][]{Heyer_etal_2001,Field_etal_2011} and other galaxies \citep[e.g.,][]{Hughes_etal_2013a,Schruba_etal_2019}.
Most of these studies adopt a ``bottom-up'' approach. That is, they segment the observed CO emission into individual clouds, and consider pressure balance between the identified clouds and the large-scale galactic environment in order to explain the observed cloud properties.
In this paper, we formulate an alternative, ``top-down'' approach. We consider all molecular gas in each kpc-size region, without employing any cloud identification algorithm (for an explicit comparison between our approach and a cloud-based approach, see Section~\ref{sec:systematics:CPROPS}). We then attempt to explain the ensemble average of molecular gas properties on fixed spatial scales in the context of a dynamical equilibrium model.

Our ``top-down'' approach captures many of the same physics as cloud-centered approaches, i.e., the balance between internal pressure, ambient pressure, self-gravity, and external gravity. In addition, it is designed to robustly treat data with a wide range of physical resolutions, even when individual gas structures are not fully resolved or cleanly separated from one another.
It also considers all detected emission, and so should yield highly reproducible results that characterize the behavior of the entire molecular gas reservoir.

In this section, we explain our methodology for estimating 1) the internal pressure in molecular gas, 2) the classic, kpc-scale dynamic equilibrium pressure, and 3) the modified, cloud-scale equilibrium pressure. These pressure estimates are also listed in Table~\ref{tab:def}.

\subsection{Internal Pressure in Molecular Gas}
\label{sec:expectation:Pturb}

Internal pressure in molecular gas includes the contributions from thermal and turbulent motion, as well as magnetic fields. Observational evidence, including super-thermal CO line widths and the size--line width relation observed within GMCs, suggest that turbulent motion dominates over thermal motion on physical scales comparable to cloud sizes \citep[also see \citealt{Heyer_Dame_2015}]{Larson_1981,Solomon_etal_1987,Heyer_Brunt_2004}. Numerical simulations of the star-forming ISM on galactic scales also find that the magnetic term is subdominant, typically reaching only $\sim 50\%$ of the kinetic term in the effective gas pressure \citep[also see observational evidence presented by \citealt{Crutcher_1999,Falgarone_etal_2008,Troland_Crutcher_2008,Thompson_etal_2019}]{Kim_Ostriker_2017,Pakmor_etal_2017}.
Motivated by these findings, we assume in this work that turbulent motion represents the primary source of internal pressure in molecular gas, and treat all other contributions as sub-dominant.
We do not differentiate between turbulent pressure and total internal pressure in molecular gas hereafter.

Turbulent pressure in molecular gas is commonly estimated from volume density and the observed (one-dimensional) velocity dispersion, under the assumption that turbulence is isotropic:
\begin{align}
    \Pturb = \rhomol\, \sigma\sbsc{turb,1D}^2~.
\end{align}

While one can use the observed velocity dispersion $\sigmol$ along the line of sight as a proxy of $\sigma\sbsc{turb,1D}$, $\rhomol$ is not usually directly observed. Here we convert cloud-scale molecular gas surface density, $\Sigmol$, into volume density, $\rhomol$. To do this, we assume a constant density spherical cloud filling each beam, with the cloud diameter $D\sbsc{cloud}$ equal to the beam FWHM\footnote{As stated by \citet{Sun_etal_2018}, this is appropriate when the beam size is comparable to or smaller than the molecular disk scale height or the turbulence driving scale, and when the beam dilution effect is not strong.} (i.e., 60 or 120~pc, but see Section~\ref{sec:systematics:CPROPS} for an alternative approach).
The inferred turbulent pressure in molecular gas can then be expressed as
\begin{align}
    \Pturb
    = \rho\sbsc{mol} \sigmol^2
    = \frac{6M\sbsc{mol}}{\pi D\sbsc{cloud}^3} \sigmol^2
    = \frac{3 \Sigmol \sigmol^2}{2D\sbsc{cloud}}~.
    \label{eq:Pturb}
\end{align}
\noindent Here $M\sbsc{mol}=\Sigmol\cdot(\pi D\sbsc{cloud}^2/4)$ is the total molecular gas mass of the spherical cloud.

We average the estimated cloud-scale turbulent pressure across each kpc-sized region following the same CO flux weighting scheme described in Section~\ref{sec:method}:
\begin{align}
    \Pturbavgkpc
    &= \frac{\int_A \Pturbpc\,\ICOpc\,\mathrm{d}S}{\int_A \ICOpc\,\mathrm{d}S} ~.
    \label{eq:Pturbavg}
\end{align}
This can be interpreted as the mass-weighted average turbulent pressure within each kpc-sized aperture $A$.

\subsection{Kpc-scale Dynamical Equilibrium Pressure}
\label{sec:expectation:PDE-kpc}

To compare with previous works, we first estimate the classic, kpc-scale dynamical equilibrium pressure, $\PDEkpc$. We follow the same basic formalism that has been adopted, with some variations, in many previous works \citep[e.g.,][]{Spitzer_1942,Elmegreen_1989,Elmegreen_Parravano_1994,Wong_Blitz_2002,Blitz_Rosolowsky_2004,Blitz_Rosolowsky_2006,Leroy_etal_2008,Koyama_Ostriker_2009b,Ostriker_etal_2010,Kim_etal_2011,Ostriker_Shetty_2011,Shetty_Ostriker_2012,Hughes_etal_2013a,Kim_etal_2013,Kim_Ostriker_2015b,Benincasa_etal_2016,Herrera-Camus_etal_2017,Gallagher_etal_2018a,Fisher_etal_2019,Schruba_etal_2019}.

This approach models the distribution of gas and stars in a galaxy disk as isothermal fluids in a plane-parallel geometry.
For the calculation here, we assume that the (single component) gas disk scale height is much smaller than the stellar disk scale height.
We also neglect gravity due to dark matter, as it represents only a minor component in the galactic environments we study here (i.e., the inner disks of relatively massive galaxies).
In this case, we can express $\PDE$ as:
\begin{align}
    \PDEkpc
    =&\; \frac{\pi G}{2}\,\Siggaskpc^2 \nonumber\\
    &+ \Siggaskpc \sqrt{2 G\rhostarkpc}\, \siggasz~.
\label{eq:PDEkpc}
\end{align}
\noindent The first term is the weight of the ISM due to the self-gravity of the ISM disk \citep[see e.g.,][]{Spitzer_1942,Elmegreen_1989}. The second term is the weight of the ISM due to stellar gravity \citep[see e.g.,][]{Spitzer_1942,Blitz_Rosolowsky_2004}.
$\Siggaskpc=\Sigmolkpc+\Sigatomkpc$ is the total gas surface density, $\rhostarkpc$ is stellar mass volume density near disk mid-plane, and $\siggasz$ is the vertical gas velocity dispersion (a combination of turbulent, thermal, and magnetic terms).

Following \citet{Blitz_Rosolowsky_2006,Leroy_etal_2008,Ostriker_etal_2010}, we estimate mid-plane stellar volume densities from the observed surface densities in each kpc-sized aperture:
\begin{equation}
    \rhostarkpc
    = \frac{\Sigstarkpc}{4\Hstar}
    = \frac{\Sigstarkpc}{0.54\Rstar}~.
\label{eq:rhostarkpc}
\end{equation}
\noindent The first step assumes an isothermal density profile along the vertical direction (i.e., $\rho_*(z) \propto \sech^2[z/(2H_*)]$) with $\Hstar$ being the stellar disk scale height \citep{vanderKruit_1988}. The second step assumes a fixed stellar disk flattening ratio $\Rstar/\Hstar = 7.3$ \citep[also see Appendix~\ref{apdx:flatratio}]{Kregel_etal_2002}. Here $\Rstar$ is the radial scale length of the stellar disk, for which we adopt the value from the \SfourG\ photometric decompositions of 3.6~\micron\ images \citep[see column~(6) in Table~\ref{tab:sample}]{Salo_etal_2015}.

For $\siggasz$, we calculate the mass-weighted average velocity dispersion of molecular and atomic phases
\begin{equation}
    \siggasz
    = f\sbsc{mol} \sigmolavgkpc + (1-f\sbsc{mol})\, \sigatom~,
    \label{eq:siggasz}
\end{equation}
where $f\sbsc{mol}=\Sigmolkpc\,/\,(\Sigmolkpc+\Sigatomkpc)$ is the fraction of gas mass in the molecular phase. We adopt a fixed atomic gas velocity dispersion $\sigatom=10\;\uV$ \citep[see][]{Leroy_etal_2008,Tamburro_etal_2009,Wilson_etal_2011,Caldu-Primo_etal_2013,Mogotsi_etal_2016}.

Our adopted assumptions for the $\rhostarkpc$ and $\siggasz$ estimation might introduce systematic biases in the derived $\PDEkpc$. In Section~\ref{sec:systematics:PDE}, we estimate $\PDEkpc$ by adopting two alternatives for estimating $\siggasz$ and $\rhostarkpc$, and compare the results with our fiducial $\PDEkpc$ estimates.

\subsection{Cloud-scale Dynamical Equilibrium Pressure}
\label{sec:expectation:PDE-cloud}

The classic, kpc-scale equilibrium pressure defined in Section~\ref{sec:expectation:PDE-kpc} does not account for gas substructure within each kpc-sized aperture. For atomic gas, surface density fluctuations on sub-kpc scale are usually moderate \citep[also see \citealt{Bolatto_etal_2011}; E.~Koch et al., in preparation]{Leroy_etal_2013a}, so this issue is likely minor. For molecular gas, however, we expect strong clumping \citep[e.g.,][]{Leroy_etal_2013a}. Therefore, gas self-gravity should be significantly enhanced in over-dense regions (e.g., in molecular clouds), and the required pressure in molecular gas to balance this enhanced gravity should exceed the classic, kpc-scale pressure estimates.

To account for this, we introduce a modified, cloud-scale dynamical equilibrium pressure, $\PDEavgkpc$. Using the classic formulation as a starting point, we treat the clumpy molecular ISM and diffuse atomic ISM separately, allowing them to have different geometry \citep[also see][for similar calculations]{Ostriker_etal_2010, Schruba_etal_2019}. We offer a brief summary of this alternative formalism here, but leave a more detailed description of the derivation and adopted assumptions to Appendix~\ref{apdx:formulation}.

In this alternative formalism, we split the total cloud-scale equilibrium pressure into two parts:
\begin{align}
    \PDEavgkpc
    &= \brktkpc{\mathcal{W}\sbsc{total,\,\theta pc}} \nonumber\\
    &= \brktkpc{\mathcal{W}\sbsc{cloud,\,\theta pc}} + \mathcal{W}\sbsc{atom,\,1kpc}~.
\label{eq:PDEavg}
\end{align}
\noindent The first part, $\brktkpc{\mathcal{W}\sbsc{cloud,\,\theta pc}}$, corresponds to the weight of molecular gas, most of which resides in clumpy structures on cloud scales. For simplicity, we assume that all molecular gas is organized into spherical, cloud-like structures. The weight within each individual structure is due to 1) its own self-gravity, 2) the gravity of other molecular structures, and 3) the gravity of stars:
\begin{align}
    \mathcal{W}\sbsc{cloud,\,\theta pc} =&\, \mathcal{W}\sbsc{cloud,\,\theta pc}\spsc{self} + \mathcal{W}\sbsc{cloud,\,\theta pc}\spsc{ext-mol} + \mathcal{W}\sbsc{cloud,\,\theta pc}\spsc{star} \nonumber\\
    =&\, \frac{3\pi}{8}G\Sigmolpc^2 + \frac{\pi}{2}G\Sigmolpc\Sigmolkpc \nonumber\\
    &+ \frac{3\pi}{4}G\rhostarkpc\Sigmolpc D\sbsc{cloud}~.
    \label{eq:W_cloud}
\end{align}
\noindent Consistent with our estimation of turbulent pressure in Equation~\ref{eq:Pturb}, we also assume the cloud diameter $D\sbsc{cloud}$ equals the beam FWHM here.

We then adopt the same averaging scheme used in Equation~\ref{eq:Pturbavg} to estimate the (CO flux-weighted) average $\mathcal{W}\sbsc{cloud,\,\theta pc}$ across each kpc-sized aperture:
\begin{align}
    \brktkpc{\mathcal{W}\sbsc{cloud,\,\theta pc}}
    &= \frac{\int_A \mathcal{W}\sbsc{cloud,\,\theta pc}\, \ICOpc\,\mathrm{d}S}{\int_A \ICOpc\,\mathrm{d}S}~.
    \label{eq:W_cloud_avg}
\end{align}

The second term in Equation~\ref{eq:PDEavg}, $\mathcal{W}\sbsc{atom,\,1kpc}$, corresponds to the weight of the smooth extended layer of atomic gas. This weight is due to the gravity of all gas (both atomic and molecular phases) plus the stars, as felt by the atomic layer. Motivated by the relative smoothness of the atomic gas distribution and the coarser resolution of the \HI\ data, we estimate this weight using only kpc-scale measurements, assuming uniform atomic gas surface density within each kpc-sized aperture:
\begin{align}
    \mathcal{W}\sbsc{atom,\,1kpc} =&\, \mathcal{W}\sbsc{atom,\,1kpc}\spsc{self} + \mathcal{W}\sbsc{atom,\,1kpc}\spsc{mol} + \mathcal{W}\sbsc{atom,\,1kpc}\spsc{star} \nonumber\\
    =&\, \frac{\pi G}{2}\Sigatomkpc^2 + \pi G\Sigatomkpc\Sigmolkpc \nonumber\\
    &+ \Sigatomkpc\sqrt{2G\rhostarkpc}\,\sigatom~.
\label{eq:W_atom}
\end{align}
\noindent Here we assume that the molecular gas disk is ``sandwiched'' by the atomic gas disk, and thus the second term above has a 2~times larger pre-factor than the first term (see Appendix~\ref{apdx:formulation} for detailed derivation). We adopt $\sigatom=10\;\uV$, consistent with Section~\ref{sec:expectation:PDE-kpc}.

If vertical dynamical equilibrium holds across multiple spatial scales, we would expect molecular gas internal pressure on cloud scales to match the dynamical equilibrium pressure on the same scale. By comparing our measured $\Pturbavgkpc$ and best-estimate $\PDEavgkpc$, we can test whether, in a statistical average sense, the molecular ISM in nearby, star-forming disk galaxies can be described by this model.


\section{Results} \label{sec:result}

\begin{deluxetable*}{lccc}
\tabletypesize{\footnotesize}
\tablecaption{List of Key Physical Properties\label{tab:def}}
\tablewidth{0pt}
\tablehead{
\colhead{Quantity Definition} &
\colhead{Symbol} &
\colhead{Unit} &
\colhead{Data Source}
}
\startdata
\multicolumn{4}{c}{Ensemble average of cloud-scale molecular gas properties (see \S\ref{sec:method:phys:mol})} \\
\multicolumn{4}{c}{(measured at $\theta=60$, 120~pc scale; averaged over each kpc-sized aperture; see \S\ref{sec:method})} \\
\hline
Average molecular gas surface density (Eq.~\ref{eq:average}\&\ref{eq:Sigmol}) & $\Sigmolavgkpc$ & \uSig\ & PHANGS-ALMA \CO21\ \\
Average molecular gas velocity dispersion (Eq.~\ref{eq:average}\&\ref{eq:sigmol}) & $\sigmolavgkpc$ & \uV\ & PHANGS-ALMA \CO21\ \\
CO flux recovery fraction (\S\ref{sec:method:fCO}) & $\fCOpc$ & - & PHANGS-ALMA \CO21\ \\
\hline
\multicolumn{4}{c}{Environmental characteristics (see \S\ref{sec:method:phys:kpc})} \\
\multicolumn{4}{c}{(measured at 1~kpc scale)} \\
\hline
kpc-scale molecular gas surface density (Eq.~\ref{eq:Sigmolkpc}) & $\Sigmolkpc$ & \uSig\ & PHANGS-ALMA \CO21\ \\
kpc-scale atomic gas surface density (Eq.~\ref{eq:Sigatomkpc}) & $\Sigatomkpc$ & \uSig\ & PHANGS-VLA \HI\ 21~cm, etc. \\
kpc-scale stellar mass surface density (Eq.~\ref{eq:Sigstarkpc}) & $\Sigstarkpc$ & \uSig\ & \SfourG\ IRAC 3.6~\micron\ \\
kpc-scale star formation rate surface density (Eq.~\ref{eq:SigSFRkpc}) & $\SigSFRkpc$ & \uSigSFR\ & z0MGS NUV+MIR \\
\hline
\multicolumn{4}{c}{Pressure estimates (see \S\ref{sec:expectation})} \\
\hline
Average turbulent pressure in molecular gas (Eq.~\ref{eq:Pturb}\&\ref{eq:Pturbavg}) & $\Pturbavgkpc$ & \uP\ & PHANGS-ALMA \CO21\ \\
kpc-scale ISM equilibrium pressure (Eq.~\ref{eq:PDEkpc}--\ref{eq:siggasz}) & $\PDEkpc$ & \uP\ & All combined \\
Average cloud-scale equilibrium pressure (Eq.~\ref{eq:PDEavg}--\ref{eq:W_atom}) & $\PDEavgkpc$ & \uP\ & All combined \\
Average weight of molecular clouds (Eq.~\ref{eq:W_cloud}\&\ref{eq:W_cloud_avg}) & $\brkt{\mathcal{W}\sbsc{cloud,\,\theta pc}}\sbsc{1kpc}$ & \uP\ & All combined \\
Average weight of clouds due to self-gravity (Eq.~\ref{eq:W_cloud}\&\ref{eq:W_cloud_avg}) & $\brkt{\mathcal{W}\spsc{self}\sbsc{cloud,\,\theta pc}}\sbsc{1kpc}$ & \uP\ & PHANGS-ALMA \CO21\ \\
\enddata
\vspace{-2em}
\end{deluxetable*}

We measure cloud-scale molecular gas properties at two resolutions, 60~pc and 120~pc. Then we derive the CO intensity-weighted average properties in every kpc-sized aperture. In total, our analysis at 120~pc resolution covers 1,762 kpc-sized apertures in all 28 galaxies, whereas the analysis at 60~pc covers a subsample of 344 apertures in 6 galaxies. A collection of key measurements in our analysis are available in tabular form online (see Table~\ref{tab:mrt} in Appendix~\ref{apdx:mrt}).

We divide our kpc-scale apertures into ``disk'' apertures and ``bulge/bar'' apertures, according to the criterion described in Section~\ref{sec:method}.
Our analysis at 120 (60)~pc resolution covers 1,445 (294) apertures in which no CO flux originates from bulge or bar regions.
When plotting our results, we represent these ``disk'' apertures in blue, and the ``bulge/bar'' apertures in orange.

\subsection{Turbulent Pressure versus Kpc-scale Dynamical Equilibrium Pressure}
\label{sec:result:PDE-kpc}

\begin{figure*}[htp]
\gridline{
\fig{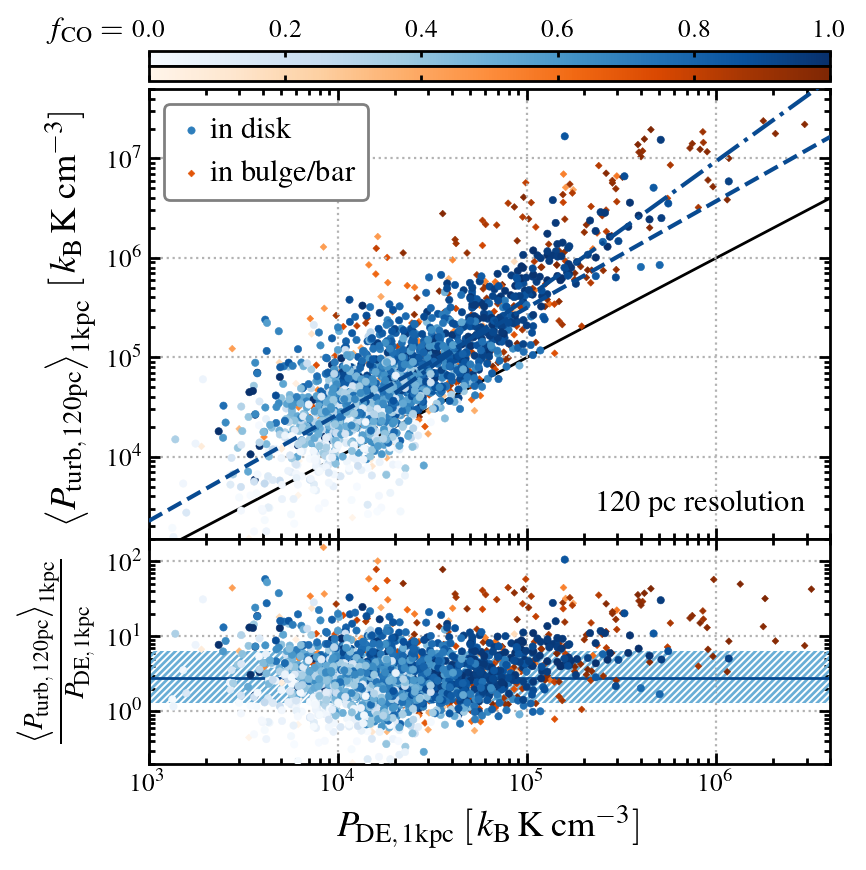}{0.49\textwidth}{}
\fig{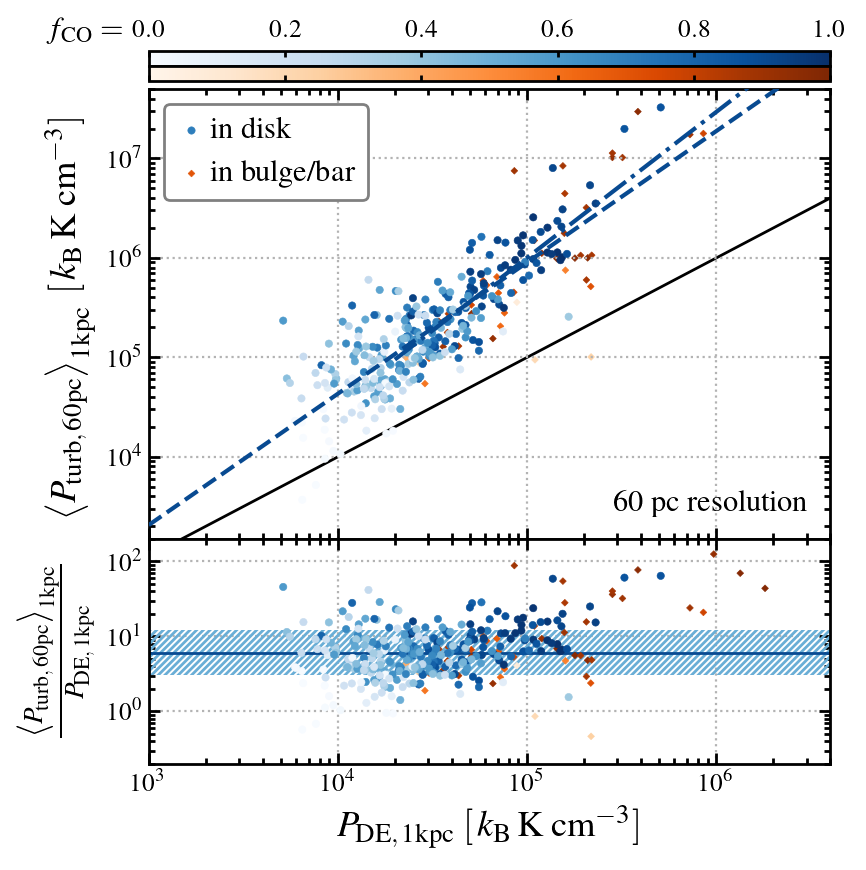}{0.49\textwidth}{}
}
\vspace{-2.5em}
\caption{
\textbf{\textit{Top:}
Average cloud-scale turbulent pressure in the molecular gas, $\Pturbavgkpc$, as a function of the kpc-scale dynamical  equilibrium pressure, $\PDEkpc$.}
Here ``cloud-scale'' means $\theta=120$~pc (left) or $60$~pc (right).
Each data point represents a kpc-sized aperture, where blue and orange symbols denote the ``disk'' and ``bulge/bar'' samples, respectively (Section~\ref{sec:method:region}). Darker color means higher CO flux recovery fraction (Section~\ref{sec:method:fCO}), and therefore less sensitivity-induced systematic uncertainty.
Black solid lines denote equality. Blue dashed and dash-dotted lines denote the best-fit power-law relations for the disk sample, with the former fitted to all data (Equation~\ref{eq:Pturbavg-PDEkpc}), and the latter fitted only to data with $\PDEkpc>2\times10^4\,\uP$ (Equation~\ref{eq:Pturbavg-PDEkpc-alt}).
\textbf{\textit{Bottom:} Ratio between $\Pturbavgkpc$ and $\PDEkpc$ (i.e., the over-pressurized factor) as a function of $\PDEkpc$.}
The blue horizontal line and shaded area denote the median and 1$\sigma$ range of this over-pressurized factor.
The figure shows that $\Pturb$ on cloud scales correlates with, but usually exceeds, the kpc-scale average $\PDE$ calculated by assuming a \textit{smooth} disk in hydrostatic equilibrium.
A logical explanation is that $\PDEkpc$ underestimates the actual ISM weight as it does not account for the locally enhanced gravity in denser sub-structures, where a significant fraction of molecular gas is hosted.
}
\label{fig:Pturb-PDE-kpc}
\end{figure*}

The top panels in Figure~\ref{fig:Pturb-PDE-kpc} show the average molecular gas turbulent pressure $\Pturbavg$\footnote{Given that there is only one averaging scale in this work (i.e., 1~kpc), we will use $\brkt{X}$ as a shorthand for $\brktkpc{X}$ hereafter.} (see Section~\ref{sec:expectation:Pturb}), measured on $\theta=60$~pc and $120$~pc scales, as a function of the kpc-scale dynamical equilibrium pressure, $\PDEkpc$ (see Section~\ref{sec:expectation:PDE-kpc}).
Each data point corresponds to one kpc-sized aperture.
Darker colors denote higher CO flux recovery fraction (see Section~\ref{sec:method}), so the $\Pturbavg$ measurements are more representative of the bulk molecular gas population within the aperture.

At a physical scale of 120~pc, we see $\brkt{P\sbsc{turb,\,120pc}}$ values spanning the range $10^4$--$10^7\,\uP$. The corresponding range in $\PDEkpc$ is $10^3$--$10^6\,\uP$. For reference, the typical GMC internal pressure (i.e., $\Pturb$) in the Solar Neighborhood is $\sim10^5\,\uP$ \citep[see e.g.,][]{Blitz_1993}, whereas the estimated local dynamical equilibrium pressure is $\sim10^4\,\uP$ \citep[see e.g.,][]{Elmegreen_1989}. Typical GMC internal pressure in the Galactic Center or nearby galaxy centers is $\sim10^5$--$10^8\,\uP$ \citep{Oka_etal_2001,DonovanMeyer_etal_2013,Colombo_etal_2014a,Leroy_etal_2015,Walker_etal_2018,Sun_etal_2018,Schruba_etal_2019}. Therefore, one may think of our data as spanning from ``outer disk'' conditions to galaxy centers.

At both 120 and 60~pc resolution, most data points lie above the equality line (solid black line). This suggests that the average internal pressure in molecular gas is usually higher than what is needed to support the weight of a smooth gas disk with its surface density equals the observed kpc-scale average value.

\subsubsection{Quantifying the Over-pressurization of the Molecular Gas}

The pressure excess in the molecular gas is better quantified in the bottom panels in Figure~\ref{fig:Pturb-PDE-kpc}. There the $y$-axis shows the ratio between $\Pturbavg$ and $\PDEkpc$ (i.e., the over-pressurized factor).
At 120~pc resolution, 90\% of the disk sample shows over-pressurized molecular gas. The majority of the remaining 10\% suffers from low CO recovery fraction, and thus we expect these data to be affected by sensitivity-related systematic effects. We find a median over-pressurized factor of 2.8, and a 1$\sigma$ range of 1.3--6.3 at this resolution.
At 60~pc resolution, 99\% of the disk sample indicates over-pressurized molecular gas. The median and 1$\sigma$ range of the over-pressurized factor is 6.0 and 3.1--12.1, respectively.
The difference between the measurements at different resolutions is likely because at 60~pc resolution one can better resolve the denser substructures in molecular gas, which have higher internal pressure.

As mentioned in Section~\ref{sec:expectation:PDE-cloud}, we expect molecular gas to be over-pressurized relative to the expectations for a smooth disk. This is because a significant fraction of molecular gas lives in denser, small-scale substructures, where gravity is locally enhanced. The actual weight of the molecular gas clouds should therefore be higher than the estimation by assuming a smooth disk with the same overall surface density. Given that molecular gas is clumpy at any instant, this argument holds even in a time-averaged sense.

\begin{figure*}[!t]
\gridline{
\fig{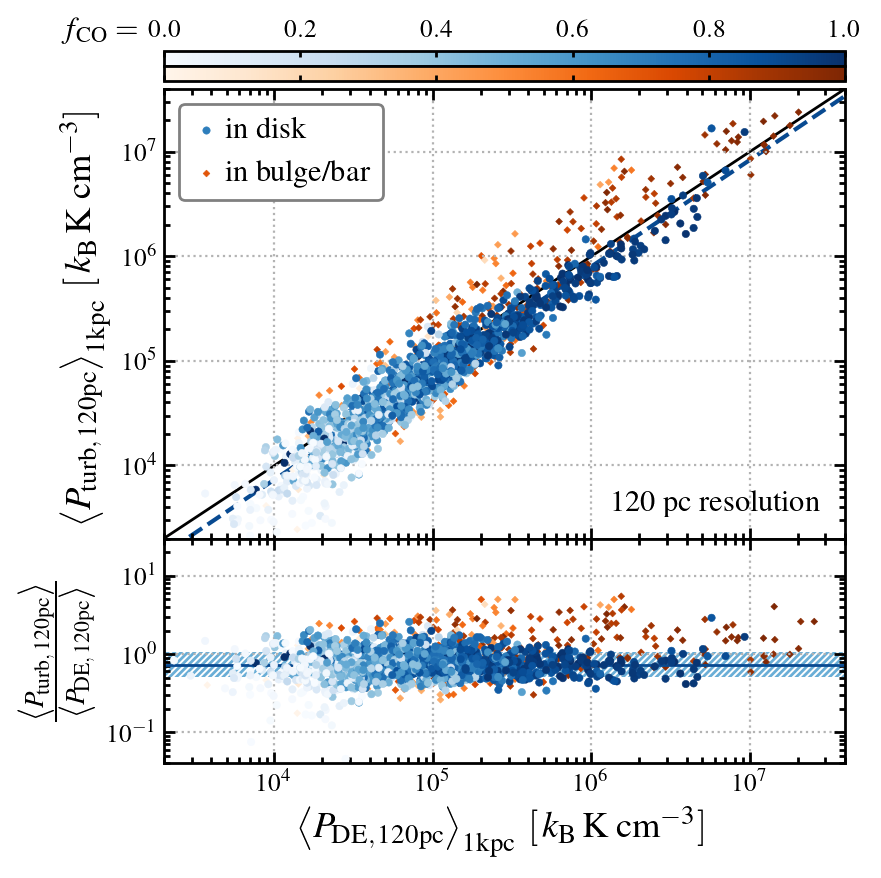}{0.49\textwidth}{}
\fig{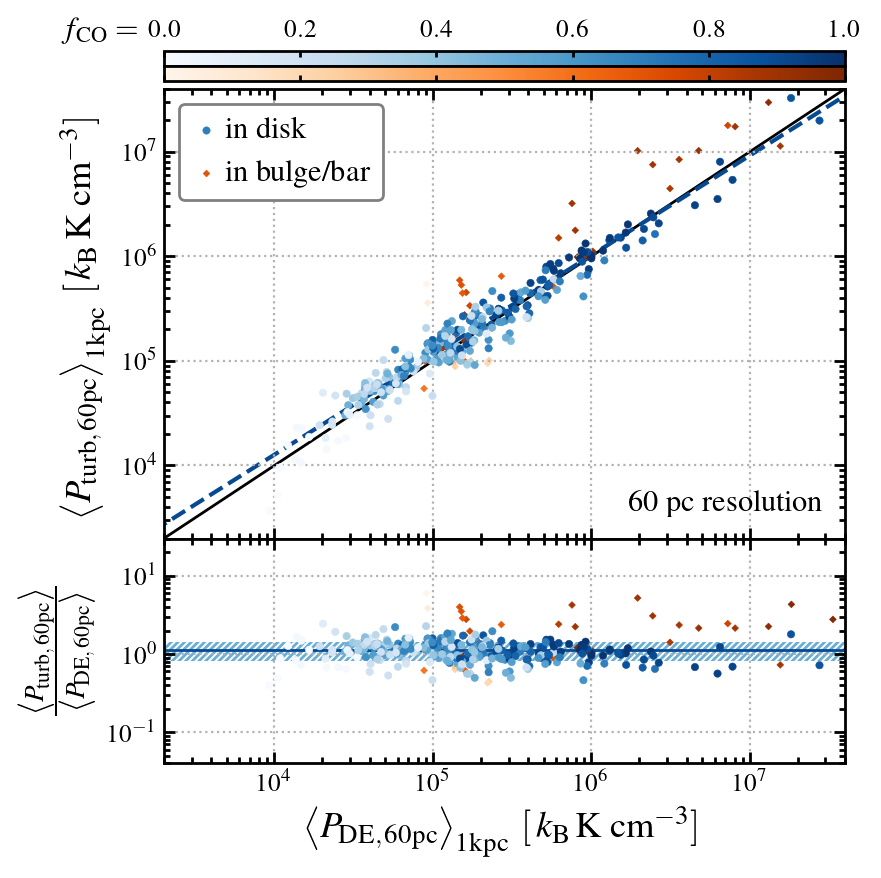}{0.49\textwidth}{}
}
\vspace{-2.5em}
\caption{
\textbf{\textit{Top:}
Average turbulent pressure, $\Pturbavgkpc$ (for $\theta=\rm 120\;or\;60$~pc), as a function of average cloud-scale dynamical equilibrium pressure, $\PDEavgkpc$.}
This $\PDEavgkpc$ is the pressure required to balance the weight of both molecular and atomic gas in the appropriate local potential, with molecular gas substructure taken into account (see Section~\ref{sec:expectation:PDE-cloud}).
In disk regions (blue symbols), $\Pturbavgkpc$ and $\PDEavgkpc$ show a tight correlation, with both slope and normalization close to the expected equality line (solid black line).
This suggests that in the disks of nearby, massive, star-forming galaxies, the turbulent pressure in the molecular gas at 60--120~pc scales agrees with the expectation from dynamical equilibrium at matched spatial scales.
\textbf{\textit{Bottom:} Ratio between $\Pturbavgkpc$ and $\PDEavgkpc$ as a function of $\PDEavgkpc$.}
Labels and lines have the same meaning as in the bottom panels in Figure~\ref{fig:Pturb-PDE-kpc}.
Across the disk sample, the ratio $\brkt{P\sbsc{turb,\,\theta pc}}\sbsc{1kpc}/\brkt{P\sbsc{DE,\,\theta pc}}\sbsc{1kpc}$ is close to unity and shows less scatter than $\brkt{P\sbsc{turb,\,\theta pc}}\sbsc{1kpc}/P\sbsc{DE,\,1kpc}$.
}
\label{fig:Pturb-PDE-cloud}
\end{figure*}

\subsubsection{Predicting Turbulent Pressure from Kpc-scale Equilibrium Pressure}

$\PDEkpc$ contains no information about the small-scale gas distribution, and thus tends to underestimate the true equilibrium pressure on cloud scales. However, calculating $\PDEkpc$ only requires knowing the kpc-scale gas and stellar mass distribution, plus assumptions on the vertical gas velocity dispersion. This makes it possible to estimate $\PDEkpc$ in low resolution observations of more distant galaxies, in low resolution numerical simulations, or even from analytic and semi-analytic models of galaxies.
If the ISM in other environments follows the same $\Pturbavgkpc$--$\PDEkpc$ relation that we observe in Figure~\ref{fig:Pturb-PDE-kpc}, then one could use an estimated $\PDEkpc$ to predict the turbulent pressure on cloud scales.

To make this prediction possible, we fit an empirical $\brkt{P\sbsc{turb,\,120pc}}\sbsc{1kpc}$--$\PDEkpc$ scaling relation. This can be seen as a benchmark relation for the ISM in local star-forming disk galaxies. We derive this relation by fitting a power-law to all the disk measurements (blue circles in Figure~\ref{fig:Pturb-PDE-kpc}), using the ordinary least square (OLS) method in logarithmic space, and treating $\PDEkpc$ as the independent variable. This yields best-fit power-law relations (blue dashed lines in Figure~\ref{fig:Pturb-PDE-kpc}, top panels):
\begin{align}
    \frac{\brktkpc{P\sbsc{turb,\,120pc}}}{10^5\;\uP}
    =&\, 3.2 \left(\frac{\PDEkpc}{10^5\;\uP}\right)^{1.07}~, \nonumber\\
    \frac{\brktkpc{P\sbsc{turb,\,60pc}}}{10^5\;\uP}
    =&\, 9.0 \left(\frac{\PDEkpc}{10^5\;\uP}\right)^{1.32}~.
    \label{eq:Pturbavg-PDEkpc}
\end{align}
\noindent We report the scatter around these best-fit relations, and the estimated statistical uncertainties on the fitting parameters in Table~\ref{tab:fit}.

We caution that Equation~\ref{eq:Pturbavg-PDEkpc} likely has a shallower slope than the actual $\Pturbavgkpc$--$\PDEkpc$ relation. This is largely due to the asymmetric data censoring on $\Pturbavgkpc$ and $\PDEkpc$. The PHANGS-ALMA CO observations have higher surface brightness sensitivity at coarser spatial resolution \citep[Section~\ref{sec:data:kpc}; also see discussion in][]{Sun_etal_2018}. This means that our cloud-scale measurements cannot probe as low molecular gas surface density as our kpc-scale measurements do, and our pressure estimates suffer from a similar censoring effect. The impact of this is even visible in the top left panel in Figure~\ref{fig:Pturb-PDE-kpc}: there are few data points with $\brkt{P\sbsc{turb,\,120pc}}\sbsc{1kpc} \lesssim 10^4\,\uP$, and those few measurements all suffer from low CO recovery fraction. Though not as easily discernible in the top right panel in Figure~\ref{fig:Pturb-PDE-kpc}, a similar censoring effect is also present at 60~pc resolution.

To quantify how this data censoring biases our empirical $\Pturbavgkpc$--$\PDEkpc$ fit, we calculate another version of the best-fit relation using the same OLS method, but only fitting data points with $\PDEkpc > 2\times10^4\,\uP$. The impact of the data censoring is much less prominent above this threshold, and thus we expect these fitting results to be less affected. With this fitting strategy, we have (dash-dotted lines in Figure~\ref{fig:Pturb-PDE-kpc}, top panels)
\begin{align}
    \frac{\brktkpc{P\sbsc{turb,\,120pc}}}{10^5\;\uP}
    =&\, 4.0 \left(\frac{\PDEkpc}{10^5\;\uP}\right)^{1.37}~, \nonumber\\
    \frac{\brktkpc{P\sbsc{turb,\,60pc}}}{10^5\;\uP}
    =&\, 10 \left(\frac{\PDEkpc}{10^5\;\uP}\right)^{1.47}~.
    \label{eq:Pturbavg-PDEkpc-alt}
\end{align}
\noindent Again, we report statistical uncertainties and residual scatters in Table~\ref{tab:fit}.

For the purpose of predicting $\Pturbavgkpc$ from $\PDEkpc$ in low resolution observations or simulations, we recommend to use Equation~\ref{eq:Pturbavg-PDEkpc-alt} in the regime where $\PDEkpc > 2\times10^4\,\uP$. More sensitive observations are needed to pin down this $\Pturbavgkpc$--$\PDEkpc$ relation in lower pressure regimes.

\subsection{Turbulent Pressure versus Cloud-Scale Dynamical Equilibrium Pressure}
\label{sec:result:Pturb-PDE-cloud}

In Section~\ref{sec:result:PDE-kpc} we find that $\Pturbavgkpc$ exceeds $\PDEkpc$ in almost all regions across our sample. Our hypothesis is that this reflects molecular gas clumping on small scales, which is left unaccounted for in the kpc-scale $\PDE$ estimate.
We directly test this hypothesis in Figure~\ref{fig:Pturb-PDE-cloud}, where we show the average molecular gas turbulent pressure $\Pturbavgkpc$ as a function of the cloud-scale dynamical equilibrium pressure $\PDEavgkpc$ (as derived in Section~\ref{sec:expectation:PDE-cloud}). This is a direct ``apples to apples'' comparison in the sense that both $\Pturbavgkpc$ and $\PDEavgkpc$ are derived on the same $\theta = 60$~pc and $120$~pc physical scales.

We observe a tight, almost linear relation between $\brkt{P\sbsc{turb,\,120pc}}$ and $\brkt{P\sbsc{DE,\,120pc}}$ across more than three orders of magnitude. We find a similarly strong correlation on 60~pc scale, though with a slightly different normalization.
Compared to Figure~\ref{fig:Pturb-PDE-kpc}, the observed distribution in Figure~\ref{fig:Pturb-PDE-cloud} shows a relationship much closer to equality (as expected), and much less scatter around the relation as well.

Figure~\ref{fig:Pturb-PDE-cloud} shows that: 1) the dynamical equilibrium model is able to predict the observed turbulent pressure in molecular gas based on the resolved gas and stellar mass distribution in galaxy disks; and 2) to correctly estimate equilibrium pressure within the more clumpy molecular component, it is crucial to account for small-scale density structures, which are only accessible in high spatial resolution observations.

\subsubsection{Differentiating Morphological Regions}

In Figure~\ref{fig:Pturb-PDE-cloud}, we differentiate the measurements in galaxy disks (blue circles) from those in bulge and bar regions (orange diamonds). While most disk measurements fall around the equality line, many measurements in bulge or bar regions show systematically higher $\Pturbavg$.
This likely reflects a stronger impact of large-scale dynamical processes on the ISM in these regions. As pointed out by \citet{Meidt_etal_2018,Meidt_etal_2019} and many others, the gravitational potential in galaxy bulges and bars often has a steeper gradient, and thus it could significantly perturb the gas motions even on $\sim$100~pc scales. In this case, the gas velocity field is strongly anisotropic, and the observed gas velocity dispersion along the line of sight is elevated by the projected in-plane motions.
This could qualitatively explain the higher $\Pturbavg$ relative to $\PDEavg$ in bulge and bar regions.

While the impact of large-scale dynamical processes on molecular gas properties is itself an interesting and important topic \citep[see][]{Kruijssen_Longmore_2013,Kruijssen_etal_2014,Meidt_etal_2018,Meidt_etal_2019,Sormani_etal_2019}, further exploration in this direction is beyond the scope of this work. Hereafter, we only focus on measurements in disk regions, in which case the in-plane orbital motions play only a minor role. Future higher resolution observations targeting the central regions of these galaxies, paired with dynamical modelling exercises, will help resolve the remaining ambiguities.

\subsubsection{Quantifying the Turbulent Pressure--Equilibrium Pressure Relation on Cloud Scales}

To get a quantitative description of the observed $\Pturbavg$--$\PDEavg$ relation, we fit a power-law to all disk measurements (blue dots). We weight each measurement by its corresponding CO recovery fraction $f\sbsc{CO}$, and perform an OLS bisector fit \citep[blue dashed line]{Isobe_etal_1990} in logarithmic space. Combining all disk measurements, we find the following best-fit relations on 120~pc and 60~pc scales:
\begin{align}
    \frac{\brktkpc{P\sbsc{turb,\,120pc}}}{10^5\;\uP}
    =&\, 0.77 \left(\frac{\brktkpc{P\sbsc{DE,\,120pc}}}{10^5\;\uP}\right)^{1.02}~, \nonumber\\
    \frac{\brktkpc{P\sbsc{turb,\,60pc}}}{10^5\;\uP}
    =&\, 1.1 \left(\frac{\brktkpc{P\sbsc{turb,\,60pc}}}{10^5\;\uP}\right)^{0.95}~.
    \label{eq:Pturbavg-PDEavg}
\end{align}
\noindent The corresponding statistical uncertainties and residual scatters are reported in Table~\ref{tab:fit}.

Thanks to both our large sample size and the tightness of the $\Pturbavg$--$\PDEavg$ relation, the statistical errors on the best-fit parameters (as quoted in Equation~\ref{eq:Pturbavg-PDEavg}) are very small. However, our estimates for $\Pturbavg$ and $\PDEavg$ do depend on several assumptions, including the CO-to-H$_2$ conversion factor, the geometry of the stellar disk, and the geometry of molecular gas structures on small scales. In Section~\ref{sec:systematics}, we investigate the systematic uncertainties associated with each of these assumptions.

The normalization of the best-fit $\Pturbavg$--$\PDEavg$ relation appears to be different (by a factor of 1.4) when estimated from data at different resolution (also see bottom panels in Figure~\ref{fig:Pturb-PDE-cloud}). This can be partly attributed to the overall dependence of CO flux recovery fraction on resolution. As mentioned in Section~\ref{sec:result:PDE-kpc}, the sensitivity is often poorer at higher resolution, which means that only the brightest CO emission remains above the detection limit. For this reason, our estimated turbulent pressure at higher resolution suffers a stronger bias towards the brightest CO peaks, which trace high density, high pressure regions. This can qualitatively explain the mildly higher normalization of the $\Pturbavg$--$\PDEavg$ relation at 60~pc resolution.

\begin{deluxetable*}{lccrc}
\tabletypesize{\footnotesize}
\tablecaption{Summary of the Best-fit Power-law Relations in Section~\ref{sec:result}\label{tab:fit}}
\tablewidth{0pt}
\tablehead{
\colhead{Relation} &
\colhead{Figure/Equation} &
\colhead{Slope} &
\colhead{Offset along y-axis} &
\colhead{Scatter along y-axis} \\[-1.5ex]
\colhead{} &
\colhead{} &
\colhead{} &
\colhead{at $10^5\,\uP$} &
\colhead{around relation}
}
\startdata
$\brkt{P\sbsc{turb,\,120pc}}$--$\PDEkpc$ & Fig.\ref{fig:Pturb-PDE-kpc}; Eq.~\ref{eq:Pturbavg-PDEkpc} & $1.07[\pm0.03]\tablenotemark{*}$ & $0.50[\pm0.02]\tablenotemark{*}$~dex & 0.36~dex \\
$\brkt{P\sbsc{turb,\,120pc}}$--$\PDEkpc$ (debias\tablenotemark{$\dagger$}) & Fig.\ref{fig:Pturb-PDE-kpc}; Eq.~\ref{eq:Pturbavg-PDEkpc-alt} & $1.37[\pm0.05]\tablenotemark{*}$ & $0.60[\pm0.02]\tablenotemark{*}$~dex & 0.32~dex \\
$\brkt{P\sbsc{turb,\,120pc}}$--$\brkt{P\sbsc{DE,\,120pc}}$ & Fig.\ref{fig:Pturb-PDE-cloud}; Eq.~\ref{eq:Pturbavg-PDEavg} & $1.02[\pm0.01]\tablenotemark{*}$ & $-0.12[\pm0.01]\tablenotemark{*}$~dex & 0.17~dex \\
\hline
$\brkt{P\sbsc{turb,\,60pc}}$--$\PDEkpc$ & Fig.\ref{fig:Pturb-PDE-kpc}; Eq.~\ref{eq:Pturbavg-PDEkpc} & $1.32[\pm0.06]\tablenotemark{*}$ & $0.96[\pm0.04]\tablenotemark{*}$~dex & 0.31~dex \\
$\brkt{P\sbsc{turb,\,60pc}}$--$\PDEkpc$ (debias\tablenotemark{$\dagger$}) & Fig.\ref{fig:Pturb-PDE-kpc}; Eq.~\ref{eq:Pturbavg-PDEkpc-alt} & $1.47[\pm0.09]\tablenotemark{*}$ & $1.00[\pm0.04]\tablenotemark{*}$~dex & 0.26~dex \\
$\brkt{P\sbsc{turb,\,60pc}}$--$\brkt{P\sbsc{DE,\,60pc}}$ & Fig.\ref{fig:Pturb-PDE-cloud}; Eq.~\ref{eq:Pturbavg-PDEavg} & $0.95[\pm0.02]\tablenotemark{*}$ & $0.06[\pm0.01]\tablenotemark{*}$~dex & 0.13~dex \\
\enddata
\tablenotetext{*}{All the quoted errors here are \textit{statistical} errors estimated from bootstrapping. However, we expect \textit{systematic} errors to dominate the total uncertainties on these parameters. See Section~\ref{sec:systematics}.}
\tablenotetext{\dagger}{These relations are derived in the range $\PDE > 2\times10^4\,\uP$. Caution should be used when extrapolating outside this range.}
\vspace{-2em}
\end{deluxetable*}

\subsubsection{Inferring the Dynamical State of Molecular Gas from the ISM Weight Budget}

\begin{figure*}[!t]
\centering
\fig{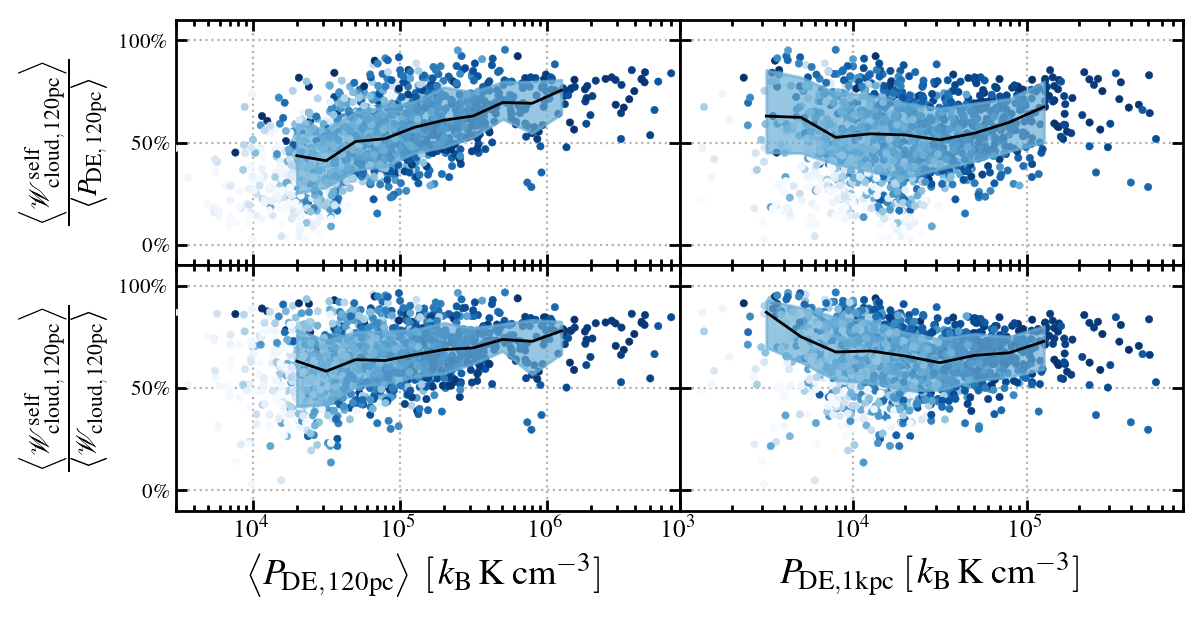}{0.85\textwidth}{}
\vspace{-2.5em}
\caption{
\textbf{\textit{Top row:}
Fractional contribution from the self-gravity of cloud-scale molecular structures to the total ISM weight budget ($\brkt{\mathcal{W}\spsc{self}\sbsc{cloud,\,120pc}}/\brkt{P\sbsc{DE,\,120pc}}$), shown as a function of dynamical equilibrium pressure estimated on cloud scales ($\brkt{P\sbsc{DE,\,120pc}}$, \textit{left}) and on kpc scales ($P\sbsc{DE,\,1kpc}$, \textit{right}), in galaxy disks.}
The running median (black line) and 16-84\% percentile trends (blue shaded region) suggest that: (a) the self-gravity of individual molecular structures typically accounts for $33$--$70$\% of the total ISM weight, whereas the remainder is attributed to gravity associated with external material and pressure in the ambient atomic gas); and (b) self-gravity is more likely to be dominant when molecular structures have higher internal pressure ($\brkt{P\sbsc{DE,120pc}}$); yet no clear trend is seen with the large-scale environment pressure ($\PDEkpc$).
\textbf{\textit{Bottom row:}
Fractional contribution of self-gravity to the internal weight of cloud-scale molecular structures ($\brkt{\mathcal{W}\spsc{self}\sbsc{cloud,\,120pc}}/\brkt{\mathcal{W}\sbsc{cloud,\,120pc}}$), again shown as a function of the two equilibrium pressure estimates.}
We conclude that (a) self-gravity dominates the internal weight felt by these structures across most of our sample; and (b) there is a mild trend of the self-gravity term being more dominant in structures with high internal pressure.
}
\label{fig:frac-PDE}
\end{figure*}

In our formulation
(see Section~\ref{sec:expectation:PDE-cloud} and Appendix~\ref{apdx:formulation}), all the terms contributing to the cloud-scale equilibrium pressure $\PDEavg$ can be grouped into three classes: (1) the weight of the cloud-scale molecular structures due to their self-gravity ($\mathcal{W}\spsc{self}\sbsc{cloud}$; referred to as the ``self-gravity term'' hereafter); (2) the weight of these molecular structures due to the gravity associated with external material (including both $\mathcal{W}\spsc{ext\text{--}mol}\sbsc{cloud}$ and $\mathcal{W}\spsc{star}\sbsc{cloud}$; referred to as the ``external gravity terms''); and (3) the weight of the ambient atomic ISM in the combined potential created by stars and gas ($\mathcal{W}\sbsc{atom}$; the ``ambient pressure term'').
The relative importance of these terms offers clues on a key question: which factor plays a more prominent role in governing the dynamical state of molecular structures like GMCs --- is it self-gravity, external gravity, or ambient pressure?

The top panels in Figure~\ref{fig:frac-PDE} show the fractional contribution of the self-gravity term in the total $\brkt{P\sbsc{DE}}$ estimate, as a function of dynamical equilibrium pressure estimated on 120~pc scale ($\brkt{P\sbsc{DE,120pc}}$, top-left panel) and on kpc-scale ($\PDEkpc$, top-right panel). We find that the self-gravity term typically accounts for $\sim$33--70\% of the total $\brkt{P\sbsc{DE}}$, and its fractional contribution exceeds $50\%$ in about half of our disk sample. In the other half of our disk sample, the combination of external gravity terms and ambient pressure term dominate the self-gravity term.
In this case, the dynamical state of molecular structures like GMCs is strongly influenced by pressure in the ambient atomic ISM, and/or the gravitational potential created by stars and gas external to a given molecular structure.

The black lines and blue shaded regions in Figure~\ref{fig:frac-PDE} represent the running median and 16--84\% percentile trends.
According to the trends shown in the top panels, the relative importance of the self-gravity term in the total ISM weight appears to increase with increasing $\brkt{P\sbsc{DE,120pc}}$ (rank correlation coefficient $\rho=0.62$, corresponding $p$-value $\ll0.001$), while it correlates less well with $\PDEkpc$ ($\rho=0.12$, $p\ll0.001$).

As discussed above, $\brkt{P\sbsc{DE,120pc}}$ reflects the pressure \textit{within} the molecular gas (as seen by its tight correlation with $\Pturbavg$), whereas $\PDEkpc$ represents the pressure in the kpc-scale environment (when neglecting the substructure in the molecular gas).
The observed trends in the top two panels in Figure~\ref{fig:frac-PDE} can thus be interpreted as follows:
across our sample, the dynamical state of cloud-scale molecular gas structures (hereafter ``molecular structures'') is strongly related to their internal pressure. Structures with high internal pressure ($\gtrsim10^5\,\uP$) are more likely to be self-gravity-dominated, whereas those with low internal pressure are more likely to be external gravity- and/or ambient pressure-dominated.
The large-scale environment pressure, however, offers less predicting power --- the correlation is much less monotonic, and the chance of finding self-gravity-dominated molecular structures is about the same in low pressure environments as in high pressure environments within our sample.

The above discussion considers the relative importance of the self-gravity term in the total ISM weight budget.
Alternatively, one could focus on the gravity felt by the molecular structures, and ask: ``What fraction of the total weight of these structures ($\mathcal{W}\sbsc{cloud,\,\theta pc}$; see Equation~\ref{eq:W_cloud}) is due to their self-gravity, as opposed to external gravity?''
To address this question, we plot the fractional contribution of the self-gravity term to the total weight of the cloud-scale molecular structures in the bottom panels in Figure~\ref{fig:frac-PDE}.
We find that the self-gravity term dominates the total internal weight of cloud-scale molecular structures in most (83\%) of our disk sample. That is, in most cases, the observed molecular structures are dense enough to significantly alter the local gravitational potential. Moreover, in the cases when dynamical equilibrium holds and the ambient pressure is negligible, most of these molecular structures would be self-gravitating.

In the cases when self-gravity fails to outweigh external gravity, however, the molecular structures in question are likely not ``significant'' over-densities.
We do not expect these molecular structures to be decoupled from large-scale dynamics, and if this remains the case, these structures might ``dissolve'' over roughly a galactic dynamical timescale.
This picture is in line with recent findings by \citet{Chevance_etal_2020}, that the lifetime of molecular clouds in some cases is driven by the timescales of galactic dynamical processes.

In the bottom panels of Figure~\ref{fig:frac-PDE}, the contribution of self-gravity to the total weight of molecular structures shows a positive correlation with $\brkt{P\sbsc{DE,120pc}}$ (bottom-left; rank correlation coefficient $\rho=0.34$, corresponding $p$-value $\ll0.001$), and a very mild negative correlation with $\PDEkpc$ (bottom-right; $\rho=-0.07$, $p=0.006$). These trends appear to indicate that molecular structures with high internal pressure and/or in low pressure environments are less likely to be external gravity-dominated.

We note that our estimates of the relative importance of molecular gas self-gravity may be biased by sensitivity and (spatial) resolution related effects.
The finite sensitivity of the CO data might introduce a selection bias against low surface density molecular gas in low density, low pressure environments (see discussed in Section~\ref{sec:result:PDE-kpc}). This selection bias offers a likely alternative explanation for the apparently high $\brkt{\mathcal{W}\spsc{self}\sbsc{cloud,\,120pc}}/\brkt{P\sbsc{DE,\,120pc}}$ and $\brkt{\mathcal{W}\spsc{self}\sbsc{cloud,\,120pc}}/\brkt{\mathcal{W}\sbsc{cloud,\,120pc}}$ ratios at the low $\PDEkpc$ end (Figure~\ref{fig:frac-PDE}, right column).

On the other hand, the finite spatial resolution of the CO data means that we do not have access to the sub-beam density distribution of molecular gas. Our assumption of a uniform density sphere filling each beam could lead to under-estimations of $\Sigmol$ and $\brkt{\mathcal{W}\spsc{self}\sbsc{cloud,\,120pc}}$ if the actual sub-beam density distribution is strongly clumped. It is not trivial to predict how this bias would affect the trends we observe in Figure~\ref{fig:frac-PDE}, because it remains unclear how different cloud/environment properties affect the clumping of molecular gas below these scales.
In the future, CO observations with higher sensitivity and higher spatial resolution targeting low pressure environments will help to eliminate these systematic effects.

\subsubsection{A Physical Picture of Molecular Gas Dynamics on Cloud Scales}

The results shown in Figure~\ref{fig:Pturb-PDE-cloud} and \ref{fig:frac-PDE} together lead to the following conclusions.
In a typical star-forming disk environment, the observed turbulent pressure in molecular gas on 60--120~pc scales can be explained by dynamical equilibrium holding down to cloud scales.
Molecular structures with high internal pressure ($\brkt{\Pturb}\approx\brkt{\PDE}\gtrsim10^5\,\uP$) appear more dominated by self-gravity; structures with lower internal pressure appear more heavily influenced by ambient pressure and/or external gravity.
We can find structures in either of these two regimes at any environment pressure in our sample.

These observations likely signal an important transition in the dynamical state of cloud-scale molecular structures --- with increasing internal pressure and thus increasing pressure contrast against the environment, molecular structures shift from existing at the ambient ISM pressure and participating in the large-scale dynamical motions to being over-pressurized and confined by the enhanced self-gravity \citep[also see][]{Field_etal_2011,Meidt_2016,Sun_etal_2018,Meidt_etal_2018,Meidt_etal_2019,Schruba_etal_2019}. The difference in the range of molecular gas internal pressure and environment pressure probed by our sample allows us to cover both regimes in our analysis.

Our data suggest that dynamical equilibrium on cloud scales holds whether the weight of molecular structures is dominated by self-gravity or not. Our sample spans both types of regimes, and the $\Pturbavg$--$\PDEavg$ correlation appears to hold across the whole sample.
This strongly supports (a) the idea that cloud-scale molecular structures do appear to exist in dynamical equilibrium (in a statistical averaged sense), and (b) that our formalism captures the relevant physics across the full range of physical conditions probed by our sample.


\section{Systematic Effects} \label{sec:systematics}

\defcitealias{Accurso_etal_2017}{A17}
\defcitealias{Narayanan_etal_2012}{N12}
\defcitealias{Bolatto_etal_2013}{B13}

\begin{deluxetable*}{lcccccl}
\tabletypesize{\footnotesize}
\tablecaption{$\brkt{P\sbsc{turb,\,120pc}}$--$\brkt{P\sbsc{DE,\,120pc}}$ Relations Derived from Various Approaches\label{tab:systematic}}
\tablewidth{0pt}
\tablehead{
\colhead{Methodology Choice} &
\multicolumn{2}{c}{16, 50, 84\% Percentiles of} &
\multicolumn{2}{c}{Best-fit Power-law\tablenotemark{*}} &
\colhead{Residual Scatter in} &
\colhead{Text/Figures} \\[-1.5ex]
\colhead{} &
\colhead{$\log\brkt{P\sbsc{turb,120pc}}$} &
\colhead{$\log\brkt{P\sbsc{DE,120pc}}$} &
\colhead{Slope $\beta$} &
\colhead{Offset $A$} &
\colhead{$\log\brkt{P\sbsc{turb,120pc}}$} &
\colhead{}
}
\startdata
Fiducial method & (4.21, 4.69, 5.26) & (4.37, 4.82, 5.36) & 1.02 & -0.12~dex & 0.17~dex & \S\ref{sec:result:Pturb-PDE-cloud}; Fig.~\ref{fig:Pturb-PDE-cloud} \\
\hline
Galactic $\alphaCO$ value & (4.12, 4.61, 5.22) & (4.27, 4.70, 5.28) & 1.01 & -0.09~dex & 0.17~dex & \S\ref{sec:systematics:alphaCO}; Fig.~\ref{fig:Pturb-PDE-alphaCO} \\
\citetalias{Narayanan_etal_2012} $\alphaCO$ prescription & (4.46, 4.92, 5.41) & (4.72, 5.19, 5.62) & 1.08 & -0.26~dex & 0.17~dex & \S\ref{sec:systematics:alphaCO}; Fig.~\ref{fig:Pturb-PDE-alphaCO} \\
\citetalias{Bolatto_etal_2013} $\alphaCO$ prescription & (4.55, 4.88, 5.29) & (4.87, 5.11, 5.41) & 1.34 & -0.26~dex & 0.17~dex & \S\ref{sec:systematics:alphaCO}; Fig.~\ref{fig:Pturb-PDE-alphaCO} \\
\hline
Flared stellar disk & (4.21, 4.69, 5.26) & (4.31, 4.77, 5.34) & 0.99 & -0.08~dex & 0.17~dex & \S\ref{sec:systematics:PDE} \\
\hline
Cloud statistics & \multirow{2}{*}{(4.39, 4.79, 5.29)} & \multirow{2}{*}{(4.44, 4.80, 5.33)} & \multirow{2}{*}{1.05} & \multirow{2}{*}{-0.01~dex} & \multirow{2}{*}{0.27~dex} & \multirow{2}{*}{\S\ref{sec:systematics:CPROPS}; Fig.~\ref{fig:Pturb-PDE-CPROPS}b} \\
(fixed l.o.s. depth) & & & & & \\
\enddata
\tablenotetext{*}{Here the power-law parameters are defined as $\log_{10}\left(\frac{\brkt{P\sbsc{turb,\,120pc}}}{10^5\;\uP}\right) = \beta\;\log_{10}\left(\frac{\brkt{P\sbsc{DE,\,120pc}}}{10^5\;\uP}\right) + A$. Across all rows, the amplitude of variations in the best-fit $\beta$ and $A$ values roughly reflect their systematic uncertainties. For a reference, the corresponding statistical errors are 0.01 for $\beta$, and 0.01~dex for $A$ (see Table~\ref{tab:fit}).}
\vspace{-2em}
\end{deluxetable*}

Our analysis involves estimates of multiple physical quantities. Deriving these from observables requires making assumptions about, for example, how CO emission traces molecular gas mass (see Section~\ref{sec:method}), or the geometry of the stellar disk (see Section~\ref{sec:expectation:PDE-kpc}).
In Section~\ref{sec:systematics:alphaCO} we consider the impact of our adopted CO-to-H$_2$ conversion factor treatment.
In Section~\ref{sec:systematics:PDE} we vary some of the assumptions that enter into $\PDE$, testing the effects of a different stellar disk geometry and a different gas velocity dispersion.

Another major methodological choice made in our analysis is that, rather than attempting to identify clouds using any segmentation algorithm, we derive cloud-scale gas properties directly from the observed CO intensity distribution by statistical analysis.
In Section~\ref{sec:systematics:CPROPS} we compare our method with an alternative method that relies on cloud segmentation.

To illustrate how these choices would impact our main results, in Table~\ref{tab:systematic} we summarize the $\brkt{P\sbsc{turb,\,120pc}}$--$\brkt{P\sbsc{DE,\,120pc}}$ relation derived from each alternative approach.
Variations of the best-fit power-law slope and intercept among these results provide us with an estimate of the systematic uncertainties on them.
The quoted 16, 50, 84\% percentiles of $\brkt{P\sbsc{turb,\,120pc}}$ and $\brkt{P\sbsc{DE,\,120pc}}$ reveal how each approach impacts these two pressure estimates individually.

\subsection{CO-to-H\textsubscript{2} Conversion Factor} \label{sec:systematics:alphaCO}

\begin{figure}[t]
\fig{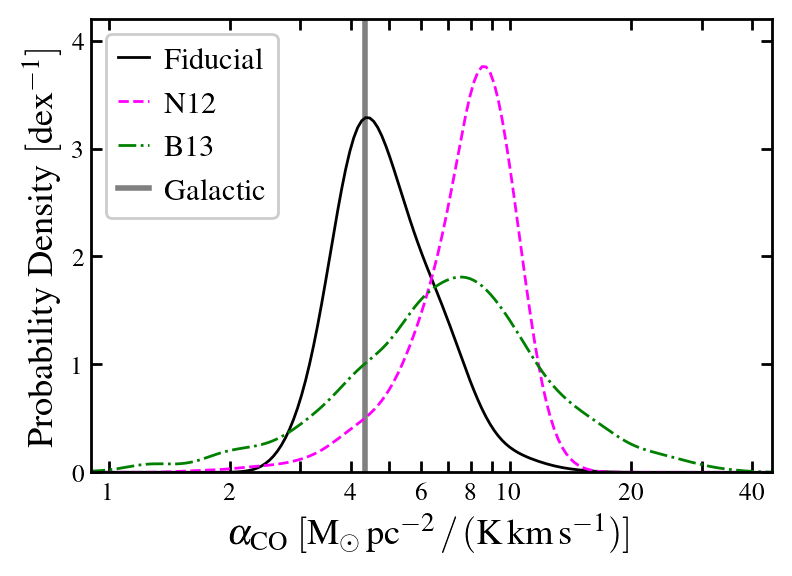}{0.45\textwidth}{}
\vspace{-2.5em}
\caption{
\textbf{Distribution of $\alphaCO$ values across our sample, as predicted by four different prescriptions.}
These probability density functions are estimated through Gaussian kernel density estimations (with 0.05~dex bandwidth). Our fiducial prescription (black solid line) leads to a distribution peaking near the Galactic value (gray vertical line). The \citetalias{Narayanan_etal_2012} prescription (magenta dashed line) predicts comparatively higher $\alphaCO$ values in most cases. The \citetalias{Bolatto_etal_2013} prescription (green dot-dashed line) leads to a much wider distribution than the fiducial and the \citetalias{Narayanan_etal_2012} prescriptions.
}
\label{fig:alphaCO-PDF}
\end{figure}

\begin{figure*}[htp]
\hspace{-0.7em}\fig{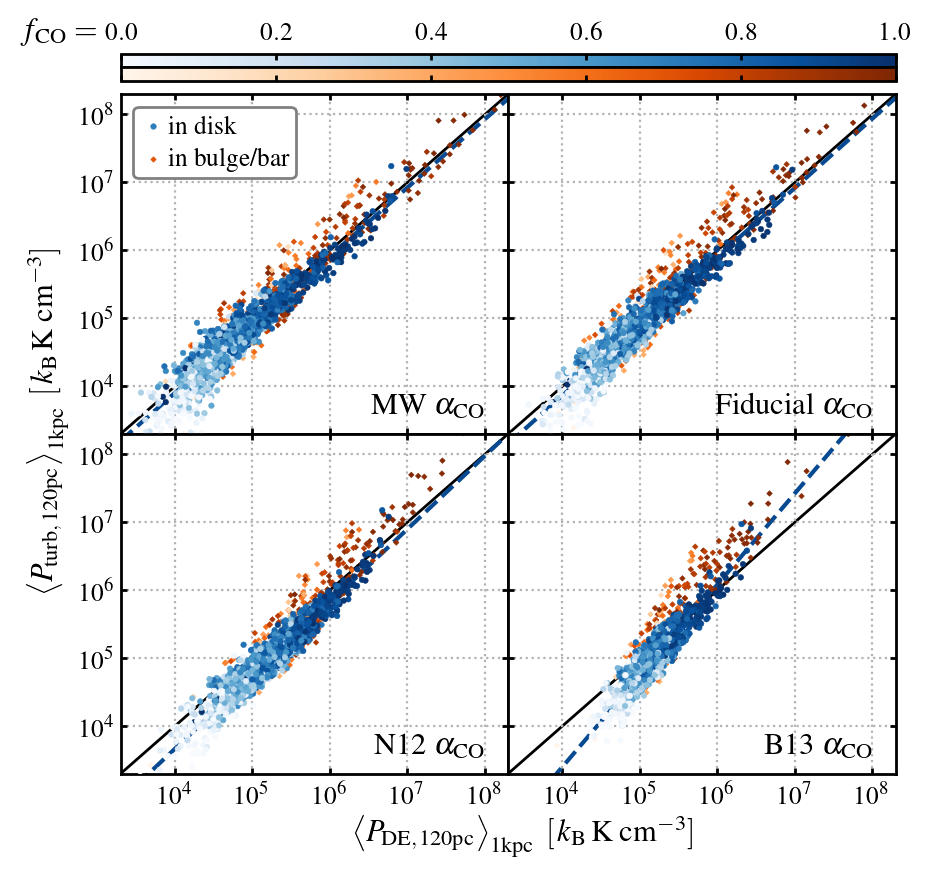}{0.65\textwidth}{}
\vspace{-2.5em}
\caption{
\textbf{The $\brkt{P\sbsc{turb,\,120pc}}$--$\brkt{P\sbsc{DE,\,120pc}}$ relations derived using four different prescriptions for the CO-to-H$_2$ conversion factor, $\alphaCO$.}
The top left panel shows result for a constant, Galactic $\alphaCO$. The top right panel corresponds to our fiducial $\alphaCO$ prescription \citep[similar to][]{Accurso_etal_2017}, and the bottom panels correspond to prescriptions suggested by \citet{Narayanan_etal_2012}, and
\citet{Bolatto_etal_2013}.
The blue dashed lines represent the best-fit power-law relations for all the disk measurements.
Adopting the Galactic $\alphaCO$, our fiducial prescription, or the \citet{Narayanan_etal_2012} prescription all lead to similar $\brkt{P\sbsc{turb,\,120pc}}$--$\brkt{P\sbsc{DE,\,120pc}}$ relations (blue dashed line) that are consistent with the equality line (black solid line), whereas adopting the \citet{Bolatto_etal_2013} prescription leads to a super-linear relation (see Table~\ref{tab:systematic}).
}
\label{fig:Pturb-PDE-alphaCO}
\end{figure*}

In this work, we adopt a metallicity-dependent $\alphaCO$ prescription (see Section~\ref{sec:method}), which is similar to the prescription suggested by \citet[hereafter \citetalias{Accurso_etal_2017}]{Accurso_etal_2017}. Quite a few alternative prescriptions exist in the literature \citep[e.g.,][]{Wolfire_etal_2010,Glover_MacLow_2011,Feldmann_etal_2012,Narayanan_etal_2012,Bolatto_etal_2013}. However, none of these prescriptions provides a concrete, observationally tested estimate of $\alphaCO$ that simultaneously captures the effects of metallicity, radiation field, and gas dynamics.

Our choice of $\alphaCO$ prescription could affect our results.
Both $\Sigmol$ and $\Pturb$ are proportional to $\alphaCO$, and the molecular gas self-gravity term in $\PDE$ is proportional to $\alphaCO^2$. To estimate the amount of systematic uncertainty associated with the choice of $\alphaCO$, we re-derive our key measurements using three alternative $\alphaCO$ prescriptions, and compare them with our fiducial prescription. The three alternative prescriptions are: 1) a constant, Galactic conversion factor, 2) a simulation-based $\alphaCO$ calibration suggested by \citet[hereafter \citetalias{Narayanan_etal_2012}]{Narayanan_etal_2012}, and 3) an empirical $\alphaCO$ prescription suggested by \citet[hereafter \citetalias{Bolatto_etal_2013}]{Bolatto_etal_2013}.

For the constant $\alphaCO$ prescription, we use the Galactic value suggested by \citetalias{Bolatto_etal_2013}:
\begin{equation}
    \alpha\sbsc{CO,\,MW} = 4.35\;\uSig\,(\uIco)^{-1}~.
\label{eq:alphaCO_MW}
\end{equation}

For the \citetalias{Narayanan_etal_2012} prescription, we predict $\alphaCO$ in each kpc-sized aperture from the metallicity ($Z'$) and the flux-weighted CO intensity ($\brkt{I\sbsc{CO(1-0)}}$), following their equation~11:
\begin{align}
    \alpha\sbsc{CO,\,N12} =&\; 8.5\;\uSig\,(\uIco)^{-1} \times Z'^{-0.65} \times \nonumber\\
    & \mathrm{min}\left[1,\;1.7\times \left(\frac{\brkt{I\sbsc{CO(1-0)}}}{\uIco}\right)^{-0.32} \right]~.
\label{eq:alphaCO_N12}
\end{align}
\noindent We note that our implementation includes a factor of 1.36 correction for the mass of heavy elements, which was not included in the original \citetalias{Narayanan_etal_2012} prescription. The quantity $\brkt{I\sbsc{CO(1-0)}}$ here is estimated from its \CO21\ counterpart, $\brkt{I\sbsc{CO(2-1),120pc}}$, assuming $R_{21}=0.7$.

For the \citetalias{Bolatto_etal_2013} prescription, we predict $\alphaCO$ from the metallicity ($Z'$), typical GMC surface density ($\Sigma\sbsc{GMC}$), and kpc-scale total surface density of both gas and stars ($\Sigma\sbsc{total}$), following their equation~31:
\begin{align}
    \alpha\sbsc{CO,\,B13} =&\; 2.9\;\uSig\,(\uIco)^{-1} \times \nonumber\\
    &\exp\left(\frac{0.4}{Z'\,\Sigma\sbsc{GMC}\spsc{100}}\right) \times \left(\Sigma\sbsc{total}\spsc{100}\right)^{-\gamma}~, \label{eq:alphaCO_B13}\\
    &\text{with}\; \gamma =
    \begin{cases}
        0.5, & \text{if}\ \Sigma\sbsc{total}\spsc{100} > 1 \\
        0. & \text{otherwise}
    \end{cases} \nonumber
\end{align}
\noindent Here $\Sigma\sbsc{GMC}\spsc{100}$ and $\Sigma\sbsc{total}\spsc{100}$ are the corresponding mass surface densities normalized to $100\;\uSig$.
Because neither of them could be derived from our observations without knowing $\alphaCO$ \textit{a priori}, we set
\begin{align}
    \Sigma\sbsc{GMC} &= \frac{\alphaCO}{R_{21}} \brkt{I\sbsc{CO(2\text{--}1),\,120pc}}~, \label{eq:SigGMC}\\
    \Sigma\sbsc{total} &= \frac{\alphaCO}{R_{21}}\, I\sbsc{CO(2\text{--}1),\,1kpc} + \Sigma\sbsc{HI,\,1kpc} + \Sigma\sbsc{star,\,1kpc}~,
\end{align}
\noindent and then iteratively solve for $\alphaCO$ in each kpc-scale aperture.
We note that this iterative approach does not guarantee convergence, and 1.5\% of the apertures in our sample do not yield a good solution. We discard the measurements for these apertures from this part of the analysis.

Figure~\ref{fig:alphaCO-PDF} shows the probability density functions (PDFs) of the predicted $\alphaCO$ across our sample for each of the four prescriptions.
Our fiducial prescription leads to an $\alphaCO$ distribution that peaks around the Galactic value. This agreement is largely by construction, as the fiducial prescription itself is normalized to the Galactic conversion factor at Solar metallicity.
The \citetalias{Narayanan_etal_2012} prescription predicts comparatively higher $\alphaCO$ values, with the distribution peaking at around $8.5\;\uSig\,(\uIco)^{-1}$. This coincides with the ``turning point'' where the dependency on $\brkt{I\sbsc{CO(1-0)}}$ switches off (see Equation~\ref{eq:alphaCO_N12}). Therefore, the location of this peak is likely determined by the prescription itself rather than the input data.
The \citetalias{Bolatto_etal_2013} prescription produces a much wider $\alphaCO$ distribution  compared to the other two distributions. This is attributable to the exponential term in Equation~\ref{eq:alphaCO_B13}, which is a stronger dependence on metallicity than any of the other prescriptions. The \citetalias{Bolatto_etal_2013} prescription also tends to predict higher-than-Galactic $\alphaCO$ values. This is likely driven by the relatively low $\Sigma\sbsc{GMC}$ values implied by Equation~\ref{eq:SigGMC} (the median value in our sample is $\Sigma\sbsc{GMC} \approx 30\;\uSig$).

We demonstrate how our adopted $\alphaCO$ prescription affects our main conclusions in Figure~\ref{fig:Pturb-PDE-alphaCO}.
We show four versions of the $\brkt{P\sbsc{turb,\,120pc}}$--$\brkt{P\sbsc{DE,\,120pc}}$ relation, each of which corresponds to a different $\alphaCO$ prescription.
In all four panels, the data clusters around the line of equality. This is because both $\brkt{P\sbsc{turb,\,120pc}}$ and $\brkt{P\sbsc{DE,\,120pc}}$ correlate positively with $\alphaCO$, and thus the choice of $\alphaCO$ prescription has less impact on their ratio. However, the choice of $\alphaCO$ prescription does have a more apparent impact on the absolute pressure values. As visible in Figure~\ref{fig:Pturb-PDE-alphaCO}, the \citetalias{Narayanan_etal_2012} and \citetalias{Bolatto_etal_2013} prescriptions both push the whole distribution towards higher values of both pressures (also see Table~\ref{tab:systematic} for quantitative results showing this trend). This is exactly what we would expect from the $\alphaCO$ PDFs: \citetalias{Narayanan_etal_2012} predicts on average higher $\alphaCO$, and thus higher pressure. \citetalias{Bolatto_etal_2013} tends to predict higher $\alphaCO$ in disk regions, which pushes points near the low pressure end up to higher pressures.

For each prescription, we fit a power-law relation to all measurements from disk regions. We report the results in Table~\ref{tab:systematic}.
All $\alphaCO$ prescriptions except \citetalias{Bolatto_etal_2013} yield almost linear $\brkt{P\sbsc{turb,\,120pc}}$--$\brkt{P\sbsc{DE,\,120pc}}$ relations. The \citetalias{Bolatto_etal_2013} prescription instead leads to a super-linear slope of 1.34, with an estimated statistical uncertainty of $\sim$0.01. This is driven by the higher predicted $\alphaCO$ in low surface density and low metallicity environments.

In summary, adopting a different $\alphaCO$ prescription does not change the conclusion that the observed $\brkt{P\sbsc{turb,\,120pc}}$--$\brkt{P\sbsc{DE,\,120pc}}$ distribution lies near the equality line. Adopting the \citetalias{Bolatto_etal_2013} prescription makes its slope significantly steeper than linear, while adopting the other three prescriptions all leads to nearly linear slopes. Adopting different prescriptions does significantly change the observed range of both pressures. These results illustrate the importance of quantifying $\alphaCO$ variations, and motivate future works to provide better constraints on the potential dependence of $\alphaCO$ on key physical properties, including metallicity, radiation field, gas (column) density and dynamics.

\subsection{Calculation of Equilibrium Pressure} \label{sec:systematics:PDE}

\begin{figure}[htp]
\gridline{
\fig{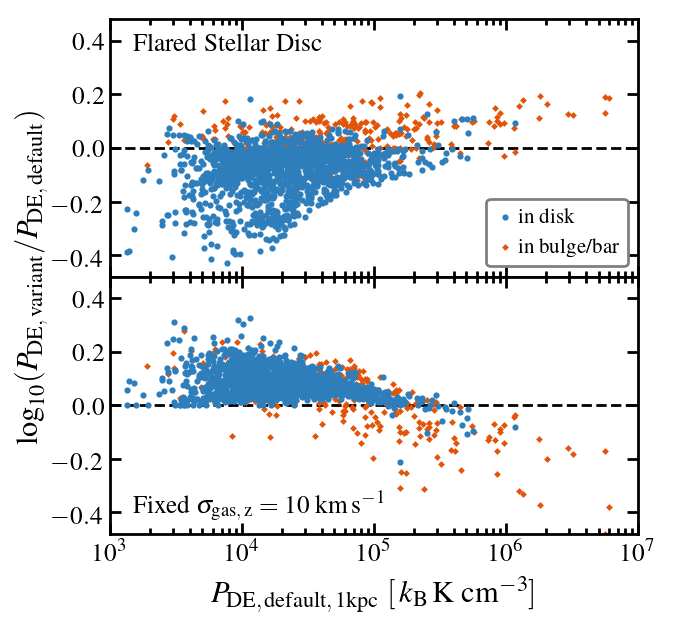}{0.45\textwidth}{}
}
\vspace{-2.5em}
\caption{
\textbf{Comparing our fiducial $\PDEkpc$ estimate to two alternative estimates.}
\textit{Top}:
Assuming a flared stellar disk \citep{Ostriker_etal_2010} results in lower $\PDEkpc$ in outer disks and slightly higher $\PDEkpc$ in bulge/bar regions.
\textit{Bottom}:
Assuming a fixed $\siggasz=10\rm\;km\,s^{-1}$ \citep{Leroy_etal_2008,Hughes_etal_2013a} leads to higher $\PDEkpc$ estimates in disk regions.
In either case, the differences in these $\PDEkpc$ estimates are often smaller than 0.2~dex, meaning that the systematic uncertainties on $\PDEkpc$ associated with these assumptions are no larger than a factor of 1.6.}
\label{fig:PDE-variants}
\end{figure}

Calculating $\PDEkpc$ and $\PDEavg$ requires knowing the three dimensional distribution of stars and gas in the galaxy disk. In Sections~\ref{sec:expectation:PDE-kpc} and \ref{sec:expectation:PDE-cloud}, we made a few assumptions to help us infer $\PDEkpc$ and $\PDEavg$ from the observed two dimensional projected quantities. In our fiducial approach, we assumed that a) the stellar disk scale height $\Hstar$ is proportional to the radial scale length $\Rstar$, and is not a function of galactocentric radius (i.e., flat stellar disk); and b) the mass-weighted ISM velocity dispersion is the relevant quantity that sets ISM disk scale height, which in turn sets $\PDE$ near the disk mid-plane.
The first assumption is partially motivated by the work by \citet{Kregel_etal_2002}, and we provide more support for this assumption in Appendix~\ref{apdx:flatratio}.
In this section, we explore the impact of modifying these two key assumptions.

\textit{Adopted Stellar Disk Geometry:} The assumption of flat stellar disk geometry is widely used in previous works on stellar disk structure \citep[e.g.,][]{vanderKruit_Searle_1981,Yoachim_Dalcanton_2006,Comeron_etal_2012}, and supported by a recent observational study of edge-on disk galaxies \citep[e.g., see figure~12 in][]{Comeron_etal_2011}. We adopt this assumption as the fiducial choice in this paper (see Section~\ref{sec:expectation:PDE-kpc}).

An alternative, commonly considered possibility is a flared disk geometry \citep{Yang_etal_2007,Ostriker_etal_2010}. Here we consider this alternative scenario, and explore whether adopting this alternative affects our conclusion. For this purpose, we re-evaluate $\rhostar$ via
\begin{equation}
    \rhostarkpc\spsc{flared} = \frac{\Sigstarkpc}{0.54\Rstar}\;\exp{\left(1-\frac{\rgal}{\Rstar}\right)}~.
\label{eq:rhostarkpc_flared}
\end{equation}
\noindent This assumes that the disk scale height flares exponentially at larger $\rgal$, which is equivalent to assuming $\Hstar \propto \Sigstar^{-1}$ \citep[corresponding to a constant stellar velocity dispersion; see][]{Ostriker_etal_2010}, and that $\Sigstar$ drops exponentially as a function of $\rgal$.

The top panel in Figure~\ref{fig:PDE-variants} shows the fractional deviation in the $\PDEkpc$ estimates, when assuming a flared disk shape instead of a flat shape, as a function of $\PDEkpc$.
We find that assuming a flared disk geometry mainly leads to lower $\PDEkpc$ at the low pressure end.
This trend makes sense given the structure of galaxy disks. Low $\PDEkpc$ generally corresponds to large $\rgal$, and the flared disk shape will imply lower $\rhostar$ in this regime.
However, the amplitude of deviation in $\PDEkpc$ is $\lesssim 0.2$~dex in most cases. This suggests that the deviation from a flat disk shape may lead to a factor of $\lesssim1.6$ uncertainty on our $\PDEkpc$ estimates.

We also re-evaluate $\brkt{P\sbsc{DE,\,120pc}}$ assuming a flared stellar disk geometry. The corresponding $\brkt{P\sbsc{turb,\,120pc}}$--$\brkt{P\sbsc{DE,\,120pc}}$ relation is quoted in Table~\ref{tab:systematic}. The changes in the 16, 50, 84\% percentiles of $\brkt{P\sbsc{DE}}$ show that the flared disk scenario gives lower $\brkt{P\sbsc{DE}}$ estimates compared to the fiducial scenario, and that this deviation is more significant at the low pressure end. This leads to a slightly shallower slope (0.99) and a higher normalization ($-0.08$~dex at $10^5\,\uP$) for the $\brkt{P\sbsc{turb}}$--$\brkt{P\sbsc{DE}}$ relation. Nonetheless, the overall impact on the best-fit parameters is not large.

Therefore, the range of $\PDE$ depends on our assumed stellar disk geometry, but the $\brkt{\Pturb}$--$\brkt{\PDE}$ relation appears reasonably robust. In the near future, work using data from the Multi Unit Spectroscopic Explorer (MUSE; PI: E.~Schinnerer) will provide direct measurements of the stellar velocity dispersion in a subset of our targets. This should help improve our knowledge of the three dimensional stellar disk structure in these targets.

\textit{Adopted Gas Velocity Dispersion:} When calculating the kpc-scale equilibrium pressure in Section~\ref{sec:expectation:PDE-kpc}, we treat the entire ISM as a single component fluid. This motivates us to use the mass-weighted velocity dispersion combining atomic and molecular gas for estimating $\PDEkpc$ (Equation~\ref{eq:siggasz}). Many previous studies have instead adopted a fixed velocity dispersion of $\sigma\sbsc{gas,z} \approx 8\text{--}11\;\uV$ \citep[e.g.,][]{Blitz_Rosolowsky_2004,Blitz_Rosolowsky_2006,Leroy_etal_2008,Hughes_etal_2013a,Ostriker_etal_2010}, which is about the mean observed value for atomic gas at moderate galactocentric radii in nearby galaxy disks \citep{Leroy_etal_2008,Tamburro_etal_2009,Caldu-Primo_etal_2013,Mogotsi_etal_2016}.

We compare our approach with the fixed $\sigma\sbsc{gas,z}$ approach, again by comparing their corresponding $\PDEkpc$ estimates. As shown in the lower panel in Figure~\ref{fig:PDE-variants}, assuming a fixed $\sigma\sbsc{gas,z}=10\;\uV$ generally leads to higher $\PDEkpc$ in disk regions relative to our fiducial estimates. This is because the observed molecular gas velocity dispersion at 60--120~pc scales is usually less than $10\rm\;km\,s^{-1}$ in the disk regions in our sample, and generally smaller at larger $\rgal$ (or at low pressure).
Nevertheless, the resulting deviation in $\PDEkpc$ values is again within 0.2~dex in most cases.

We note that our $\PDEavg$ estimates treat the molecular and atomic gas separately, and thus do not rely directly on the mass-weighted velocity dispersion (see Section~\ref{sec:expectation:PDE-cloud}). Therefore, the discussion above does not apply to these cloud-scale estimates.

\subsection{Intensity Statistics versus Cloud Segmentation}
\label{sec:systematics:CPROPS}

\begin{figure*}[thp]
\gridline{
\fig{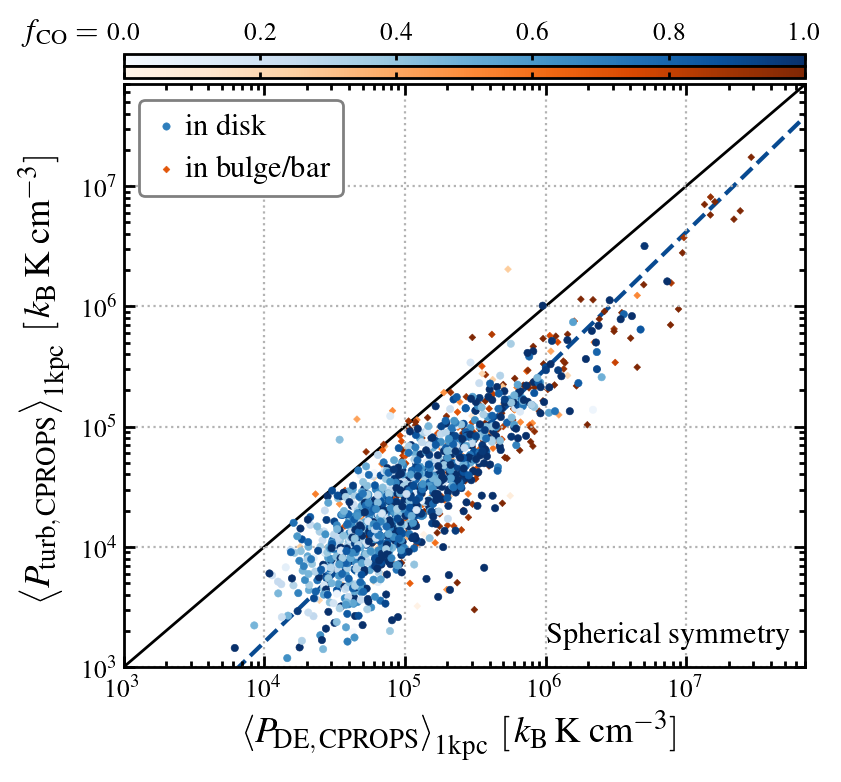}{0.47\textwidth}{}
\fig{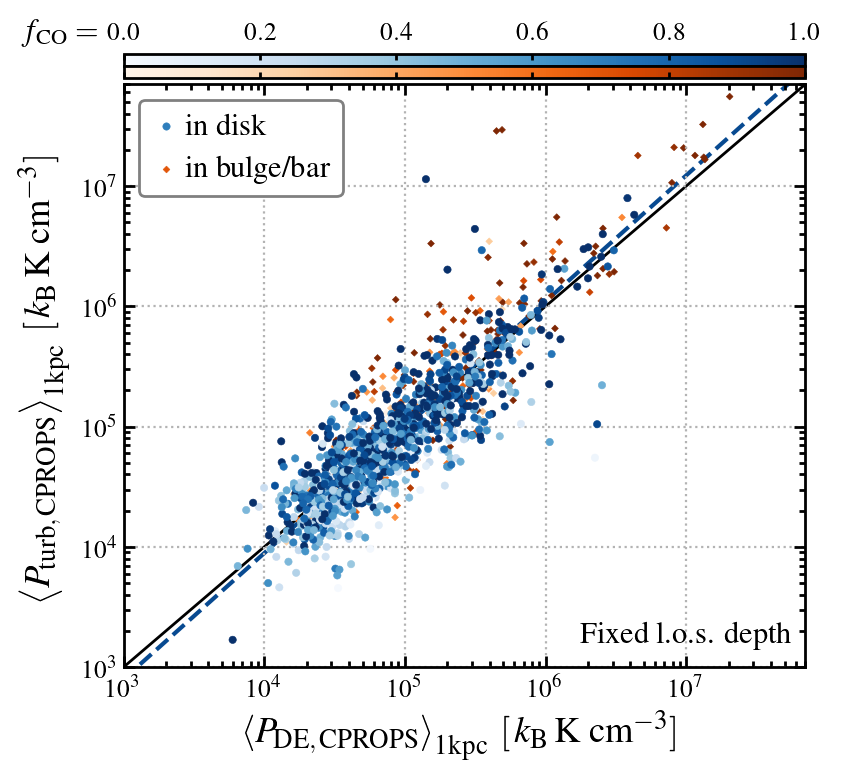}{0.47\textwidth}{}
}
\vspace{-3.5em}
\caption{
\textbf{$\brkt{P\sbsc{turb,\,120pc}}$--$\brkt{P\sbsc{DE,\,120pc}}$ relation derived using the cloud statistics approach.}
The left panel shows the estimates by assuming spherical symmetry for each CPROPS identified cloud. Equating the large cloud sizes estimated by CPROPS ($\brkt{2R\sbsc{cloud}}\approx400$~pc) to the line-of-sight depth leads to a systematic over-estimation of $\brkt{\PDE}$ and under-estimation of $\brkt{\Pturb}$.
The right panel shows the corresponding estimates derived by assuming a fixed line-of-sight depth of $120$~pc for all clouds (i.e., matching the beam size). This instead results in a much better agreement between $\brkt{\Pturb}$ and $\brkt{\PDE}$, consistent with the results derived from the pixel statistics approach.
}
\label{fig:Pturb-PDE-CPROPS}
\end{figure*}

In Sections~\ref{sec:method} and \ref{sec:expectation}, we adopt an approach that treats the gas in each pixel separately. This approach, which we refer to as the ``pixel statistics approach,'' preserves information from the smallest recoverable scale. We use this approach to derive mean cloud-scale gas properties (e.g., $\Pturbavg$) across our sample.

Another popular approach is to segment the observed gas distribution into regions that likely correspond to coherent physical objects. For example, many cloud segmentation algorithms (e.g., CLUMPFIND, \citealt{Williams_etal_1994}; CPROPS, \citealt{Rosolowsky_Leroy_2006}) group individual voxels into cloud-like objects associated with local maxima. Using the voxels associated with each cloud, one can derive cloud size, velocity dispersion, and luminosity, as well as other higher order properties.

From such a cloud catalog, one can derive the mass-weighted mean pressure and similar quantities for clouds in each region of each galaxy. We refer to this segmentation-based calculation as the ``cloud statistics'' approach. The cloud statistics approach accesses the same physics as our pixel statistics measurements, but differs in two important ways. First, the cloud-based approach treats identified objects, rather than resolution elements, as the fundamental structural unit. Second, the cloud-based approach yields size measurements for each of these objects. Most of the cloud literature assumes spherical symmetry, i.e., the projected size of the objects on the sky is assumed to reflect the depth of the objects along the line of sight.

Our data allow both approaches. In this Section, we re-derive our key measurements using a cloud statistics approach, and then compare the results to those from our pixel statistics measurements.

We use the PHANGS-ALMA CPROPS cloud catalogs, which are derived from the same CO dataset that we use (E.~Rosolowsky et al. in preparation; A.~Hughes et al. in preparation). Similar to our calculations, these begin with a set of fixed, 120~pc physical resolution data. The algorithm identifies significant, independent local maxima and then associates emission with each maximum. Then it measures the size, CO luminosity, and line width associated with each cloud using moment methods, and corrects for biases due to finite sensitivity and resolution. We note that the PHANGS-ALMA CPROPS application uses the ``seeded CLUMPFIND'' assignment option in CPROPS. This assigns all significant emission to nearby local maxima, and so represents a hybrid between the default CPROPS assignment and the CLUMPFIND assignment schemes. Otherwise, the calculations, detailed in E.~Rosolowsky et al. (in preparation), follow the original CPROPS approach.

Within each kpc-sized aperture, we derive the CO-flux-weighted average turbulent pressure and dynamical equilibrium pressure for all objects identified by CPROPS. To do this, we first write down the expressions for these two quantities using cloud mass, size, and velocity dispersion (i.e., the CPROPS counterparts of Equations~\ref{eq:Pturb} \& \ref{eq:PDEavg}):
\begin{align}
    \brkt{P\sbsc{turb,\,CPROPS}}
    =&\, \brkt{\frac{3 M\sbsc{cloud} \sigma\sbsc{cloud}^2}{4\pi R\sbsc{cloud}^3}}~, \label{eq:Pturb-CPROPS}\\
    \brkt{P\sbsc{DE,\,CPROPS}}
    =&\, \brkt{\frac{3\,G M\sbsc{cloud}^2}{8\pi R\sbsc{cloud}^4}} \nonumber\\
    &+ \brkt{\frac{G M\sbsc{cloud}}{2 R\sbsc{cloud}^2}}\Sigmolkpc \nonumber\\
    &+ \brkt{\frac{3\,G M\sbsc{cloud}}{2 R\sbsc{cloud}}}\rhostarkpc \nonumber\\
    &+ \mathcal{W}\sbsc{atom,\,1kpc}~. \label{eq:PDE-CPROPS}
\end{align}
\noindent The ``$\brkt{}$'' symbol here denotes a CO-flux-weighted average over all CPROPS clouds that have their central coordinate inside the kpc-sized aperture in question.

When substituting the measured cloud parameters from CPROPS into Equations~\ref{eq:Pturb-CPROPS} \& \ref{eq:PDE-CPROPS}, we pay special attention to two caveats.
First, we adopt the same metallicity dependent conversion factor for the cloud-based analysis as we do for the pixel statistics (see Section~\ref{sec:method:phys:mol}).
Second, the cloud radius quoted in the CPROPS catalogs is defined as 1.91 times the one-dimensional rms size calculated based on the object's projected intensity distribution on the sky (i.e., following the \citealt{Solomon_etal_1987} convention to account for clouds being centrally condensed). To enforce better consistency between the cloud and pixel measurements, we convert the radius quoted by CPROPS ($R\sbsc{CPROPS}$) to the radius of a hypothesized, constant density spherical cloud via
\begin{equation}
    R\sbsc{cloud} = \sqrt{5}\;\frac{R\sbsc{CPROPS}}{1.91} = 1.17\,R\sbsc{CPROPS}~.
    \label{eq:R_CPROPS}
\end{equation}
\noindent Here the factor of $\sqrt{5}$ is the ratio between the radius of a spherical, constant density cloud and its projected rms size on the sky \citep[see equations~11--13 in][]{Rosolowsky_Leroy_2006}.

Using the $\alphaCO$-corrected cloud mass, the adjusted radius, and the measured velocity dispersion, we derive estimates of $\brkt{P\sbsc{turb,\,CPROPS}}$ and $\brkt{P\sbsc{DE,\,CPROPS}}$ via Equations~\ref{eq:Pturb-CPROPS} and \ref{eq:PDE-CPROPS}.
The left panel in Figure~\ref{fig:Pturb-PDE-CPROPS} shows the relation between these two quantities across our sample.
We find that almost all data points lie below the equality line. That is, using the cloud statistics approach and assuming spherical symmetry for the objects, we find ubiquitously lower $\brkt{P\sbsc{turb,\,CPROPS}}$ than $\brkt{P\sbsc{DE,\,CPROPS}}$. On average, this offset is about 0.66~dex.

To understand this apparent discrepancy between the results from cloud statistics and from pixel statistics, we look into the actual measured sizes of the objects in the CPROPS catalogs.
With Equation~\ref{eq:R_CPROPS} applied, the median value of estimated object diameters across PHANGS-ALMA is $\brkt{2\Rcloud} \approx 400$~pc, or about 3 times ($\sim$0.5~dex) larger than the beam size (which is 120~pc in this case).

These apparently large cloud sizes are not out of expectation. Just like many other segmentation algorithms designed to find ``clumps,'' CPROPS tends to recover structures with sizes comparable to or larger than the beam size. This effect has been long noticed, and discussed by many previous works \citep[see][]{Verschuur_1993,Hughes_etal_2013a,Leroy_etal_2016}. These objects may be real physical structures (e.g., giant molecular associations or filaments). However, with such large sizes, the assumption of spherical symmetric is unlikely to hold since the $\sim 400$~pc diameters are much larger than the $\sim 100$~pc vertical FWHM of the Milky Way molecular gas disk \citep[see][]{Heyer_Dame_2015}.

As an ad-hoc correction, we re-derive the values of $\brkt{P\sbsc{turb,\,CPROPS}}$ and $\brkt{P\sbsc{DE,\,CPROPS}}$ assuming a modified object geometry. We still use $R\sbsc{cloud}$ as the projected size of the object on the sky, but now we assume the line of sight depth of the object to be $120$~pc. This matches the assumption used for the pixel statistics estimates.
This effectively assumes a cylindrical geometry for the identified objects, with their projected shapes on the sky kept the same, but their depth fixed to a constant value.
In practice, this means that we derive the cloud surface density via $\Sigma\sbsc{cloud} = M\sbsc{cloud}/(\pi R\sbsc{cloud}^2)$, and use the cloud surface density, velocity dispersion, and a fixed $D\sbsc{cloud}=120$~pc in Equations~\ref{eq:Pturb}, \ref{eq:PDEavg}, and \ref{eq:W_cloud} to estimate $\brkt{P\sbsc{turb,\,CPROPS}}$ and $\brkt{P\sbsc{DE,\,CPROPS}}$.

The right panel in Figure~\ref{fig:Pturb-PDE-CPROPS} shows the relation between the ``corrected'' $\brkt{P\sbsc{turb,\,CPROPS}}$ and $\brkt{P\sbsc{DE,\,CPROPS}}$ estimates.
In contrast to the results shown in the left panel, we find much better agreement between these ``corrected'' pressure estimates. The best-fit power-law relation (see Table~\ref{tab:systematic}) also becomes much more consistent with the results derived from the pixel statistics approach.
These findings suggest that, compared to the spherical symmetry assumption, the assumption of a fixed 120~pc line-of-sight depth is a much better description of the actual geometry of the CPROPS identified objects in the PHANGS-ALMA CO maps.

In summary, the derived $\brkt{\Pturb}$--$\brkt{\PDE}$ relation from the cloud statistics approach shows consistency with the pixel statistics results, provided that one adopts an appropriate assumption for the geometry of the identified objects. We also emphasize that, when analysing data with marginal spatial resolution, extra caution should be used when interpreting results of cloud identification algorithms like CPROPS.


\section{Discussion} \label{sec:discussion}

In Section~\ref{sec:result} we show that the prediction from the dynamical equilibrium model quantitatively matches the observed turbulent pressure within the molecular gas. Here we put this dynamical equilibrium consideration into the broader context of star formation and ISM evolution in galaxies.

Motivated by our findings that dynamical equilibrium seems to hold across spatial scales, an obvious next question is how the ISM and the molecular clouds within it maintain such an equilibrium state. What are the underlying mechanisms that regulate turbulent pressure in the ISM, and keep it at a level just enough to support the weight of the gas? Several possibilities have been suggested in the literature, including momentum injection due to stellar feedback \citep[e.g.,][]{Spitzer_1941b,Thompson_etal_2005,Ostriker_Shetty_2011,Faucher-Giguere_etal_2013} and/or gravitational instability \citep[e.g.,][]{Krumholz_Burkhart_2016,Ibanez-Mejia_etal_2017,Krumholz_etal_2018}. In particular, the former mechanism has been proven successful in explaining many aspects of massive star-forming disks in the local Universe \citep[e.g.,][]{Leroy_etal_2008,Ostriker_etal_2010,Krumholz_etal_2018}. In Section~\ref{sec:discussion:SFR}, we compare our new observations with the predictions from a family of feedback-regulated models developed by \citet{Ostriker_etal_2010} and \citet{Ostriker_Shetty_2011}, and synthesized in \citet{Kim_etal_2011}.

Beside its major role in regulating the intensity of star formation in galaxy disks, the pressure in the ISM might also affect the evolution of the ISM itself.
It has long been suggested that the molecular-to-atomic gas ratio ($R\sbsc{mol}\equiv\Sigmol/\Sigatom$) of the ISM is partly determined by the ambient ISM pressure \citep[e.g., see][]{Elmegreen_1993}. Many observational works use the dynamical equilibrium pressure $\PDE$ as a tracer of this ambient pressure, and indeed find a positive correlation between $R\sbsc{mol}$ and $\PDE$ \citep{Wong_Blitz_2002,Blitz_Rosolowsky_2006,Leroy_etal_2008}.
In Section~\ref{sec:discussion:Rmol}, we revisit this topic by characterizing this correlation in our sample, and comparing it to results in previous works \citep{Blitz_Rosolowsky_2006,Leroy_etal_2008}.

\subsection{Link to the Self-regulated Star Formation Model}
\label{sec:discussion:SFR}

\begin{figure*}[htp]
\gridline{
\fig{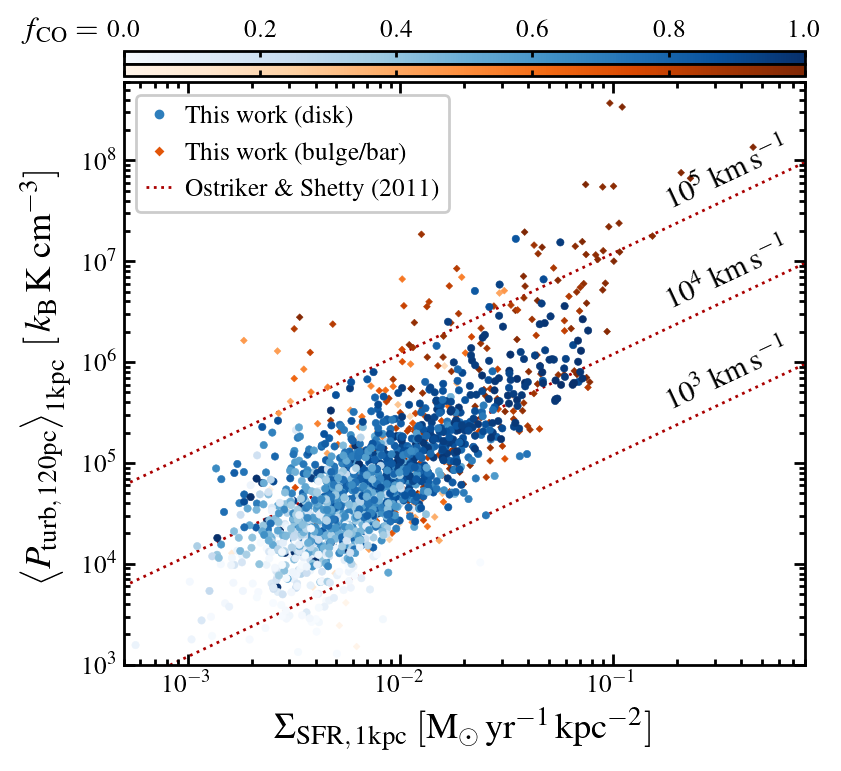}{0.47\textwidth}{}
\fig{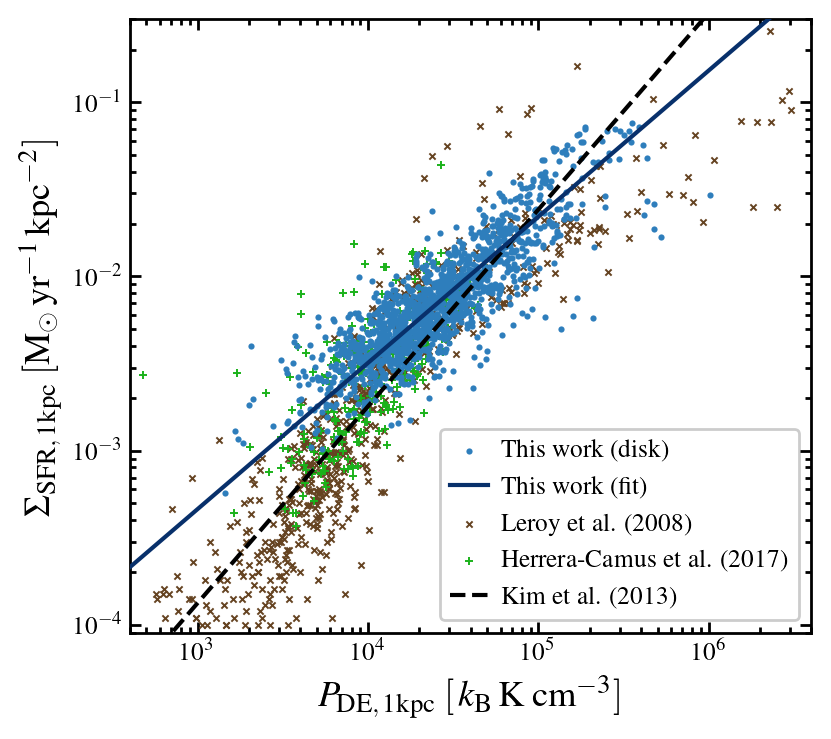}{0.48\textwidth}{}
}
\vspace{-3.0em}
\caption{
\textbf{\textit{Left:} Cloud-scale molecular gas turbulent pressure as a function of kpc-scale SFR surface density.}
The dotted lines represent linear relations parameterized as $\Pturb = (p_\star/4m_\star)\,\SigSFR$ \citep{Ostriker_Shetty_2011}, where $p_\star/m_\star$ is the feedback momentum injection rate
(see Section~\ref{sec:discussion:SFR:Pturb}).
\textbf{\textit{Right:} kpc-scale SFR surface density as a function of kpc-scale ISM dynamical equilibrium pressure.}
Blue points show our measurements in galaxy disks;
brown crosses show measurements in 23 nearby galaxies \citep[including 11 low-mass, \HI\ dominated galaxies;][]{Leroy_etal_2008};
green plus symbols show measurements in the \HI\ dominated regions in 31 KINGFISH galaxies \citep{Herrera-Camus_etal_2017}.
We find overall consistency between our results and literature measurements.
The blue solid line shows the power-law fit on our disk sample, which has a slightly shallower slope compared to the predicted relation by a hydrodynamic simulation \citep[black dashed line;][]{Kim_etal_2013}.
}
\label{fig:SigSFR}
\end{figure*}

In the self-regulated star formation model of \citet{Ostriker_etal_2010,Ostriker_Shetty_2011}, an actively star-forming disk is viewed as a (quasi-)steady state system.
Stellar feedback of in the form of radiation, winds, and supernovae offsets (in a time-averaged sense) losses of energy and pressure due to cooling and turbulent dissipation. At the same time, pressure maintains (again, in a time-averaged sense) support for the gas against collapse in the gravitational field, balancing both gas self-gravity and external disk gravity. However, a small fraction of the interstellar gas can collapse and form stars locally in regions where support against gravity is insufficient. The result of this localized collapse is the star formation feedback that pressurizes the rest of the ISM, providing internal support. Although this is likely a violent process with alternating episodes of collapse and expansion \citep[see e.g.,][]{Kruijssen_etal_2019b,Rahner_etal_2019,Schinnerer_etal_2019,Chevance_etal_2020}, dynamical equilibrium is expected when considering the entire ISM across large spatial scales ($\gg$ typical size of GMCs and star-forming regions) and long time scale ($\gg$ typical lifetime of SF cycle).  Numerical simulations indeed show that a well-defined quasi-steady state exists on spatial scales of order $\sim$1~kpc and timescales of a few hundred Myr, even though the star formation rate and the ISM properties strongly vary on short space- and time-scales \citep[see e.g.][]{Kim_Ostriker_2017,Semenov_etal_2017,Kim_Ostriker_2018,Orr_etal_2018,Semenov_etal_2018}.

In this framework, turbulent pressure ($\Pturb$) and dynamical equilibrium pressure ($\PDE$) are both closely related to a third variable --- the local star formation rate surface density, $\SigSFR$. $\Pturb$ should be directly proportional to $\SigSFR$ given a fixed momentum injection per star formed  \citep[e.g.,][]{Ostriker_Shetty_2011}.  Simultaneously, the equilibrium-state $\SigSFR$ is determined by requiring the sum of turbulent, thermal and magnetic pressures (individually proportional to $\Sigma_\mathrm{SFR}$ due to feedback) to balance $\PDE$ \citep[e.g.,][]{Ostriker_etal_2010,Kim_etal_2011}. Here we explore this scenario by showing the relationships between $\Pturb$ and $\SigSFR$, and between $\SigSFR$ and $\PDE$.

\subsubsection{SFR Surface Density versus Turbulent Pressure} \label{sec:discussion:SFR:Pturb}

The left panel in Figure~\ref{fig:SigSFR} shows the measured cloud-scale average turbulent pressure in molecular gas, $\Pturbavgkpc$, as a function of the kpc-scale average SFR surface density, $\SigSFRkpc$.
We observe strong correlation between these two quantities. For all measurements in disk regions (blue dots), we find a rank correlation coefficient of $\rho = 0.71$ (corresponding p-value~$\ll0.001$).

In the case that the ISM pressure is feedback-driven, the ratio between ISM turbulent pressure and SFR surface density -- where these are averages computed over the same area -- reflects to the momentum injection per unit mass of stars formed, $p_\star/m_\star$, via
\begin{equation}
    \Pturb = \frac{1}{4}\frac{p_\star}{m_\star}\,\SigSFR~.
\end{equation}
\noindent
The pre-factor of $1/4$ assumes spherical expansion sites centered on the disk midplane and that the momentum flux to upper and lower halves of the ISM disk translates directly to ISM turbulent pressure \citep{Ostriker_Shetty_2011}. While the above picture is idealized and should be modified by details of turbulent injection and dissipation, it quantitatively agrees with the measured relationship between $\Pturb$ and $\SigSFR$ in disk simulations of the star-forming multi-phase ISM \citep{Kim_etal_2013,Kim_Ostriker_2015b,Kim_Ostriker_2017}.

The ratio $p_*/m_*$ is predicted to range between $10^3$--$10^4\;\uV$ for supernova feedback, depending on the ISM properties, spatial and temporal clustering of supernovae, and energy losses due to interface mixing \citep[e.g.,][]{Iffrig_Hennebelle_2015,Kim_Ostriker_2015a,Martizzi_etal_2015,Walch_Naab_2015,Kim_etal_2017,El-Badry_etal_2019,Gentry_etal_2019}.
In Figure~\ref{fig:SigSFR}, we show the predicted $\Pturb$--$\SigSFR$ relation for a range of $p_\star/m_\star$ values (dotted lines).
Our observed $\brkt{P\sbsc{turb,\,120pc}}$--$\SigSFRkpc$ relation has a normalization that would correspond to a high momentum injection.

One possible explanation for this apparently high momentum injection rate is the clumping of the star formation distribution.
Unlike $\brkt{P\sbsc{turb,\,120pc}}$, which is estimated on 120~pc scales and then averaged over the kpc-scale aperture, our $\SigSFRkpc$ measurements are derived directly on kpc scale. Just like the molecular gas, we expect star formation to cluster on sub-kpc scales \citep[e.g.,][]{Grasha_etal_2018,Grasha_etal_2019,Schinnerer_etal_2019,Chevance_etal_2020}. This will cause the $\SigSFRkpc$ values in Figure~\ref{fig:SigSFR} to appear lower due to the inclusion of area without star formation, and thus it underestimates the actual SFR surface density relevant to feedback momentum injection.

To account for this issue, we introduce a dimensionless prefactor $C\sbsc{fb}\gtrsim1$, which corrects for the artificial dilution of $\SigSFRkpc$ compared to $\Sigma\sbsc{SFR,\,\theta pc}$. Then we have
\begin{equation}
    \brkt{P\sbsc{turb,\,120pc}}\sbsc{1kpc} = C\sbsc{fb}\,\frac{p_\star}{4m_\star}\,\SigSFRkpc~.
\end{equation}
\noindent The median value of $C\sbsc{fb}\,p_\star/m_\star$ among the disk sample is $6.3\times10^3\;\uV$, and its 16--84\% percentile range is 0.73~dex.

A direct estimation of $C\sbsc{fb}$ requires high resolution data tracing SFR on matched $\sim10\text{--}100$~pc scales. This is currently not available for our entire sample \citep[c.f.,][]{Kreckel_etal_2018,Schinnerer_etal_2019,Chevance_etal_2020}. However, we can obtain a rough estimate of $C\sbsc{fb}$ based on our high resolution CO data, if we assume that the clustering of star formation matches that of the molecular gas. Given a median area filling fraction of $\approx1/3$ in our CO data, we expect $C\sbsc{fb} \approx 3$. In this case, most of the $\Pturbavg$ and $\SigSFRkpc$ measurements in the disk sample are consistent with $p_\star/m_\star = 10^3\text{--}10^4\;\uV$.

There are many caveats related to this approximate estimation of $C\sbsc{fb}$. The spatial and temporal distribution of gas and star formation could change the amount of momentum injected into the gas, and the spatial scales on which most momentum is deposited also matters. Given these complications, a logical next step on this topic will be a detailed, multi-scale comparison between observations of gas and star formation at high resolution and numerical simulations with realistic feedback prescriptions.

An alternative explanation for the high $\Pturb/\SigSFR$ ratio measurements in Figure~\ref{fig:SigSFR} (left panel) is that other turbulence driving mechanisms might also play a role, at least within a subset of our sample. For example, many measurements in bulge or bar regions show the highest $\Pturb$ at fixed $\SigSFR$ (orange symbols in Figure~\ref{fig:SigSFR}, left panel). While part of this can be explained by beam smearing effects, we also expect bar-induced radial inflow and the corresponding conversion of gravitational potential energy to be an additional source of turbulence in the ISM \citep[e.g., through shocks near the center of the stellar bar; see][]{Binney_etal_1991,Sormani_etal_2015}. More quantitative comparisons between our results and other turbulent driving mechanisms will be carried out in future works.

\subsubsection{Equilibrium Pressure versus SFR Surface Density}
\label{sec:discussion:SFR:PDE}

The right panel in Figure~\ref{fig:SigSFR} shows kpc-scale SFR surface density, $\SigSFRkpc$, as a function of kpc-scale dynamical equilibrium pressure, $\PDEkpc$. We find a strong correlation between the two quantities for all measurements in disk regions. This supports the idea that in steadily star-forming disks, the three dimensional star and gas distribution on kpc-scale largely determines the average SFR surface density on the same scale. In any part of the disk where the ISM weight is higher, the steady-state SFR must also increase for feedback to maintain the pressure that matches the dynamical equilibrium pressure.

To quantify this observed correlation, we derive a best-fit power-law relation for all disk measurements, using the OLS bisector method in logarithmic space:
\begin{equation}
    \frac{\SigSFR}{10^{-3}\;\uSigSFR} = 3.2\, \left(\frac{\PDE}{10^4\;\uP}\right)^{0.84}~.
    \label{eq:SigSFR-PDE}
\end{equation}
\noindent The rms scatter in $\SigSFR$ around this relation is 0.20~dex. The statistical errors estimated from bootstrapping are 0.01 for the slope and 0.01~dex for the normalization, but we expect systematic errors (on both $\SigSFRkpc$ and $\PDEkpc$) to be a larger source of uncertainty on the fitting results.

The same correlation has been observed in various types of galaxies by many previous studies \citep[e.g.,][]{Leroy_etal_2008,Genzel_etal_2010,Ostriker_etal_2010,Ostriker_Shetty_2011,Herrera-Camus_etal_2017,Fisher_etal_2019}. To put our measurements into context, we also include in Figure~\ref{fig:SigSFR} two other datasets derived also from observations of nearby galaxies.
\citet{Leroy_etal_2008} (brown crosses) measure this relation in a sample of 23 nearby galaxies, with each independent measurement representing one kpc-wide radial bin in a galaxy.
\citet{Herrera-Camus_etal_2017} (green plus symbols) investigate the same relation in the \HI\ dominated regions in 31 KINGFISH galaxies \citep{Kennicutt_etal_2011}, treating each kpc-sized region independently.

These previous works adopted slightly different assumptions when deriving $\PDE$ from observables. To make a fair comparison between our measurements and theirs, our $\PDEkpc$ measurements shown in Figure~\ref{fig:SigSFR} are instead derived by assuming $\siggasz = 11\;\uV$, following exactly the same prescription in \citet{Leroy_etal_2008} and \citet{Herrera-Camus_etal_2017} (also see Section~\ref{sec:systematics:PDE}). We find overall consistency, but our measurements concentrate more towards the high $\PDE$, high $\SigSFR$ end. This is because the PHANGS-ALMA CO observations primarily target the star-forming inner part of high mass disk galaxies, while the \citet{Herrera-Camus_etal_2017} sample focuses on the \HI-dominated outer disks on purpose, and the \citet{Leroy_etal_2008} sample includes many dwarf galaxies.

In addition to previous observational results, we also show in Figure~\ref{fig:SigSFR} the predicted $\PDE$--$\SigSFR$ relation from a hydrodynamic simulation \citep[][black dashed line in the right panel]{Kim_etal_2013}:
\begin{equation}
    \frac{\SigSFR}{10^{-3}\;\uSigSFR} = 1.8 \,\left(\frac{\PDE}{10^4\;\uP}\right)^{1.13}~.
    \label{eq:SigSFR-PDE-Kim13}
\end{equation}
\noindent This relation has a steeper slope than our best-fit power-law relation (Equation~\ref{eq:SigSFR-PDE}), while it agrees with previous observations in the low $\SigSFR$ regime. We point out that the simulations in \citet{Kim_etal_2013} cover a $\SigSFR$ range from $10^{-4}$ to $10^{-2}\;\uSigSFR$, whereas our sample covers from $10^{-3}$ to $10^{-1}\;\uSigSFR$. The shallower slope found in our sample might then reflect some systematic change in the properties of the ISM in massive galaxies and inner disk environments, relative to the ISM in dwarf galaxies and/or outer disk environments.

A shallower $\PDE$--$\SigSFR$ relation in high $\SigSFR$ environments is seen in several previous works. In the \citet{Leroy_etal_2008} sample, some hint of a shallower $\SigSFR$--$\PDE$ relation is visible from the high $\SigSFR$ measurements. For a sample of local ultra luminous infra-red galaxies (ULIRGs) and high-redshift star-forming galaxies, \citet{Ostriker_Shetty_2011} found a $\SigSFR$--$\PDE$ relation with a slope of 0.95 at $\PDE > 10^5\,\uP$, which is closer to our result\footnote{\citet{Ostriker_Shetty_2011} assume that the gas self-gravity term dominates in $\PDE$, and that $\alphaCO \propto I_\mathrm{CO}^{-0.3}$}. More recently, \citet{Fisher_etal_2019} report a much shallower slope of 0.77 at $\PDE \gtrsim 10^5\,\uP$ for a sample of local turbulent disk galaxies\footnote{The $\PDE$ estimates in \citet{Fisher_etal_2019} assume a Galactic $\alphaCO$ and use the ionized gas velocity dispersion for $\siggasz$. Newly obtained CO velocity dispersion for the same galaxies implies systematically lower $\PDE$ (by $\sim$0.3~dex; D.~Fisher, priv. comm.), but no significant change in the $\SigSFR\text{--}\PDE$ relation slope.}, which are believed to resemble typical star-forming galaxies at $z \sim 1\text{--}2$. Future studies on the ISM in more local ULIRGs and galaxy centers can provide better constraints on the slope of the $\SigSFR$--$\PDE$ relation at the high $\SigSFR$ end, and potentially help unveil the physics regulating star formation in the ``starburst'' regime \citep[see e.g.,][]{Thompson_etal_2005,Ostriker_Shetty_2011,Shetty_Ostriker_2012,Crocker_etal_2018,Krumholz_etal_2018}.

\subsection{Link to the Molecular-to-Atomic Gas Ratio}
\label{sec:discussion:Rmol}

Following early suggestions by \citet{Elmegreen_1989}, the ISM dynamical equilibrium pressure has also been viewed as a determinant of the molecular/atomic phase balance in the ISM. In this scenario, $\PDE$ relates closely to the molecular-to-atomic gas ratio, $R\sbsc{mol}$, and thus influences the fraction of the ISM in the dense, star-forming phase.
This idea has been tested by many subsequent works \citep[e.g.,][]{Wong_Blitz_2002,Blitz_Rosolowsky_2006,Leroy_etal_2008}, and commonly adopted as a prescription for determining the molecular gas fraction in semi-analytic models of galaxy evolution \citep[e.g.,]{Lagos_etal_2018}.
Here we report the observed scaling relation between molecular-to-atomic ratio and the ISM dynamical equilibrium pressure across our sample.

Following previous studies, we use the kpc-scale dynamical equilibrium pressure $\PDEkpc$ to trace the average ambient pressure in the ISM.
To allow a quantitative comparison with previous results, we again use the $\PDEkpc$ values estimated by assuming a fixed $\siggasz=11\,\uV$ (similar to Section~\ref{sec:discussion:SFR:PDE}).
We determine the molecular-to-atomic gas ratio from the ratio of our measured molecular and atomic gas surface densities on kpc-scale:
\begin{equation}
    R\sbsc{mol,1kpc} \equiv \Sigmolkpc / \Sigatomkpc~.
\end{equation}

\begin{figure}[t!]
\gridline{
\fig{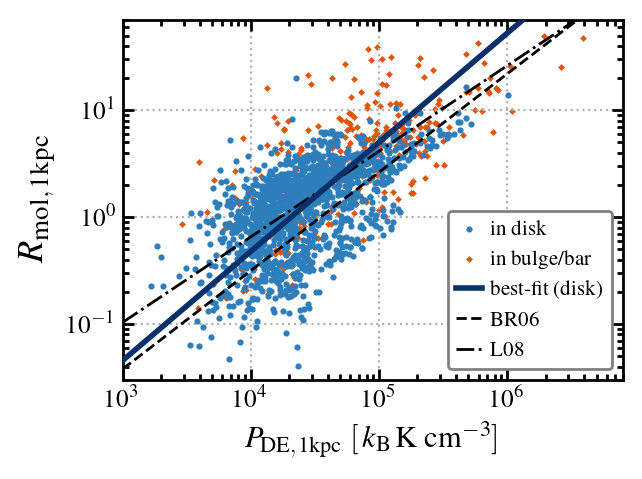}{0.47\textwidth}{}
}
\vspace{-2.5em}
\caption{
\textbf{Molecular to atomic gas ratio $R\sbsc{mol,1kpc}$ as a function of $\PDEkpc$.}
The best-fit power-law relation for the disk measurements (blue solid line) has a slope of $\alpha=1.02$, and it crosses the $R\sbsc{mol}=1$ threshold at $P_0 = 2.1\times10^4\,\uP$.
Given the systematic uncertainties associated with the choice of methodology, our best-fit $R\sbsc{mol}$--$\PDE$ relation is consistent with those reported in previous studies \citep[black lines;][]{Blitz_Rosolowsky_2006,Leroy_etal_2008}.
}
\label{fig:Rmol-PDE}
\end{figure}

We show the relation between $\PDEkpc$ and $R\sbsc{mol,1kpc}$ in Figure~\ref{fig:Rmol-PDE}.
Our sample spans nearly two orders of magnitude in $R\sbsc{mol,\,1kpc}$, with most measurements clustering around or above the atomic-to-molecular transition threshold (i.e., $R\sbsc{mol,\,1kpc}=1$). We find a positive and statistically significant correlation between $R\sbsc{mol,\,1kpc}$ and $\PDEkpc$ across our whole sample (Spearman's rank correlation coefficient $\rho=0.58$; corresponding p-value~$\ll0.001$). This strong, positive correlation is qualitatively consistent with previous observations \citep{Wong_Blitz_2002,Blitz_Rosolowsky_2006,Leroy_etal_2008}, even though the adopted CO-to-H$_2$ conversion factor, stellar mass-to-light ratio, and stellar disk geometry vary among studies.

We perform an OLS bisector fit on all our disk measurements over the range $0.1 < R\sbsc{mol,\,1kpc} < 10$, following \citet{Leroy_etal_2008}. This yields a best-fit power-law relation (blue solid line in Figure~\ref{fig:Rmol-PDE}) of
\begin{equation}
    R\sbsc{mol,\,1kpc} = \left( \frac{\PDEkpc}{2.1\times10^4\;\uP} \right)^{1.02}~.
    \label{eq:Rmol-PDE}
\end{equation}
\noindent The scatter in $R\sbsc{mol,\,1kpc}$ around this relation is 0.36~dex.
The formal statistical errors in the fit are small: 0.02 for the slope $\alpha$, and 0.01~dex for the threshold pressure $P_0=2.1\times10^4\,\uP$ (at which the ISM transitions from being predominantly atomic to molecular, or vice versa).
However, when varying the choice of $\alphaCO$ prescriptions and other assumptions, the best-fit $\alpha$ and $P_0$ appear systematically uncertain by $\sim0.20$ and $\sim0.15$~dex, respectively.
Given this level of systematic uncertainty, Equation~\ref{eq:Rmol-PDE} quantitatively agrees with the $R\sbsc{mol}$--$\PDEkpc$ relations reported in previous works \citep[$\alpha=0.73\text{--}1.05$, $P_0=1.5\text{--}4.5\times10^4\,\uP$; see][also see black lines in Figure~\ref{fig:Rmol-PDE}]{Wong_Blitz_2002,Blitz_Rosolowsky_2006,Leroy_etal_2008}.


\section{Summary} \label{sec:summary}

For a sample of 28 nearby star-forming disk galaxies, we estimate the pressure needed to support the ISM against its own weight at a range of spatial scales, taking into account the combined gravity of all gas components and the stellar disk.
We compare this estimated ``dynamical equilibrium pressure'' to the observed turbulent pressure in molecular gas.
This tests the common hypothesis that the ISM in star-forming galaxy disks is in dynamical equilibrium.

For this purpose, we create a multi-wavelength dataset, which includes high resolution PHANGS-ALMA \CO21\ imaging data (A.~K.~Leroy et al., in preparation) and GMC catalogs (E.~Rosolowsky et al., in preparation; A.~Hughes et al., in preparation), new and archival \HI\ 21~cm data (D.~Utomo et al., in preparation), processed \textit{Spitzer} IRAC 3.6~\micron\ data \citep[\SfourG;][]{Querejeta_etal_2015}, and combined GALEX near-UV and WISE mid-IR data \citep[Z0MGS;][]{Leroy_etal_2019}. These data provide us with a comprehensive picture of the molecular gas distribution and kinematics on $\sim60$--$120$~pc scales (i.e., cloud scales), and the distribution of gas mass, stellar mass, and star formation rate on $\sim1$~kpc scale.

We divide the 28 galaxies in our sample into 1,762 independent, kpc-sized hexagonal regions, covering the PHANGS-ALMA CO footprint. In each kpc-sized region, we use the high resolution CO data to calculate the mass-weighted mean turbulent pressure in molecular gas, $\Pturbavg$ ($\theta=60$ and $120$~pc; see Section~\ref{sec:expectation:Pturb}). We then compare $\Pturbavg$ to the required pressure to balance the weight of the ISM disk in the galaxy gravitational potential, a.k.a., the dynamical equilibrium pressure, $\PDE$.

Building on earlier works \citep{Hughes_etal_2013a,Schruba_etal_2019}, we compute two different measures of $\PDE$. One measure, widely adopted in previous studies, represents the expected mean mid-plane ISM pressure averaged over all gas, and implicitly assumes no bound sub-structures. The other measure considered in this work represents the pressure expected within individual resolved molecular structures on 60--120~pc scales, allowing them to be self-gravitating.

The first estimate assumes uniform gas surface density within each kpc-sized aperture. This provides a measure of equilibrium pressure on kpc-scale, $\PDEkpc$ (see Section~\ref{sec:expectation:PDE-kpc}).
With this measure, we find that:
\begin{enumerate}
    \setcounter{enumi}{0}
    \item $\PDEkpc$ ranges from $10^3$ to $10^6\,\uP$ across our sample. This agrees well with previous estimates of the average ISM pressure in galaxy disks. The lower bound roughly corresponds to Solar Neighborhood-like environments, whereas the higher bound corresponds to conditions found in gas-rich galaxy centers.
    \item The mass-weighted turbulent pressure $\Pturbavg$ ranges from $10^4$ to $10^7\,\uP$. $\Pturbavg$ correlates with $\PDEkpc$, but it almost always exceeds the $\PDEkpc$ estimate in the same region (Figure~\ref{fig:Pturb-PDE-kpc}).
    At 120~pc scale, we measure an average $\Pturbavg/\PDEkpc$ ratio of $\sim 2.8$ across our sample.
    That is, molecular gas appears highly over-pressurized compared to the mean $\PDEkpc$ calculated assuming a \textit{smooth} ISM distribution in each kpc-sized aperture.
    \item We fit a relation between $\Pturbavg$ and $\PDEkpc$ (Equation~\ref{eq:Pturbavg-PDEkpc-alt}) that can be used to predict cloud-scale molecular gas properties from kpc-resolution observation of distant galaxies, from low resolution galaxy simulations, or from analytic or semi-analytic models of star-forming galaxy disks.
\end{enumerate}

In reality, the molecular ISM is highly clumped, and self-gravity plays an important role in dynamical equilibrium.
Indeed, the molecular gas distribution traced by PHANGS-ALMA on 60--120~pc scales does display this rich substructure.
The presence of this substructure means that $\PDEkpc$ under-estimates the required pressure for the clumpy molecular gas to achieve dynamical equilibrium, because it does not account for the enhancement of gas self-gravity in over-densities.

We thus introduce a formalism which explicitly incorporates knowledge about molecular gas clumping, and considers the self-gravity of the molecular gas, the external gravitational potential, and the pressure in the ambient ISM in a unified framework. This provides a measure of the equilibrium pressure on cloud scales, $\PDEavg$ (see Section~\ref{sec:expectation:PDE-cloud}).

Applying this formalism, we find that:
\begin{enumerate}
    \setcounter{enumi}{3}
    \item Accounting for the enhanced gas self-gravity due to clumping at small scales, we estimate that the pressure needed to support the gas is $\PDEavg=10^4$--$10^8\,\uP$, systematically higher than the kpc-scale estimates of $\PDEkpc$.
    \item $\Pturbavg$ and $\PDEavg$ are nearly equal across most regions in our sample (Figure~\ref{fig:Pturb-PDE-cloud}). This is consistent with the idea that molecular clouds have internal pressures close to the value needed to balance the sum of their own internal weight and the weight of the ambient atomic gas.
    \item
    In our sample, the self-gravity of cloud-scale molecular gas structures accounts for $\sim 33-70\%$ of the total $\PDEavg$. For the molecular gas at high internal pressure ($\gtrsim10^5\,\uP$), $\PDEavg$ is more likely to be dominated by the self-gravity term. Gas at low internal pressure is more likely to be affected by ambient pressure and/or external gravity (Figure~\ref{fig:frac-PDE}). However, we observe large scatter about this general trend, and we see examples for both scenarios across all environments.
\end{enumerate}

We explore the systematic effects associated with key assumptions (Section~\ref{sec:systematics}).
We (1) vary the adopted CO-to-H$_2$ conversion factors in a reasonable range, (2) choose different assumptions for the stellar disk geometry and/or gas vertical velocity dispersion, and (3) adopt an alternative, cloud statistic approach utilizing the PHANGS-ALMA CPROPS cloud catalogs. We find that varying these assumptions changes the range of the derived pressure estimates, but does not affect our qualitative conclusion of a strong, nearly unity correlation between $\PDEavg$ and $\Pturbavg$. Our quantitative measurements do show mild variations due to these systematic effects, which we report in Table~\ref{tab:systematic}.

Based on our analysis and tests of systematic effects, our most general conclusion is that:
\begin{enumerate}
    \setcounter{enumi}{6}
    \item A close-to-unity $\Pturbavg$--$\PDEavg$ relation holds across different physical regimes, and is robust against many systematic effects. In other words, the molecular gas in the disk regions of nearby, massive, star-forming galaxies does appear to be in or near a state of dynamical equilibrium.
\end{enumerate}

Beside testing the assumption of dynamical equilibrium, we also investigate the driving mechanism of this equilibrium.
We find that the observed kpc-scale SFR surface density $\SigSFR$ shows a strong correlation with both $\Pturbavg$ and $\PDEkpc$ (Figure~\ref{fig:SigSFR}).
The ratio between $\Pturbavg$ and $\SigSFRkpc$ is generally consistent with the expected range of momentum injection from supernova feedback, if one considers the clumping of star formation on small scales.
The nearly linear relationship between $\SigSFRkpc$ and $\PDEkpc$ is consistent with the feedback-regulated scenario, and in quantitative agreement with previous observational and theoretical studies where the parameter regimes overlap.
We show that $\PDEkpc$ correlates positively with the molecular to atomic ratio, $R\sbsc{mol,\,1kpc}$ (Figure~\ref{fig:Rmol-PDE}). The best-fit relation we find (Equation~\ref{eq:Rmol-PDE}) is consistent with those reported in previous studies \citep{Blitz_Rosolowsky_2006,Leroy_etal_2008}.

We publish the estimated $\Pturb$, $\PDE$, $\SigSFR$, and $R\sbsc{mol}$ across our full sample in machine-readable form (see Table~\ref{tab:mrt} in Appendix~\ref{apdx:mrt}). We encourage the use of this dataset as a benchmark for future observations and numerical simulations.

In the near future, it will be possible to extend these measurements from the disks of massive spiral galaxies to a wider set of environments. Sensitive CO observations targeting dwarf galaxies and the outer disks of spiral galaxies will reveal whether a similar equilibrium holds in the atomic gas- and/or external pressure-dominated regimes. A more careful treatment of in-plane motions and a better modeling of the three-dimensional distribution and kinematics of the stellar component will provide better understanding of dynamical equilibrium in early-type galaxies, galaxy bulges, and central regions.


\acknowledgments
{We thank the anonymous referee for helpful comments that improved the quality of the paper. J.S. would like to thank R.~Herrera-Camus and C.~Murugeshan for kindly sharing their data, and D.~Fisher and T.~A.~Thompson for helpful discussions.

This work was carried out as part of the PHANGS collaboration.
The work of J.S., A.K.L., and D.U. is partially supported by the National Science Foundation under Grants No. 1615105, 1615109, and 1653300.
The work of E.C.O. was partly supported by NASA under ATP Grant NNX17AG26G.
A.H., C.N.H., and J.P. acknowledge funding from the Programme National ``Physique et Chimie du Milieu Interstellaire (PCMI)'' of CNRS/INSU with INC/INP, co-funded by CEA and CNES, and from the ``Programme National Cosmology et Galaxies (PNCG)'' of CNRS/INSU with INP and IN2P3, co-funded by CEA and CNES.
E.R. acknowledges the support of the Natural Sciences and Engineering Research Council of Canada (NSERC), funding reference number RGPIN-2017-03987.
E.S., C.F., and T.S. acknowledges funding from the European Research Council (ERC) under the European Union's Horizon 2020 research and innovation programme (grant agreement No. 694343).
J.M.D.K. and M.C. gratefully acknowledge funding from the Deutsche Forschungsgemeinschaft (DFG) through an Emmy Noether Research Group (grant number KR4801/1-1) and the DFG Sachbeihilfe (grant number KR4801/2-1). J.M.D.K. gratefully acknowledges funding from the European Research Council (ERC) under the European Union's Horizon 2020 research and innovation programme via the ERC Starting Grant MUSTANG (grant agreement number 714907).
F.B. acknowledges funding from the European Research Council (ERC) under the European Union's Horizon 2020 research and innovation programme (grant agreement No. 726384).
S.C.O.G. acknowledges support by the Deutsche Forschungsgemeinschaft (DFG, German Research Foundation) -- Project-ID 138713538 -- SFB 881 (``The Milky Way System'', sub-projects B01, B02, B08), and by the Heidelberg cluster of excellence EXC 2181-390900948 ``STRUCTURES: A unifying approach to emergent phenomena in the physical world, mathematics, and complex data'', funded by the German Excellence Strategy.
A.U. acknowledges support from the Spanish funding grants AYA2016-79006-P (MINECO/FEDER) and PGC2018-094671-B-I00 (MCIU/AEI/FEDER).

This paper makes use of the following ALMA data: ADS/JAO.ALMA\# 2012.1.00650.S, 2015.1.00925.S, 2015.1.00956.S, 2017.1.00392.S, 2017.1.00886.L, and 2018.1.01321.S. ALMA is a partnership of ESO (representing its member states), NSF (USA), and NINS (Japan), together with NRC (Canada), NSC and ASIAA (Taiwan), and KASI (Republic of Korea), in cooperation with the Republic of Chile. The Joint ALMA Observatory is operated by ESO, AUI/NRAO, and NAOJ. The National Radio Astronomy Observatory is a facility of the National Science Foundation operated under cooperative agreement by Associated Universities, Inc.

This work is based in part on observations made with NSF's Karl~G.~Jansky Very Large Array (Legacy ID: AS 1303, AS1387, AS1434, AU157). VLA is also operated by the National Radio Astronomy Observatory.

This work is based in part on observations made with the Australia Telescope Compact Array (ATCA). ATCA is part of the Australia Telescope National Facility, which is funded by the Australian Government for operation as a National Facility managed by CSIRO.

This work is based in part on observations made with the Spitzer Space Telescope, which is operated by the Jet Propulsion Laboratory, California Institute of Technology under a contract with NASA.

This work is based in part on observations made with the Galaxy Evolution Explorer (GALEX). GALEX is a NASA Small Explorer, whose mission was developed in cooperation with the Centre National d'Etudes Spatiales (CNES) of France and the Korean Ministry of Science and Technology. GALEX is operated for NASA by the California Institute of Technology under NASA contract NAS5-98034.

This publication makes use of data products from the Wide-field Infrared Survey Explorer (WISE), which is a joint project of the University of California, Los Angeles, and the Jet Propulsion Laboratory/California Institute of Technology, funded by the National Aeronautics and Space Administration.

We acknowledge the usage of the Extragalactic Distance Database\footnote{\url{http://edd.ifa.hawaii.edu/index.html}} \citep{Tully_etal_2009}, the HyperLeda database\footnote{\url{http://leda.univ-lyon1.fr}} \citep{Makarov_etal_2014}, the NASA/IPAC Extragalactic Database\footnote{\url{http://ned.ipac.caltech.edu}}, and the SAO/NASA Astrophysics Data System\footnote{\url{http://www.adsabs.harvard.edu}}.}

\facilities{ALMA, VLA, ATCA, Spitzer, WISE, GALEX}

\software{
\texttt{CASA} \citep{McMullin_etal_2007},
\texttt{Astropy} \citep{AstropyCollaboration_etal_2013,AstropyCollaboration_etal_2018}
}


\appendix

\section{Dynamical Equilibrium on Cloud Scales} \label{apdx:formulation}

\setcounter{equation}{0}

In Section~\ref{sec:expectation:PDE-cloud}, we determine the cloud-scale dynamical equilibrium pressure in molecular gas, $\PDEavg$, by adding up the weight of molecular clouds and the ambient atomic gas in the total gravitational potential. Here we provide a detailed derivation for each of the constituent terms included in this calculation.

\begin{figure}[thp]
\fig{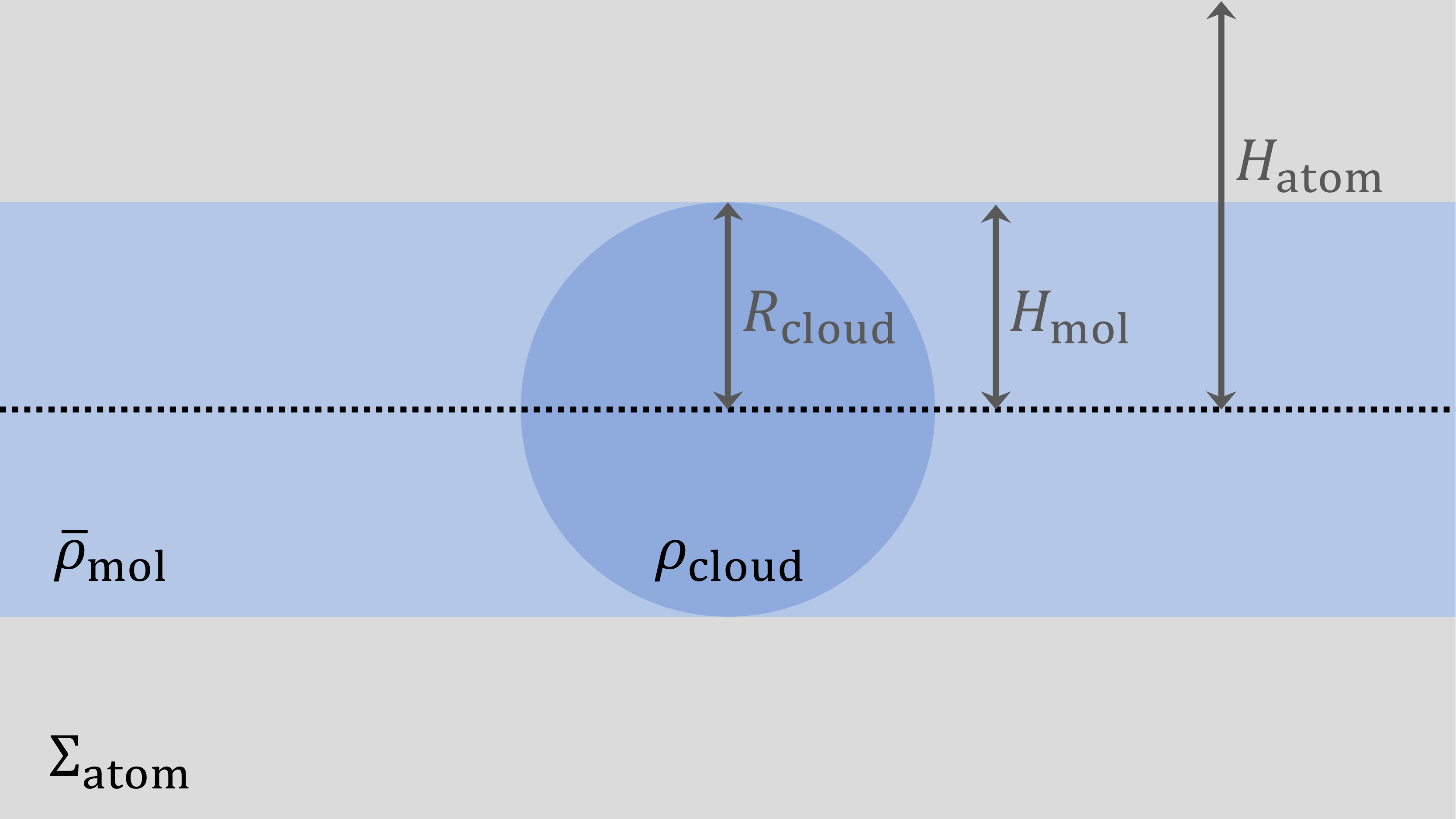}{0.47\textwidth}{}
\vspace{-1.5em}
\caption{
\textbf{A sketch showing the adopted geometrical model for the dynamical equilibrium calculation.}
We consider an arbitrary molecular cloud (dark blue) with radius $R\sbsc{cloud}$ and volume density $\rho\sbsc{cloud}$. This molecular cloud lives in a molecular gas layer (light blue), which has a half width of $H\sbsc{mol}=R\sbsc{cloud}$ and an average volume density of $\bar{\rho}\sbsc{mol}$ \textit{outside} the molecular cloud in focus. The molecular layer is ``sandwiched'' by an outer atomic gas layer (light gray), which has a half width of $H\sbsc{atom} > H\sbsc{mol}$ and an integrated surface density of $\Sigma\sbsc{atom}$. This entire multi-phase ISM is centered near the disk mid-plane (dotted black line), and its overall scale height is much smaller than that of the galaxy stellar disk.
}
\label{fig:model}
\end{figure}

As illustrated by Figure~\ref{fig:model}, we approximate the ISM in a galaxy disk as comprised of two components: 1) a thin, clumpy layer of molecular gas near the disk mid-plane, which includes many denser molecular clouds in it, and 2) a smooth, plane-parallel atomic gas outer layer ``sandwiching'' the molecular gas layer. We further assume that the vertical scale height of the molecular gas ($H\sbsc{mol}$) and the atomic gas ($H\sbsc{atom}$) are both much smaller than that of the stellar disk ($\Hstar$), with $H\sbsc{mol}<H\sbsc{atom}$. With this setup, we can consider an arbitrary molecular cloud in the molecular layer. To compute the equilibrium pressure at the center of the cloud, we first integrate the weight of the cloud from its center to its edge along the vertical direction, and then integrate the weight of the atomic gas above the cloud.

\subsection{The Weight of a Molecular Cloud}

The weight of a cloud, $\mathcal{W}\sbsc{cloud}$, includes three constituent parts:
\begin{equation}
    \mathcal{W}\sbsc{cloud} = \mathcal{W}\sbsc{cloud}\spsc{self} + \mathcal{W}\sbsc{cloud}\spsc{ext\text{-}mol} + \mathcal{W}\sbsc{cloud}\spsc{star}~.
\end{equation}
\noindent These three terms represent the weights due to the cloud's own self-gravity ($\mathcal{W}\sbsc{cloud}\spsc{self}$), to the gravity of external molecular gas outside the cloud ($\mathcal{W}\sbsc{cloud}\spsc{ext\text{-}mol}$), and to the gravity of stars ($\mathcal{W}\sbsc{cloud}\spsc{star}$). By symmetry, the outer atomic gas layers exert no gravity on any structure in the molecular layer.

For the cloud self-gravity term, we integrate the weight of a sphere with radius $R\sbsc{cloud}$ and constant density\footnote{Given our assumption that turbulent motions dominate the pressure budget inside molecular clouds, these clouds could still achieve an internal pressure-gravity balance by having a scale-dependent turbulent velocity field.} $\rho\sbsc{cloud}$:
\begin{align}
    \mathcal{W}\sbsc{cloud}\spsc{self}
    &= \int_0^{R\sbsc{cloud}} \rho\sbsc{cloud} \frac{G\rho\sbsc{cloud} (4/3)\pi r^3}{r^2}\,\mathrm{d}r \nonumber\\
    &= \frac{2\pi}{3}G\rho\sbsc{cloud}^2 R\sbsc{cloud}^2 \nonumber\\
    &= \frac{3\pi}{8}G\Sigma\sbsc{cloud}^2~.
\end{align}
\noindent The last step re-expresses $\mathcal{W}\sbsc{cloud}\spsc{self}$ in terms of the cloud surface density $\Sigma\sbsc{cloud} \equiv M\sbsc{cloud}/(\pi R\sbsc{cloud}^2)=(4/3)\rho\sbsc{cloud}R\sbsc{cloud}$.

For the external molecular gas gravity term, we approximate the distribution of all the molecular gas outside the cloud as a slab centered at the disk mid-plane. This slab has constant volume density $\bar\rho\sbsc{mol}$ anywhere outside the cloud, and has zero density within the extent of the cloud (i.e., it has a spherical ``hole'').
For simplicity, we further assume that the vertical half width of this slab is equal to the cloud radius, $H\sbsc{mol}=R\sbsc{cloud}$.
The weight of the cloud due to the gravity of this component is thus:
\begin{align}
    \mathcal{W}\sbsc{cloud}\spsc{ext\text{-}mol}
    &= \mathcal{W}\sbsc{cloud,slab} - \mathcal{W}\sbsc{cloud,hole} \nonumber\\
    &= \int_0^{R\sbsc{cloud}} \rho\sbsc{cloud}\, (4\pi G \bar\rho\sbsc{mol}z)\, \mathrm{d}z - \int_0^{R\sbsc{cloud}} \rho\sbsc{cloud} \frac{G\bar\rho\sbsc{mol} (4/3)\pi r^3}{r^2}\, \mathrm{d}r \nonumber\\
    &= 2\pi G \bar\rho\sbsc{mol} \rho\sbsc{cloud} R\sbsc{cloud}^2 - \frac{2\pi}{3} G \bar\rho\sbsc{mol} \rho\sbsc{cloud} R\sbsc{cloud}^2 \nonumber\\
    &= \frac{4\pi}{3} G \bar\rho\sbsc{mol} \rho\sbsc{cloud} R\sbsc{cloud}^2 \nonumber\\
    &= \frac{\pi}{2} G \bar\Sigma\sbsc{mol} \Sigma\sbsc{cloud}~.
\end{align}
\noindent The last step re-expresses  $\mathcal{W}\sbsc{cloud}\spsc{ext\text{-}mol}$ in terms of the cloud surface density $\Sigma\sbsc{cloud}$ and the average surface density of the slab $\bar\Sigma\sbsc{mol}=\bar\rho\sbsc{mol}\,(2H\sbsc{mol})=\bar\rho\sbsc{mol}\,(2R\sbsc{cloud})$.

For the stellar gravity term, given that $R\sbsc{cloud} \ll \Hstar$, we treat the stellar disk as having a uniform density near the mid-plane, $\rhostar$. We thus have:
\begin{align}
    \mathcal{W}\sbsc{cloud}\spsc{star}
    &= \int_0^{R\sbsc{cloud}} \rho\sbsc{cloud}\, (4\pi G \rhostar z)\, \mathrm{d}z \nonumber\\
    &= 2\pi G \rhostar \rho\sbsc{cloud} R\sbsc{cloud}^2 \nonumber\\
    &= \frac{3\pi}{2}G \rhostar \Sigma\sbsc{cloud} R\sbsc{cloud}~.
\end{align}

\subsection{The Weight of the Outer Atomic Gas Layer}

The weight of the atomic gas layer, $\mathcal{W}\sbsc{atom}$, also consists of three parts:
\begin{equation}
    \mathcal{W}\sbsc{atom} = \mathcal{W}\sbsc{atom}\spsc{self} + \mathcal{W}\sbsc{atom}\spsc{mol} + \mathcal{W}\sbsc{atom}\spsc{star}~.
\end{equation}
\noindent These terms represent the weights due to the atomic layer's self-gravity ($\mathcal{W}\sbsc{atom}\spsc{self}$), to the gravity of the inner molecular gas layer ($\mathcal{W}\sbsc{atom}\spsc{mol}$), and to the gravity of stars ($\mathcal{W}\sbsc{atom}\spsc{star}$). The calculation of the first and last terms is very similar to that of $\PDEkpc$ in Section~\ref{sec:expectation:PDE-kpc}. We simply quote the results here:
\begin{align}
    \mathcal{W}\sbsc{atom}\spsc{self} &= \frac{\pi}{2}G \Sigatom^2 \\
    \mathcal{W}\sbsc{atom}\spsc{star} &= \Sigatom \sqrt{2G\rhostar}\,\sigma\sbsc{atom,\,z};
\end{align}
Note that the second expression adopts the assumption that $H_\mathrm{mol} \ll H_\mathrm{atom}$.

The molecular gravity term, however, is slightly different from the version in the $\PDEkpc$ calculation. As the entire molecular gas inner layer is assumed to be ``sandwiched'' by the atomic gas outer layer in this calculation, the latter feels the full gravity of the former, such that:
\begin{equation}
    \mathcal{W}\sbsc{atom}\spsc{mol} = \pi G \bar\Sigma\sbsc{mol} \Sigatom~.
\end{equation}
\noindent In reality, the molecular and atomic medium (or at least the cold atomic phase) are often well mixed, so this estimate represents an upper limit for the true $\mathcal{W}\sbsc{atom}\spsc{mol}$ value.

\subsection{The Total Weight}

Combining all the above derivations together, we have:
\begin{align}
    \mathcal{W}\sbsc{total}
    =&\, \mathcal{W}\sbsc{cloud} + \mathcal{W}\sbsc{atom} \nonumber\\
    =&\, \Sigma\sbsc{cloud} \left( \frac{3\pi}{8}G\Sigma\sbsc{cloud} + \frac{\pi}{2}G \bar\Sigma\sbsc{mol} + \frac{3\pi}{2}G \rhostar R\sbsc{cloud} \right) \nonumber\\
    &+ \Sigma\sbsc{atom} \left( \frac{\pi}{2}G\Sigma\sbsc{atom} + \pi G \bar\Sigma\sbsc{mol} + \sqrt{2G\rhostar}\,\sigma\sbsc{atom,\,z} \right)~.
    \label{eq:W_total}
\end{align}
\noindent This gives the estimated total weight of a molecular cloud and the atomic gas above it in a galaxy disk, or equivalently, the required pressure in this molecular cloud to keep it under dynamical equilibrium.

\section{Stellar Disk Flattening Ratio} \label{apdx:flatratio}

\setcounter{equation}{0}

In this paper, we assume a constant stellar disk flattening ratio $\Rstar/\Hstar=7.3$ when estimating the stellar mass volume density near the disk mid-plane (see Section~\ref{sec:expectation:PDE-kpc}). This value is suggested by \citet{Kregel_etal_2002}, and has been widely adopted in recent studies on similar topics \citep[e.g.,][]{Leroy_etal_2008,Ostriker_etal_2010,Gallagher_etal_2018a}.

\citet{Kregel_etal_2002} derived this average $\Rstar/\Hstar$ value from careful analysis of 34 nearby, edge-on galaxies. To provide an improved estimate with better statistics, here we do a similar calculation for a much larger sample of 313 edge-on galaxies selected from the \SfourG\ sample \citep{Sheth_etal_2010}.

We select 313 edge-on galaxies from the \SfourG\ parent sample based on the structural decomposition results published by \citet{Salo_etal_2015}.
We pick galaxies in which \citet{Salo_etal_2015} fit the shape of an edge-on disk component (a.k.a., ``edgedisk'') that accounts for at least 50\% of the light of the galaxy.
We then calculate disk flattening ratios based on the exponential scale length and scale height of this flux-dominating ``edgedisk'' component, using the measurement in their data table. Note that in their structural decomposition analysis, \citet{Salo_etal_2015} assumed exponential density profile along both the radial and vertical direction. This is different from the isothermal vertical density profile assumed in this work (see Section~\ref{sec:expectation:PDE-kpc}). Rather than redoing their entire structural decomposition analysis, we instead use 0.5 times their (exponential) scale height values to approximate the scale height we would have measured assuming isothermal profile. This correction factor of 0.5 comes from the fact that the mid-plane stellar volume density is $\rhostar=\Sigstar/4\Hstar$ for an isothermal profile, and $\rhostar=\Sigstar/2\Hstar$ for an exponential profile \citep[see e.g.,][]{vanderKruit_1988}.

\begin{figure}[htp]
\fig{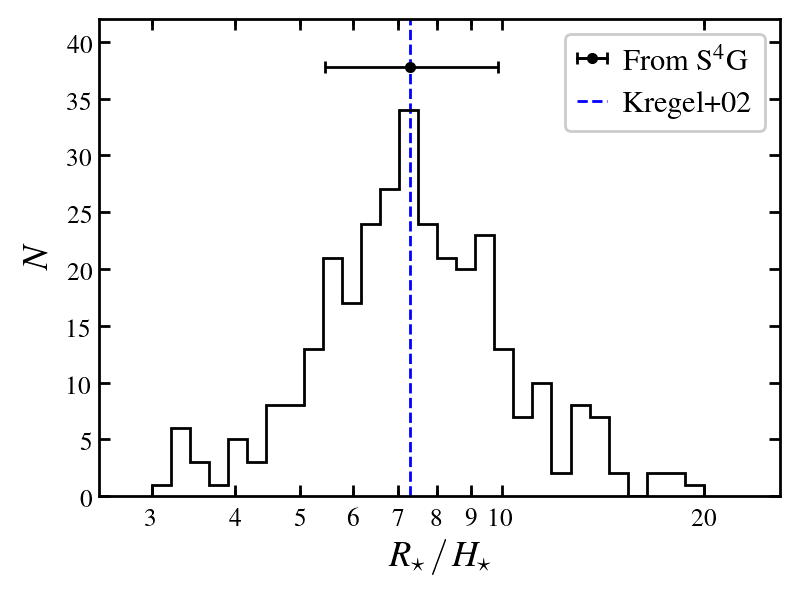}{0.47\textwidth}{}
\vspace{-2.5em}
\caption{
\textbf{Distribution of stellar disk flattening ratios ($\Rstar/\Hstar$), as derived for 313 edge-on galaxies based on the \SfourG\ galaxy structural decomposition catalog} \citep{Salo_etal_2015}.
The black dot with an error bar shows the median and 16--84\% range of the same distribution: $\Rstar/\Hstar=7.3^{+2.6}_{-1.9}$.
This agrees well with the $\Rstar/\Hstar=7.3$ value derived by \citet[blue dashed line]{Kregel_etal_2002} in a much smaller galaxy sample.
}
\label{fig:flattening_ratio}
\end{figure}

Figure~\ref{fig:flattening_ratio} shows the histogram of $\Rstar/\Hstar$ measured from the 313 \SfourG\ edge-on galaxies. We find that the median and 16--84\% range of this distribution (black dot with an error bar) is
\begin{equation}
    \Rstar/\Hstar=7.3^{+2.6}_{-1.9}~.
\end{equation}
This agrees perfectly with the $\Rstar/\Hstar=7.3$ value (blue dashed line) suggested by \citet{Kregel_etal_2002}, and supports the appropriateness of adopting this $\Rstar/\Hstar$ value for nearby disk galaxies.
The 16--84\% range of this distribution can also be translated to a 1$\sigma$ scatter of 0.13~dex in logarithmic space. This is the corresponding systematic uncertainty associated with this fixed $\Rstar/\Hstar$ ratio in all derived quantities that depend linearly on it (e.g., $\rhostarkpc$).

\section{Table of Key Measurements} \label{apdx:mrt}

\setcounter{equation}{0}

Table~\ref{tab:mrt} presents a collection of the key measurements derived in this work. Due to the space limit, in the print version we only include measurements in 10 apertures in the galaxy NGC~628. The full data table, which includes all measurements across the entire sample, is available in the online journal.

The content of the individual columns is as follows.
(1) \texttt{`Galaxy'}: name of the host galaxy;
(2) \texttt{`inDisk'}: whether each row corresponds to a ``disk'' aperture (see Section~\ref{sec:method:region});
(3) \texttt{`fCO120pc'}: CO flux recovery fraction on 120~pc scale (see Section~\ref{sec:method:fCO} and Figure~\ref{fig:Pturb-PDE-cloud}a);
(4) \texttt{`Pturb120pc'}: turbulent pressure in the molecular gas, measured on 120~pc scale (see Section~\ref{sec:expectation:Pturb} and Figure~\ref{fig:Pturb-PDE-cloud}a);
(5) \texttt{`PDE120pc'}: dynamical equilibrium pressure estimated on 120~pc scales (see Section~\ref{sec:expectation:PDE-cloud} and Figure~\ref{fig:Pturb-PDE-cloud}a);
(6) \texttt{`fCO60pc'}: CO flux recovery fraction on 60~pc scales (see Section~\ref{sec:method:fCO} and Figure~\ref{fig:Pturb-PDE-cloud}b);
(7) \texttt{`Pturb60pc'}: turbulent pressure in the molecular gas, measured on 60~pc scales (see Section~\ref{sec:expectation:Pturb} and Figure~\ref{fig:Pturb-PDE-cloud}b);
(8) \texttt{`PDE60pc'}: dynamical equilibrium pressure estimated on 60~pc scales (see Section~\ref{sec:expectation:PDE-cloud} and Figure~\ref{fig:Pturb-PDE-cloud}b);
(9) \texttt{`PDEkpc'}: dynamical equilibrium pressure estimated on kpc scales (see Section~\ref{sec:expectation:PDE-kpc} and Figure~\ref{fig:Pturb-PDE-kpc});
(10) \texttt{`PDEkpc11'}: dynamical equilibrium pressure estimated on kpc scales, assuming $\siggasz=11\,\uV$ (see Section~\ref{sec:discussion:SFR:PDE}, Figure~\ref{fig:SigSFR}b and Figure~\ref{fig:Rmol-PDE});
(11) \texttt{`SigSFRkpc'}: star formation rate surface density estimated on kpc scales (see Section~\ref{sec:method:phys:kpc} and Figure~\ref{fig:SigSFR});
(12) \texttt{`Rmolkpc'}: molecular-to-atomic gas ratio estimated on kpc scales (see Section~\ref{sec:discussion:Rmol} and Figure~\ref{fig:Rmol-PDE}).

\movetabledown=12em
\begin{rotatetable}
\begin{deluxetable}{cccccccccccc}
\tabletypesize{\footnotesize}
\tablecaption{Table of Key Measurements\label{tab:mrt}}
\colnumbers
\tablehead{
\colhead{\texttt{Galaxy}} &
\colhead{\texttt{inDisk}} &
\colhead{\texttt{fCO120pc}} &
\colhead{\texttt{Pturb120pc}} &
\colhead{\texttt{PDE120pc}} &
\colhead{\texttt{fCO60pc}} &
\colhead{\texttt{Pturb60pc}} &
\colhead{\texttt{PDE60pc}} &
\colhead{\texttt{PDEkpc}} &
\colhead{\texttt{PDEkpc11}} &
\colhead{\texttt{SigSFRkpc}} &
\colhead{\texttt{Rmolkpc}} \\[-1.5ex]
\colhead{ } &
\colhead{ } &
\colhead{ } &
\colhead{$[\uP]$} &
\colhead{$[\uP]$} &
\colhead{ } &
\colhead{$[\uP]$} &
\colhead{$[\uP]$} &
\colhead{$[\uP]$} &
\colhead{$[\uP]$} &
\colhead{$[\uSigSFR]$} &
\colhead{ }
}
\startdata
NGC0628 & 0 & 0.983 & 8.131e+04 & 1.717e+05 & 0.919 & 1.545e+05 & 1.581e+05 & 6.593e+04 & 1.048e+05 & 1.903e-02 & 1.749e+01 \\
NGC0628 & 0 & 0.962 & 4.969e+04 & 1.017e+05 & 0.896 & 1.280e+05 & 1.255e+05 & 3.600e+04 & 6.047e+04 & 1.207e-02 & 9.148e+00 \\
NGC0628 & 0 & 0.925 & 5.763e+04 & 1.011e+05 & 0.861 & 1.298e+05 & 1.136e+05 & 4.345e+04 & 6.780e+04 & 1.288e-02 & 6.907e+00 \\
NGC0628 & 0 & 0.923 & 1.195e+05 & 1.489e+05 & 0.829 & 2.743e+05 & 1.959e+05 & 3.832e+04 & 5.569e+04 & 1.383e-02 & 7.988e+00 \\
NGC0628 & 0 & 0.922 & 8.926e+04 & 1.288e+05 & 0.838 & 1.778e+05 & 1.388e+05 & 3.792e+04 & 5.817e+04 & 1.156e-02 & 9.519e+00 \\
NGC0628 & 0 & 0.966 & 1.095e+05 & 1.460e+05 & 0.866 & 2.306e+05 & 1.768e+05 & 3.416e+04 & 5.070e+04 & 1.043e-02 & 8.183e+00 \\
NGC0628 & 0 & 0.869 & 4.755e+04 & 7.255e+04 & 0.721 & 9.907e+04 & 8.932e+04 & 3.018e+04 & 4.894e+04 & 1.113e-02 & 8.417e+00 \\
NGC0628 & 1 & 0.817 & 1.530e+04 & 2.377e+04 & 0.491 & 3.771e+04 & 2.922e+04 & 1.776e+04 & 3.061e+04 & 6.582e-03 & 5.322e+00 \\
NGC0628 & 1 & 0.812 & 1.451e+04 & 2.383e+04 & 0.500 & 3.025e+04 & 2.799e+04 & 2.136e+04 & 3.360e+04 & 7.466e-03 & 3.542e+00 \\
NGC0628 & 1 & 0.819 & 2.781e+04 & 4.654e+04 & 0.662 & 7.480e+04 & 6.645e+04 & 2.127e+04 & 3.508e+04 & 8.680e-03 & 4.681e+00 \\
... & ... & ... & ... & ... & ... & ... & ... & ... & ... & ... & ... \\
\enddata
\tablecomments{This table is available in its entirety in the \href{https://www.canfar.net/storage/list/phangs/RELEASES/Sun_etal_2020}{PHANGS CADC storage}.}
\end{deluxetable}
\end{rotatetable}


\bibliography{main.bib}


\end{document}